%% file: 00mainAIP.tex
\documentclass[aip, amsmath, amssymb, reprint, longbibliography]{revtex4-2}
\usepackage{graphicx}
\usepackage{dcolumn}
\usepackage{bm}
\usepackage[utf8]{inputenc}
\usepackage[T1]{fontenc}
\usepackage{mathptmx}
\usepackage{etoolbox}
\usepackage{caption}
\usepackage{subcaption}
\usepackage{amsmath}
\usepackage[version=4]{mhchem}
\usepackage{siunitx}
\usepackage{natbib, hyperref}
\hypersetup{colorlinks,allcolors=black}
\usepackage{color,soul}
\usepackage{subcaption}
\usepackage{cleveref}

\makeatletter
\def\@email#1#2{%
 \endgroup
 \patchcmd{\titleblock@produce}
  {\frontmatter@RRAPformat}
  {\frontmatter@RRAPformat{\produce@RRAP{*#1\href{mailto:#2}{#2}}}\frontmatter@RRAPformat}
  {}{}
}

\begin{document}
%%%%%%%%%%%%%%%%%%%%%%%%%%%%%%%%%%%%%%%%%%%%%%%%%%%%%%%%%%%%%%%%%%%%%%%%%%%%%%%%%%%%%%%%%%%%%%%%%%%
\title{Mutual Interactions Between a Thin Flexible Panel and Supersonic Flows}

\author{Himakar Ganti}
\affiliation{Department of Aerospace Engineering, University of Cincinnati, Cincinnati, OH 45221-0070}
\affiliation{Hypersonics Laboratory, Digital Futures, University of Cincinnati, Cincinnati, OH 45206}
\author{Luis Bravo}
\affiliation{Army Research Directorate, DEVCOM Army Research Laboratory, Aberdeen Proving Ground, MD, 21005}
\author{Anindya Ghoshal}
\affiliation{Army Research Directorate, DEVCOM Army Research Laboratory, Aberdeen Proving Ground, MD, 21005}
\author{Prashant Khare}
\email{Prashant.Khare@uc.edu}
\affiliation{Department of Aerospace Engineering, University of Cincinnati, Cincinnati, OH 45221-0070}
\affiliation{Hypersonics Laboratory, Digital Futures, University of Cincinnati, Cincinnati, OH 45206}

\date{\today}

%%%%%%%%%%%%%%%%%%%%%%%%%%%%%%%%%%%%%%%%%%%%%%%%%%%%%%%%%%%%%%%%%%%%%%%%%%%%%%%%%%%%%%%%%%%%%%%%%%%
\begin{abstract}
\input{01abstract.tex}
\end{abstract}

%%%%%%%%%%%%%%%%%%%%%%%%%%%%%%%%%%%%%%%%%%%%%%%%%%%%%%%%%%%%%%%%%%%%%%%%%%%%%%%%%%%%%%%%%%%%%%%%%%%
\maketitle
%\tableofcontents

%%%%%%%%%%%%%%%%%%%%%%%%%%%%%%%%%%%%%%%%%%%%%%%%%%%%%%%%%%%%%%%%%%%%%%%%%%%%%%%%%%%%%%%%%%%%%%%%%%%
\section{Nomenclature}
\input{02nomen.tex}

%%%%%%%%%%%%%%%%%%%%%%%%%%%%%%%%%%%%%%%%%%%%%%%%%%%%%%%%%%%%%%%%%%%%%%%%%%%%%%%%%%%%%%%%%%%%%%%%%%%
\newpage
\section{Introduction}\label{sec:intro}
\input{03intro.tex}
\section{Governing Equations and Numerical Methods}
\input{04govern.tex}

%%%%%%%%%%%%%%%%%%%%%%%%%%%%%%%%%%%%%%%%%%%%%%%%%%%%%%%%%%%%%%%%%%%%%%%%%%%%%%%%%%%%%%%%%%%%%%%%%%%
\section{Model Validation}
\input{05exptvalgridsens.tex}

%%%%%%%%%%%%%%%%%%%%%%%%%%%%%%%%%%%%%%%%%%%%%%%%%%%%%%%%%%%%%%%%%%%%%%%%%%%%%%%%%%%%%%%%%%%%%%%%%%%
\section{Effect of Mach Number on Fluid Structure Interactions}
\input{06results.tex}
\section{Conclusions}
\input{07conclusion.tex}

%%%%%%%%%%%%%%%%%%%%%%%%%%%%%%%%%%%%%%%%%%%%%%%%%%%%%%%%%%%%%%%%%%%%%%%%%%%%%%%%%%%%%%%%%%%%%%%%%%%
\section{Acknowledgments}
This research was supported by the DEVCOM Army Research Laboratory grant, W911NF-22-2-0058.  The authors appreciate the High-Performance Computing Modernization Program (HPCMP) resources and support provided by the Department of Defense Supercomputing Resource Center (DSRC) as part of the 2022 Frontier Project, Large-Scale Integrated Simulations of Transient Aerothermodynamics in Gas Turbine Engines. The views and conclusions contained in this document are those of the authors and should not be interpreted as representing the official policies or positions, either expressed or implied, of the DEVCOM Army Research Laboratory or the U.S. Government. The U.S. Government is authorized to reproduce and distribute reprints for Government purposes notwithstanding any copyright notation herein.

%%%%%%%%%%%%%%%%%%%%%%%%%%%%%%%%%%%%%%%%%%%%%%%%%%%%%%%%%%%%%%%%%%%%%%%%%%%%%%%%%%%%%%%%%%%%%%%%%%%
\section{References}
\bibliography{refs/00mainAIP}

%%%%%%%%%%%%%%%%%%%%%%%%%%%%%%%%%%%%%%%%%%%%%%%%%%%%%%%%%%%%%%%%%%%%%%%%%%%%%%%%%%%%%%%%%%%%%%%%%%% 
\end{document}

%% file: 01abstract.tex
%%%%%%%%%%%%%%%%%%%%%%%%%%%%%%%%%%%%%%%%%%%%%%%%%%%%%%%%%%%%%%%%%%%%%%%%%%%%%%%%%%%%%%%%%%%%%%%%%%%
% Abstract:
%%%%%%%%%%%%%%%%%%%%%%%%%%%%%%%%%%%%%%%%%%%%%%%%%%%%%%%%%%%%%%%%%%%%%%%%%%%%%%%%%%%%%%%%%%%%%%%%%%%
This paper discusses the mutual interactions between a thin flexible aluminum plate and supersonic flow using two-dimensional (2D) numerical simulations. Calculations are performed using an open source library, SU2, that solves partial differential equations governing fluid and structural dynamics.  The configuration considered in this research effort is based on an experiment in which a thin flexible panel of $1.02$ $mm$ with a $50.8$ $mm$ overhang at the outer edge of a backward facing step is exposed to Mach 2 flow. The computational framework was first validated against measurements for both the initial transients of $10$ $ms$ and the fully started conditions at $0.4$ $s$. Then, numerical studies were performed to analyze the fluid-structure interactions at  4 different Mach numbers between $0.5$ and $3$. The flow behavior revealed distinct phenomena, including shear layer separation for subsonic and transonic flows, and a fully enclosed recirculation region under the overhang in supersonic cases. The time-averaged flow field identified potential temperature hotspots during the initial transients, which intensified as time evolved. For Mach $0.50$, the amplitude of the thin panel oscillations increased as the flow transitioned from transient to steady-state conditions. In the transonic case (M = $0.95$), the oscillation amplitude became significantly larger, potentially leading to resonant behavior and structural failure (we did not model failure). However, in the supersonic cases, the oscillations stabilized and were sustained after the initial transients. The research quantitatively identifies the influence of the Mach number on the fluid-structure interaction phenomena, which affect pressure loads and the development of thermal hotspots, which are crucial elements in engineering design.

%%%%%%%%%%%%%%%%%%%%%%%%%%%%%%%%%%%%%%%%%%%%%%%%%%%%%%%%%%%%%%%%%%%%%%%%%%%%%%%%%%%%%%%%%%%%%%%%%%%

%% file: 02nomen.tex
%%%%%%%%%%%%%%%%%%%%%%%%%%%%%%%%%%%%%%%%%%%%%%%%%%%%%%%%%%%%%%%%%%%%%%%%%%%%%%%%%%%%%%%%%%%%%%%%%%%
% Nomenclature:
%%%%%%%%%%%%%%%%%%%%%%%%%%%%%%%%%%%%%%%%%%%%%%%%%%%%%%%%%%%%%%%%%%%%%%%%%%%%%%%%%%%%%%%%%%%%%%%%%%%
\begin{table}[h]
\begin{tabular}{lcl}
FSI    & & Fluid-Structure-Interaction \\
2D     & & 2-dimensional \\
3D     & & 3-dimensional \\
SWBLI  & & shock-wave boundary-layer interactions \\
ROM    & & reduced order modeling \\
POD    & & Proper Orthogonal Decomposition \\
SPOD   & & Spectral Proper Orthogonal Decomposition \\
RANS   & & Reynolds-Averaged Navier-Stokes \\
LES    & & Large Eddy Simulations \\
WMLES  & & Wall Modeled Large Eddy Simulations \\
SGS    & & Sub-Grid-Scale \\
SM     & & Smagorinsky Model \\
CM     & & Clark Model \\
DCM    & & Dynamic Clark Model \\
DNS    & & Direct Numerical Simulation \\
VSGSM  & & Vreman Sub-Grid-Scale Model \\
DDES   & & Delayed Detached Eddy Simulation \\
SA     & & Spalart-Allmaras \\
FEM    & & Finite Element Method \\
FVM    & & Finite Volume Method \\
MUSCL  & & Monotonic Upwinding Centered Scalar Conservation Law \\
FGMRES & & Flexible Generalized Minimum Residual \\
FFT    & & Fast Fourier Transform \\

\end{tabular}
\end{table}

%%%%%%%%%%%%%%%%%%%%%%%%%%%%%%%%%%%%%%%%%%%%%%%%%%%%%%%%%%%%%%%%%%%%%%%%%%%%%%%%%%%%%%%%%%%%%%%%%%%

%% file: 03intro.tex
%%%%%%%%%%%%%%%%%%%%%%%%%%%%%%%%%%%%%%%%%%%%%%%%%%%%%%%%%%%%%%%%%%%%%%%%%%%%%%%%%%%%%%%%%%%%%%%%%%%
% Introduction:
%%%%%%%%%%%%%%%%%%%%%%%%%%%%%%%%%%%%%%%%%%%%%%%%%%%%%%%%%%%%%%%%%%%%%%%%%%%%%%%%%%%%%%%%%%%%%%%%%%%
Computational investigation of mutual interaction between structures and fluids (FSI), sometimes also referred to as aeroelasticity, has been a topic of interest for many decades \citep{dowell2001modeling, griffith2020immersed, smith2000evaluation, hou2012numerical}, especially in weakly compressible or incompressible low-speed flows relevant to a range of applications, including physiological flows \citep{wang2022fluid, heil2011fluid, kleinstreuer2007fluid, hirschhorn2020fluid}, turbines \citep{trivedi2017fluid} and flow in pipes \citep{tijsseling1996fluid}. There is also a rich literature that details aeroelastic interactions in aircraft and helicopters \citep{schuster2003computational, ajaj2021recent, garrick1976aeroelasticity, friedmann1999renaissance, friedmann2004rotary} that move primarily in the subsonic/transonic flow regime. Recent interest in supersonic and hypersonic flight has not only led to increased research activities on high-speed fluid dynamics \citep{redding2024analysis, redding2023thermochemical, gamertsfelder2023flow} and gaseous/liquid-fueled combustion processes \citep{kamin2023nexgen, kamin2024nexgen, kamin2024large, tripathi2022interactions, redding2022computational, kamin2022effect, valencia2024flow, illacanchi2023numerical}, but also spurred FSI research in the supersonic and hypersonic flow regime, and several experiments and computational studies have been conducted to investigate this phenomenon \citep{sullivan2020direct, hoy2022fluid, boustani2022immersed}. In particular, canonical configurations, such as supersonic/hypersonic flow over thin flexible panels that resemble the interactions between high-speed flow and the inlet of scramjet engines, have been a popular area of investigation in the recent past because of their importance to practical systems of interest; a brief summary of the literature that discusses these studies is summarized in the following paragraphs. This is also the topic of the current article, which details the interactions between a thin panel and a range of flow conditions from sub-, trans-, to supersonic flow. Unlike many studies in the past that investigated laminar-flow conditions while investigating FSI in supersonic-flow conditions, we took into account turbulent flow using a wall-modeled large-eddy simulation (WMLES) framework, loosely coupled with a structural solver to elucidate the governing physical processes in both the fluid and solid domains.

% EXPERIMENTS - CURRENT PROGRESS:
As mentioned in the previous paragraph, much progress has been made over the past decade in understanding the fundamental FSI phenomenon related to high-speed flows through well-controlled experiments. These experiments can generally be classified into two kinds depending on how the structure is held in place as it interacts with supersonic/hypersonic flow: (1) cantilevered rigid compliant panel\cite{RN27, RN28, RN31, RN19}, and (2) edge-clamped panel \cite{RN26, RN29, RN30, RN32, RN33, RN34, RN35, RN36, RN38, RN39, RN40}. While the first configuration consists of shock-free flow parallel to the cantilevered panel (shocks may appear as a result of geometric or boundary/shear-layer effects), the second configuration is designed to have a shock impinge on the surface of the panel; thus both setups investigate slightly different processes. This research effort focuses on the former and investigates the detailed flow and structural dynamics.
\begin{figure}
        \centering
        \includegraphics[width=\linewidth]{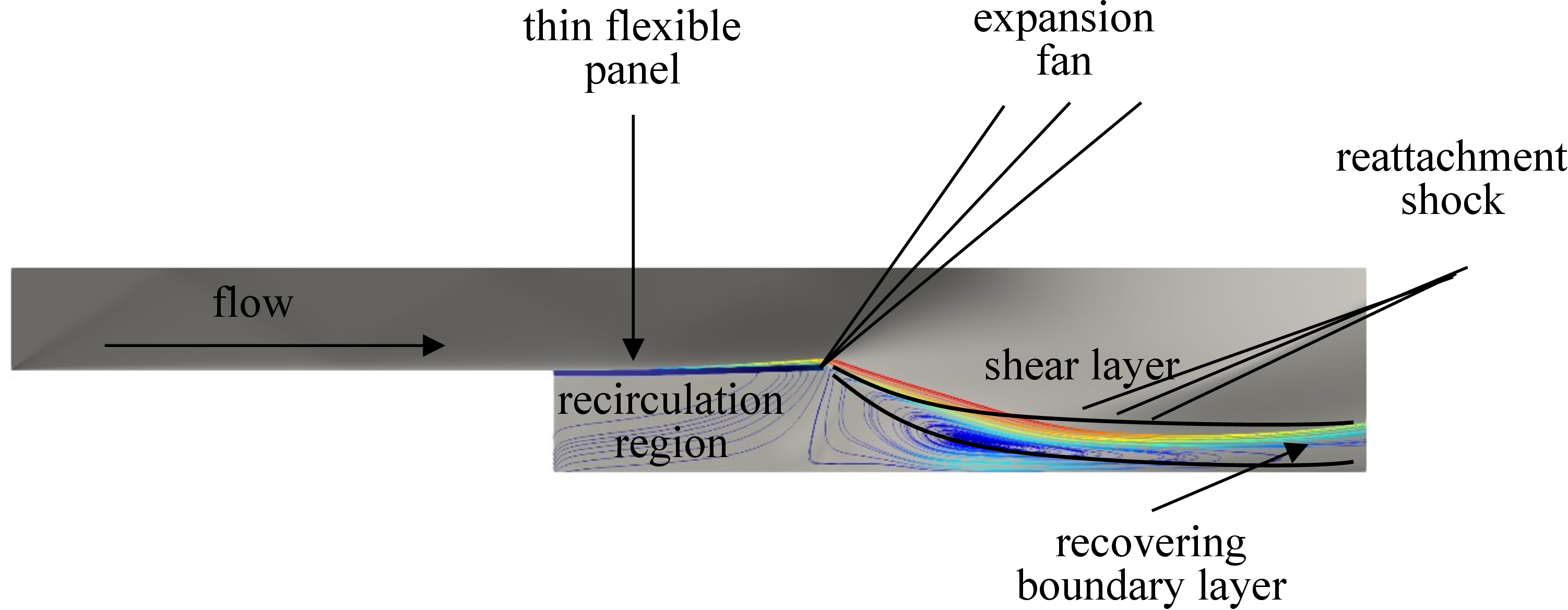}  
        \caption{Schematic of the experimental configuration used to study the interactions between a cantilevered panel and supersonic flows.}
    \label{fig:bojan_schematic}
\end{figure}

The experimental setup of \citet{RN19} represents the interactions between high-speed flow and a cantilevered-compliant panel. In this experiment, a schematic of which is shown in figure \ref{fig:bojan_schematic}, a 50 mm aluminum panel of 1.02 mm width is attached as a cantilever and exposed to a flow at a Mach number of  2.11. This canonical configuration encompasses the rich physics relevant to practical high-speed vehicles, including but not limited to shock and expansion waves, shear layers, shock-boundary layer interactions, recirculation zones, and the mutual interactions between these processes and the solid structure. High-speed digital imaging and particle image velocimetry (PIV) were used to capture flow and structural responses. The flow data were time-averaged for both transient and fully started conditions, while the thin panel's behavior was documented through the time evolution of the beam's displacement. Oscillatory behavior was monitored using pressure probes placed in the separated recirculation region. 
\begin{figure}
        \centering
        \includegraphics[width=\linewidth]{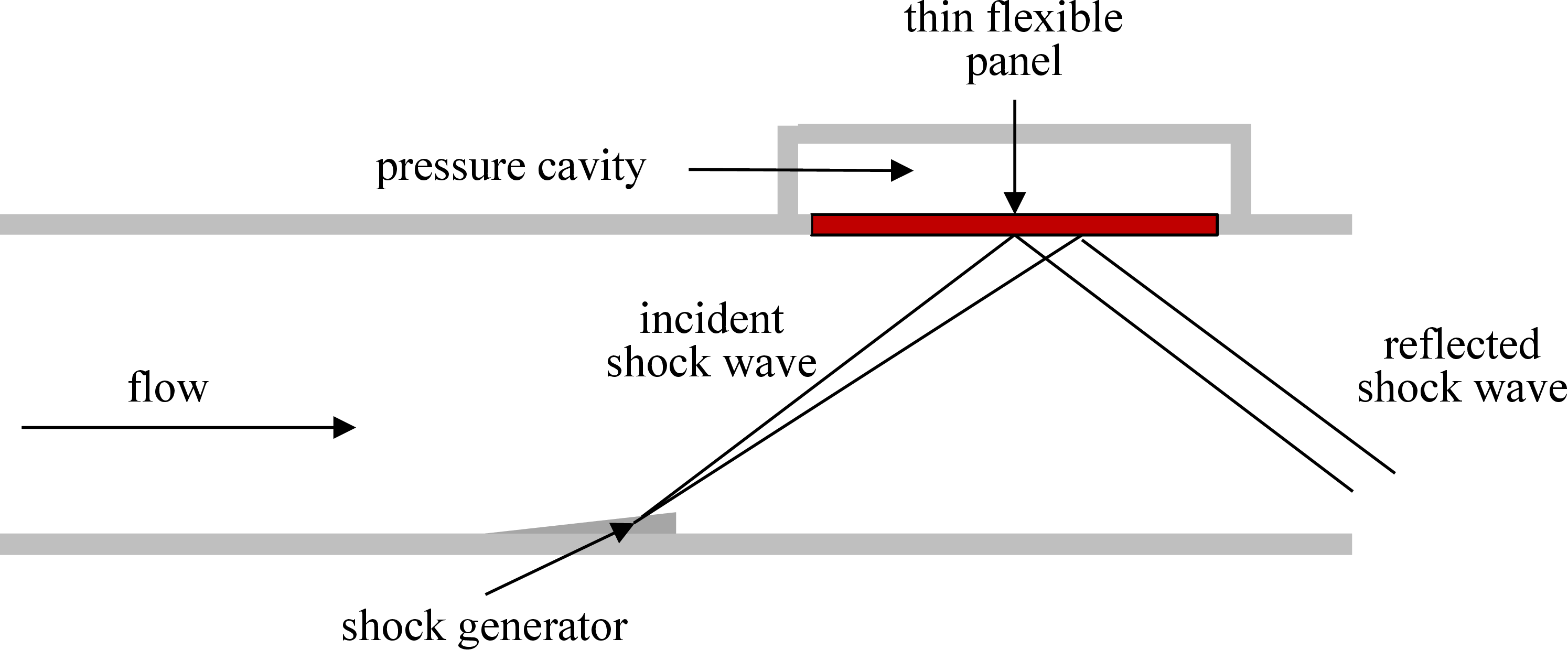}
    \caption{Schematic of the experimental configuration used to study the interactions between an edge-clamped panel and supersonic flows.}
    \label{fig:brouwer-schematic}
\end{figure}

\citet{RN36} conducted experiments on an edge-clamped, thin compliant panel in a Mach 1.5 to 3.0 supersonic wind tunnel, schematically shown in figure \ref{fig:brouwer-schematic}. The panel featured a pressure cavity on the opposite side of the flow tunnel, equipped with multiple transducers along its length to measure pressure and temperature profiles. The turbulent supersonic flow excited the panel, and a shock generator induced shock impingement at various points along the span. The study provides a detailed analysis of the panel's behavior, utilizing multiple pressure and temperature measurements and modal decomposition to determine mode shapes and oscillation frequencies. As a follow-up to this study, recently, \citet{RN40} conducted numerical simulations using the Reynolds-averaged Navier-Stokes (RANS) approach and developed a reduced-order model of the structural behaviors of the thin plate. Although the mean transport fields effectively captured bulk flow properties, the authors highlighted that higher-fidelity scale-resolving models are necessary to accurately capture boundary layer interactions and improve predictions of pressure loading on the panel.

To fully grasp such interactions in question, it is essential to simultaneously measure or model the key physical processes involved, such as structural dynamics, including buckling, conjugate heat transfer, highly compressible turbulent flow, thermochemical non-equilibrium (relevant in hypersonic flow), and shock-boundary layer interactions. Moreover, the complexity of investigating these phenomena is heightened because of the broad range of lengths (from sub-micrometer to meter) and time scales (from nanosecond to second) that dictate them. Although several studies have focused on oscillation, buckling, and deflection dynamics under supersonic \citep{RN26, RN36} or hypersonic \citep{RN32, RN27, RN28, RN31, RN33} conditions based on cantilevered or edge-clamped configurations, most of these studies, irrespective of the mode of investigation (experimental measurements or computations), discussed the structural deformation modes, primarily based on pressure and/or thermal loading (primarily for hypersonic flow conditions) loading due to the flow conditions.

A common limitation in experimental or numerical research on this topic has been the reliance on time-averaged data for fluid and solid domains, overlooking the dynamics and turbulent flow. In this paper, we attempt to address these gaps by conducting computations based on unsteady wall-modeled large eddy simulations to investigate the interactions between a cantilevered thin beam as it is exposed to a range of Mach numbers from 0.5 to 3.

The article is organized as follows: the equations that govern the fluid and structural dynamics are described in detail, including the subgrid-scale (SGS) models that we use for turbulence closure. This is followed by a detailed validation and grid sensitivity study to establish the appropriateness and accuracy of our numerical framework. To do so, we simulate the experiment of \citet{RN19} and compare our results with measurements during both the initial transients and the fully started conditions. For this case, the flow Mach number is $2.11$. Next, mutual interactions between flow and structure at a range of flow Mach numbers in the subsonic (M = 0.5), transonic (M = 0.95), and supersonic (M = 2.11 and 3.0) regimes are discussed. In the last section, we summarize the major conclusions from the study.

%% file: 04govern.tex
%%%%%%%%%%%%%%%%%%%%%%%%%%%%%%%%%%%%%%%%%%%%%%%%%%%%%%%%%%%%%%%%%%%%%%%%%%%%%%%%%%%%%%%%%%%%%%%%%%%
% Governing Equations:
%%%%%%%%%%%%%%%%%%%%%%%%%%%%%%%%%%%%%%%%%%%%%%%%%%%%%%%%%%%%%%%%%%%%%%%%%%%%%%%%%%%%%%%%%%%%%%%%%%%
The SU2 multiphysics library \citep{RN11} is used to investigate the FSI phenomenon described above. In this section, the equations that govern the processes in the fluid and solid domains, their interfacial interactions, and associated numerical methods are described. 

%%%%%%%%%%%%%%%%%%%%%%%%%%%%%%%%%%%%%%%%%%%%%%%%%%%%%%%%%%%%%%%%%%%%%%%%%%%%%%%%%%%%%%%%%%%%%%%%%%%
\subsection{Fluid Domain}
The Favre averaged conservation equations for mass, momentum and energy are given by:
\begin{equation}
\begin{gathered}
\frac{{\partial \bar \rho }}{{\partial t}} + \frac{{\partial \bar \rho \tilde u_i }}{{\partial x_i }} = 0 \\
\frac{{\partial \bar \rho \tilde u_i }}{{\partial t}} + \frac{{\partial \bar \rho \tilde u_i \tilde u_j }}{{\partial x_i }} = - \frac{{\partial \bar p}}{{\partial x_i }} + \frac{{\partial (\bar \tau _{ij}  - \tau _{ij}^{SGS} )}}{{\partial x_i }} \\
\frac{{\partial \bar \rho \tilde E}}{{\partial t}} + \frac{{\partial [(\bar \rho \tilde E + \bar p)\tilde u_i ]}}{{\partial x_i }} = \frac{{\partial ( - \bar q_i  + \tilde u_j \bar \tau _{ji}  - \sigma _i^{SGS}  - H_i^{SGS} )}}{{\partial x_i }} \\
\end{gathered}
\end{equation}

\noindent where $\rho$ is the density, $u_{i}$ is the velocity along the direction $x_i$, $p$ is the pressure. The filtered mean shear stress is $\bar \tau _{ij}$, and $\bar q_{i}$ is the heat flux or thermal energy flux. The filtered total energy, $\bar E$ is the sum of filtered internal energy, $\bar e$, resolved kinetic energy, $\frac{1}{2}\bar u_i \bar u_i$, and the SGS kinetic energy, $k^{SGS}$ are given as
\begin{equation}
\begin{gathered}
\bar E = \bar e + \frac{1}{2}\bar u_i \bar u_i  + k^{SGS} \\
k^{SGS}  = \frac{1}{2}\left( {\overline {u_i u_i }  - \bar u_i \bar u_i } \right) \\
\end{gathered}
\end{equation}

\noindent Conduction, $q_i$ is modeled using Fourier's law, where the thermal conductivity is calculated using the local temperature-dependent molecular viscosity and a Prandtl number of 0.72.  
% and thermal conductivity, $\kappa$, are given in terms of temperature difference, $\Delta T$, effective viscosity, $\mu$ and the Prandtl number, $Pr$ as:
% \begin{equation}
% \begin{gathered}
% q_i  = \kappa \Delta T \\
% \kappa  = \frac{{\mu _d c_p }}{{Pr }} \\
% \end{gathered}
% \end{equation}

% \noindent The effect of turbulence is through an increase in viscosity, leading to an effective viscosity with a dynamic and turbulent component. $Pr_d$ is the dynamic or resolved Prandtl number and $Pr_t$ is the unresolved turbulent Prandtl number. The dynamic and turbulent viscosities lead to the an increase in the thermal conductivity leading to a contribution from both as:
% \begin{equation}
% \begin{gathered}
% \mu  = \mu _d  + \mu _t \\
% \kappa  = \frac{{\mu _d c_p }}{{Pr_d }} + \frac{{\mu _t c_p }}{{Pr_t }} \\
% \end{gathered}
% \end{equation}

\noindent The temperature-dependent dynamic viscosity is evaluated using the Sutherland model. %, while the turbulent viscosity $\mu_t$ is is obtained from the SGS turbulence model. 
For a reference temperature $T_0$, reference viscosity $\mu_0$, and Sutherland constant $S_{\mu}$, the Sutherland model for dynamic viscosity is:
\begin{equation}
\mu  = \mu _0 \left( {\frac{T}{{T_0 }}} \right)^{{3 \mathord{\left/
 {\vphantom {3 2}} \right.
 \kern-\nulldelimiterspace} 2}} \left( {\frac{{T_0  + S_\mu  }}{{T + S_\mu  }}} \right)
\end{equation}

The unclosed SGS terms appearing in the Favre averaged equations are the SGS stress $\tau _{i,j}^{SGS}$, the energy flux $H_{i,j}^{SGS}$. The term $\sigma _{i,j}^{SGS}$ is obtained from the correlations of the velocity field with the viscous stress tensor.
\begin{equation}
\begin{gathered}
\tau _{i,j}^{SGS} = (\overline {\rho u_i u_j }  - \overline \rho  \widetilde{u_i }\widetilde{u_j }) \\
H_{i,j}^{SGS} = (\overline {\rho E_i u_i }  - \overline \rho  \widetilde{E_i }\widetilde{u_i }) + (\overline {pu_i }  - \overline p \widetilde{u_i }) \\
\sigma _{i,j}^{SGS} = (\overline {u_j \tau _{ij} }  - \widetilde{u_j }\widetilde{\tau _{ij} }) \\
\end{gathered}
\end{equation}

 % The Vreman SGS CLosure MOdel:
$\tau_{i,j}^{SGS}$ is closed using using the the Vreman SGS Model (VSGSM) \cite{Vreman1994direct}, given by:
\begin{equation}
\begin{gathered}
\tau _{ij}  =  - 2\nu _e S_{ij}  + \frac{1}{3}\tau _{kk} \delta _{ij} \\
\bar S_{ij}  = \frac{1}{2}\left( {\frac{{\partial u_j }}{{\partial x_i }} + \frac{{\partial u_i }}{{\partial x_j }}} \right) \\
\nu _e  = c\sqrt {\frac{{B_\beta  }}{{\alpha _{ij} \alpha _{ij} }}} \\
\alpha _{ij}  = \partial _i u_j  = \frac{{\partial u_j }}{{\partial x_i }} \\
\beta _{ij}  = \Delta _m^2 \alpha _{mi} \alpha _{mj} \\
B_\beta   = \beta _{11} \beta _{22}  + \beta _{11} \beta _{33}  + \beta _{22} \beta _{33}  - \beta _{12}^2  - \beta _{13}^2  - \beta _{23}^2 \\
\end{gathered}
\end{equation}

\noindent where the model constant, $c$, is related to the Smagorinsky constant \cite{RN17}, $C_s$, as $c \approx 2.5C_s^2$. This model needs only a local filter width and first-order derivatives of the velocity field. $\alpha$ is the matrix of derivatives of the filtered velocity, $\bar u$, the filter width is $\Delta _m$, $B_{\beta}$ is an invariant of the tensor, $\beta$. This implies that if $\Delta _i  = \Delta$ then $\beta  = \Delta ^2 \alpha ^T \alpha$. Although we realize the importance of $H_{i,j}^{SGS}$ and $\sigma _{i,j}^{SGS}$, especially for highly compressible flows, because of the second to third order effects of these terms on structural dynamics, in this study we neglect these SGS terms.

To model the boundary layer, the LES framework described above is coupled to the algebraic wall stress model of \citet{reichardt1951}. However, we realize that this and other such models (e.g., \citet{spalding1961single, musker1979explicit}) start to break down when Mach numbers are high; since this study is limited to Mach numbers of up to 3, we use the model of \citet{reichardt1951} in this research investigation. For higher Mach-number flows, the topic of our upcoming manuscript, we are implementing the Van Driest transformation in conjunction with one of the aforementioned wall models to model the boundary layer accurately.

\subsection{Solid Domain}
The solid domain is governed by elasticity equations for large deformations \citep{RN18}, implemented in the SU2 library by \citet{RN9}. A short description of the Venant-Kirchhoff model is given here. This model can handle large deformations of an isotropic and homogeneous solid in a Lagrangian framework. The elasticity equations are solved in differential form as:
\begin{equation}
\rho _s \frac{{\partial ^2 \bf{u}}}{{\partial t^2 }} = \nabla \left( {{\bf{F}} \cdot {\bf{S}}} \right) + \rho _s f
\end{equation}

\noindent The boundary conditions are imposed on the interface degrees of freedom, which are categorized into essential (e) and natural (n) conditions. The essential boundary conditions impose the solution $u_{s,e}$ on the interface nodes and later apply the tractions $\lambda_{s,n}$ on the boundary. The boundary conditions for the solid are:
\begin{equation}
\left\{ \begin{array}{l}
 {\rm{   }}{\mathop{\rm u}\nolimits} _s  = {\mathop{\rm u}\nolimits} _{s,e} \quad \qquad {\rm{ on }}\quad \Gamma _{{\rm{s,e}}}  \\ 
 \sigma _s {\rm{n}}_{\rm{s}}  = \lambda _{s,n} \qquad {\rm{ on }}\quad \Gamma _{{\rm{s,n}}}  \\ 
 \end{array} \right.
\end{equation}

\noindent where $\rho_s$ is the density of the solid, $\bf{u}$ are the displacements of the solid, $t$ is the time, $\bf{F}$ is the deformation gradient of the material and $f$ is the volume force. $\bf{S^{PK}}$ is the second Piola-Kirchoff stress tensor:
\begin{equation}
S^{PK} _{ij}  = \lambda _s E_{kk} \delta _{ij}  + 2\mu _s E_{ij} 
\end{equation}

\noindent $\delta_{ij}$ is the Kronecker delta, $\lambda$ and $\mu$ are Lam\'e's constants. The Lagrangian stress tensor, $E_{ij}$, is given as:
\begin{equation}
E_{ij}  = \frac{1}{2}\left( {\frac{{\partial u_i }}{{\partial x_j }} + \frac{{\partial u_j }}{{\partial x_i }}} \right) + \frac{1}{2}\frac{{\partial u_k }}{{\partial x_i }}\frac{{\partial u_k }}{{\partial x_j }}
\end{equation}

\noindent Lam\'e's constants $\lambda_S$ and $\mu_s$ are given in terms of Young's modulus, $Y$, and the Poisson's ratio, $\nu_s$, as:
\begin{equation}
\mu _s  = \frac{Y}{{2\left( {1 + \nu _s } \right)}}, \qquad 
\lambda _s  = \frac{{Y\nu _s }}{{\left( {1 + \nu _s } \right)\left( {1 - 2\nu _s } \right)}}
\end{equation}

\noindent For further details, the reader is referred to the literature \citep{RN18, RN9}.

%%%%%%%%%%%%%%%%%%%%%%%%%%%%%%%%%%%%%%%%%%%%%%%%%%%%%%%%%%%%%%%%%%%%%%%%%%%%%%%%%%%%%%%%%%%%%%%%%%%
\subsection{Fluid-Solid Interfacial Conditions and Numerical Methods}
The fluid and solid domain interface will have imposed boundary conditions for a physically correct flow. For a viscous fluid flow, the non-slip condition specifies that the fluid velocity at the boundary must be the same as the boundary velocity itself \cite{RN10, RN9}. Continuity is imposed at the interface as:
\begin{equation}
{\rm{u}}_f  = {\rm{u}}_s  = {\rm{u}}_\Gamma
\end{equation}

\noindent Imposing equilibrium at the interface leads to the following condition:
\begin{equation}
\lambda _f  + \lambda _s  = 0
\end{equation}

\noindent $\lambda _f$ is the fluid traction which maps the fluid displacements on the interface with a Dirichlet-to-Neumann non-linear operator, $F_f$, for the fluid domain as:
\begin{equation}
\lambda _f  = {\mathop{\rm F _f}\nolimits} \left( {u_{\Gamma _f } } \right){\rm{ \qquad on \quad}}\Gamma _{{\rm{f,i}}} {\rm{ }}
\end{equation}

\noindent and $\lambda_s$ is the solid traction and is related to the solid displacements  with a Dirichlet-to-Neumann non-linear operator, $F_s$, for the solid-structural domain as:
\begin{equation}
\lambda _s  = {\mathop{\rm F _s}\nolimits} \left( {u_{\Gamma _s } } \right){\rm{ \qquad on \quad }}\Gamma _{{\rm{s,n}}} {\rm{ }}
\end{equation}

\noindent Above equations lead to a Steklov-Poincar\'{e}  equation which can be rewritten as a fixed point equation, 
\begin{equation}
\begin{gathered}
{\mathop{\rm F}\nolimits} _f \left( {u_\Gamma  } \right) + {\mathop{\rm F}\nolimits} _s \left( {u_\Gamma  } \right) = 0 \\
{\mathop{\rm F}\nolimits} _s ^{ - 1} \left( { - {\mathop{\rm F}\nolimits} _f \left( {u_\Gamma  } \right)} \right) = u_\Gamma \\
\end{gathered}
\end{equation}

\noindent The partitioning algorithm then consists of a single solution to the Dirichlet-to-Neumann operator for the fluid domain, and an inverse Neumann-to-Dirichlet operator for the solid domain per time step. 

Note that while conjugate heat transfer is important, especially at elevated Mach numbers, in this manuscript to isolate aerodynamic loads from thermal loads, we do not consider heat transfer from the fluid to any solid walls in the computational domain. All walls, including the thin panel, are maintained at an isothermal temperature of 300 K. The combined effect of thermal and aerodynamic loading on fluid structure interactions will be discussed in a subsequent manuscript.

The fluid domain was spatially discretized using a first-order ROE scheme; an approximate Riemann solver in a Finite-Volume formulation. Time-marching was achieved with a second-order, dual-time stepping scheme. Turbulence was modeled with the Vreman SGS model. The key parameter for WMLES is the distance from the wall where the algebraic logarithmic wall function should be applied. In our simulations, the log-law was applied at $y^{+}$ of $5$, and was fixed for all simulations. A flexible generalized minimal residual method, (FGMRES), is used for the linear solver iterator of the fluid domain. The fluid domain had non-reflecting boundary conditions, and forces due to pressure along the interface are transferred to the solid domain for loading. The solid domain used simple loading to replicate pressure, and the elasticity equations are solved with a first-order finite element Newton-Raphson scheme. The Newmark algorithm is used for time integration. A conjugate-gradient method was used for iterating the linear solver iterator for the solid domain solutions. Due to the significant differences between the time scales of the fluid and the structural dynamics, the equations that describe the two domains are loosely coupled; i.e., the structural solver is called once for each time step in the fluid domain. Each of the two solvers uses a second-order dual time-stepping technique, with $101$ inner iterations per outer iteration.

%%%%%%%%%%%%%%%%%%%%%%%%%%%%%%%%%%%%%%%%%%%%%%%%%%%%%%%%%%%%%%%%%%%%%%%%%%%%%%%%%%%%%%%%%%%%%%%%%%%

%% file: 05exptvalgridsens.tex
%%%%%%%%%%%%%%%%%%%%%%%%%%%%%%%%%%%%%%%%%%%%%%%%%%%%%%%%%%%%%%%%%%%%%%%%%%%%%%%%%%%%%%%%%%%%%%%%%%%
% Grid Sensitivity & Experimental Validation:
%%%%%%%%%%%%%%%%%%%%%%%%%%%%%%%%%%%%%%%%%%%%%%%%%%%%%%%%%%%%%%%%%%%%%%%%%%%%%%%%%%%%%%%%%%%%%%%%%%%
Before using the computational framework to elucidate the fluid and structural dynamics of a thin flexible panel as it interacts with subsonic, transonic, and supersonic flows, it is first validated against experiments of \citet{RN19}. The experimental setup (see figure \ref{fig:bojan_schematic}) and instrumentation was described in Section \ref{sec:intro}. \citet{RN19} organized their measurements and associated statistics in two categories: first 10 ms and then for up to 0.4 s. The operating conditions and material properties for both air and aluminum are tabulated in table \ref{tab:exptcomp}. The natural frequency and the first three harmonics of the panel are 303, 389, 661 and 1159 Hz, respectively. In addition to the frequency of oscillation of the panel, several flow characteristics were identified, including a recirculation region underneath the plate; an expansion fan at the outer top edge of the plate; a shear layer that originates at the outer edge of the plate and descends to the bottom wall with boundary layer recovery further downstream; a reattachment shock attached to the top of the shear layer downstream of the plate -- these are shown in figure \ref{fig:bojan_schematic}. Depending on the flow Mach number, all or some features (not described above) are expected to be observed. 
\begin{table}[hbt!]
\captionof{table}{Fluid and solid properties used to simulate the experiment of \citet{RN19}.}
\centering
    \begin{tabular}{lcr}
    \hline
        \multicolumn{3}{c}{Air} \\
        \hline \hline
        Total Pressure & $kPa$ & 277.6 \\
        Total Temperature & $K$ & 277 \\
        Mach & & 2.11 \\
        Freestream Pressure & $kPa$ & 29.885 \\
        Viscosity at $273$ $K$ & $N-s/m^{2}$ & 1.173 $\times$ $10^{-5}$ \\
        Reynolds Number & & 3927788 \\
        \hline
        \multicolumn{3}{c}{Thin Aluminum Panel} \\
        \hline \hline
        Density & $kg/m^{3}$ & 2700 \\
        Elasticity & $GPa$ & 72.0 \\
        Poisson's Ratio & & 0.33 \\
        Thickness & $mm$ & 1.02 \\
        Length & $mm$ & 50.8 \\
        \hline
    \end{tabular}
    \label{tab:exptcomp}
\end{table}

%%%%%%%%%%%%%%%%%%%%%%%%%%%%%%%%%%%%%%%%%%%%%%%%%%%%%%%%%%%%%%%%%%%%%%%%%%%%%%%%%%%%%%%%%%%%%%%%%%%
\subsection{Grid Sensitivity Study}
\begin{figure*}
    \centering
    \begin{subfigure}[h]{0.48\textwidth}
        \centering
        \includegraphics[width=1\linewidth]{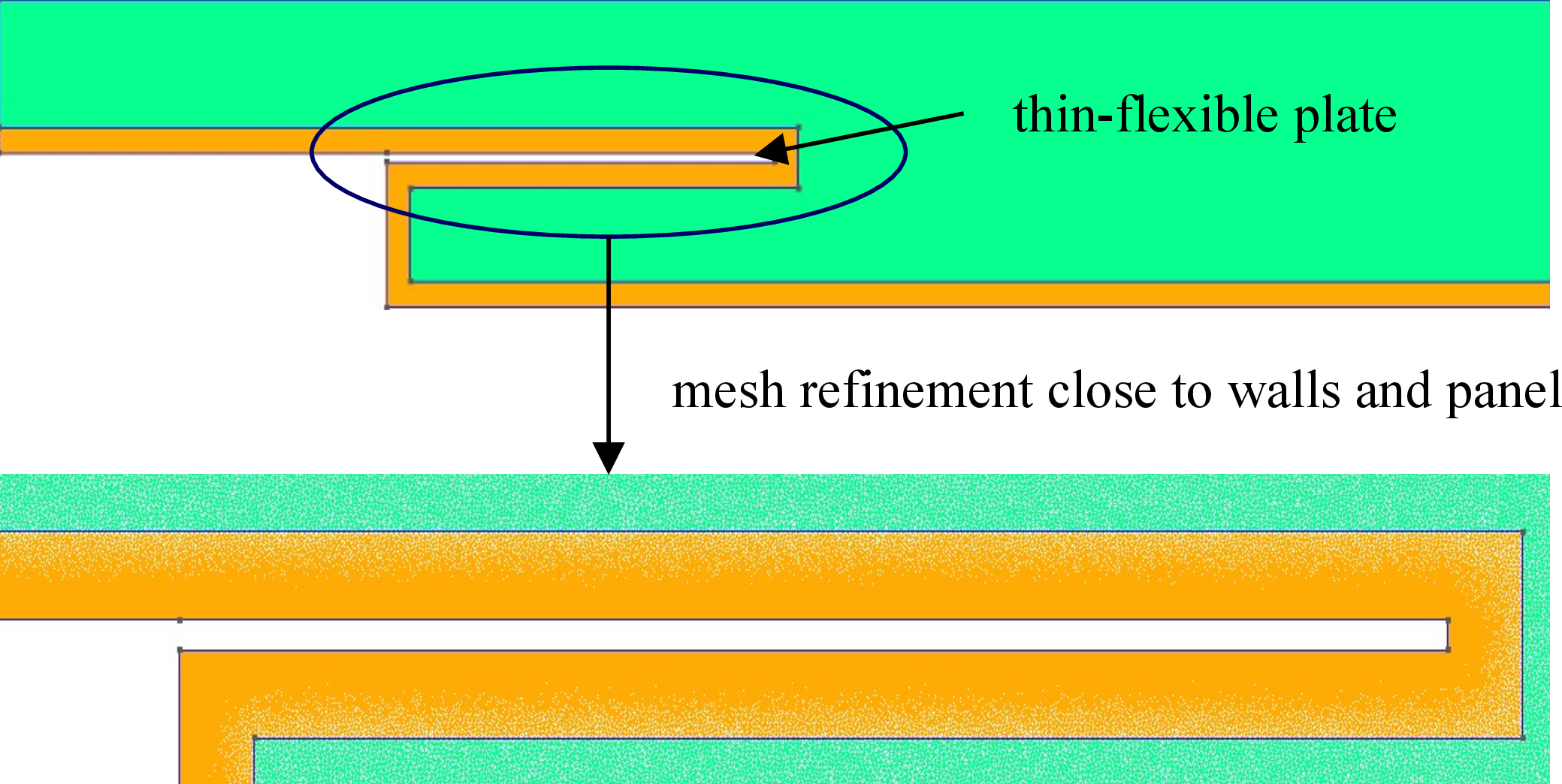}
        \caption{L2 grid, showing the grid in green for the core-flow, with refinement close to the wall boundaries and thin-panel in yellow.}
    \end{subfigure}
    \hspace{2mm}
    \begin{subfigure}[h]{0.48\textwidth}
        \includegraphics[width=1\linewidth]{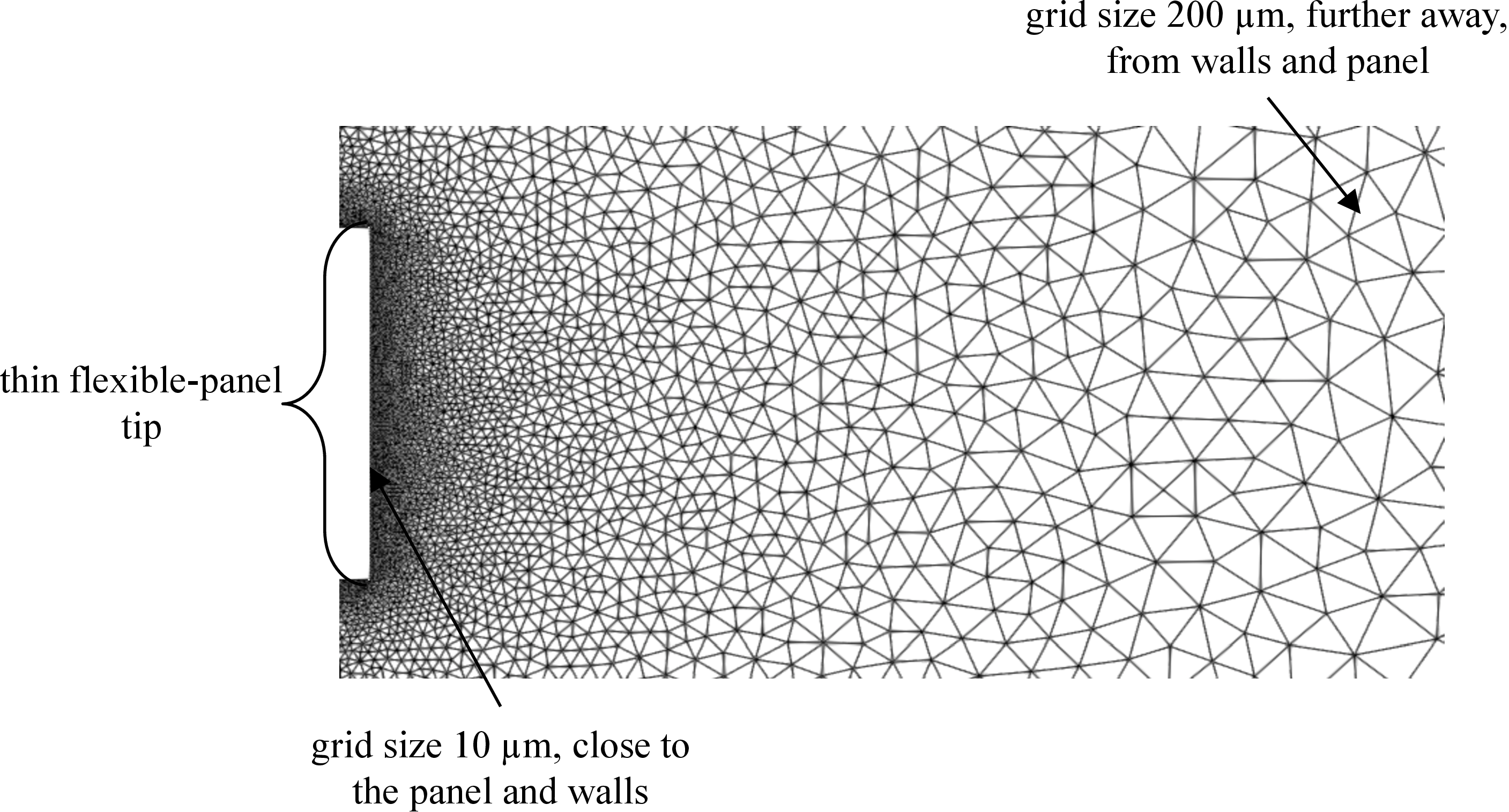}
        \caption{L2 grid refinement from 200 $\mu m$ in the core flow to 10 $\mu m$ close to the thin-panel tip  for $L2$ grid.}
    \end{subfigure}
    \caption{L2 grid details, showing refinement close to the walls and thin-flexible panel.}
    \label{fig:griddetails}
\end{figure*}

Before comparing the computational results with measurements, a grid sensitivity study was conducted. Three different unstructured meshes were generated, details of which are listed in table \ref{tab:allgrids}. All three grids are designed to be refined near the walls of the thin panel and then stretch to an isotropic grid in the core flow region. The minimum grid size in each case is 10 \textmu m, near the walls. The grid in the core region differs in each case; 200, 300 and 400 \textmu m for L2, L3 and L3, respectively. Figure \ref{fig:griddetails} shows an example of the discretized domain, where the yellow color represents the refinement region, stretching from the smallest to the largest, while the green region represents the isotropic grid away from the wall. The mesh for the solid domain is unstructured and uniform with a size of $100$ $\mu m$ for all three cases. 

% Grid movement and deformation:
Mesh deformation is handled with a pseudostructural approach as a linear-elasticity problem with boundary conditions that determine the displacements of the interior nodes, \citet{RN9}. The interface displacements are set as structural displacements and the boundaries are set as static. To prevent elements from having negative areas, the stiffness of each cell is set independently as an inverse function of the area or volume of the elements and Lam\'e's constants are determined from the area or volume. 

\begin{table}[hbt!]
    \caption{Details of the grids for both fluid and solid domains for the grid sensitivity analysis.}
    \centering
    \begin{tabular}{lcccr}
    \hline
        \multicolumn{5}{c}{Fluid Domain} \\
        \hline \hline
         & \multicolumn{2}{c}{Grid Size, $\mu m$} & \multicolumn{2}{c}{Number of} \\
        Grid & Maximum & Minimum & Elements & Grid Points \\
        \hline
        L2 & $200$ & $10$ & $1.92$ $\times$ $10^{6}$ & $980$ $\times$ $10^{3}$ \\
        L3 & $300$ & $10$ & $1.20$ $\times$ $10^{6}$ & $620$ $\times$ $10^{3}$ \\
        L4 & $400$ & $10$ & $865$ $\times$ $10^{3}$ & $452$ $\times$ $10^{3}$ \\
        \hline
        \multicolumn{5}{c}{Solid Domain} \\
        \hline \hline
        Solid & $100$ &  & $13.4$ $\times$ $10^{3}$  & $7.2$ $\times$ $10^{3}$ \\
        \hline
    \end{tabular}
        \label{tab:allgrids}
 \end{table}   

The results of the grid sensitivity analysis are discussed in terms of time-averaged flow and structural behaviors over 0.4 s. %  in the next two subsections focusing first on flow physics and then on the structural dynamics. % flow behavior captured by the three grids, $L2$, $L3$ and $L4$ specified above, is compared in the following section. It is followed by the comparison of solid behavior from the FSI simulations with the three different grids next.

\begin{figure*}[htbp!]
    \centering
    \includegraphics[width=0.96\textwidth]{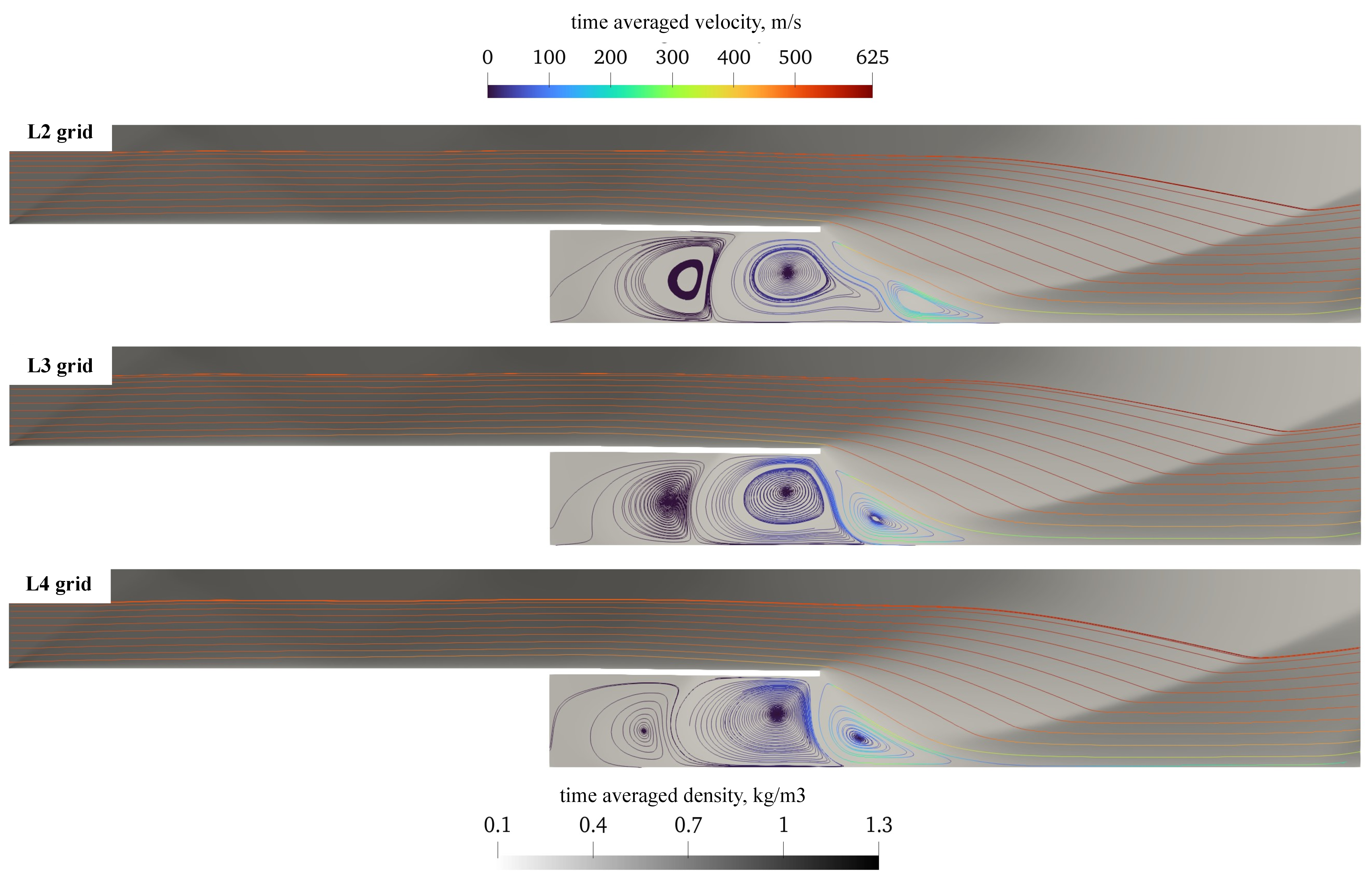}
    \caption{Comparison of time averaged flow of the grids - $L2$, $L3$ and $L4$.}
    \label{fig:gridflows}
\end{figure*}

%%%%%%%%%%%%%%%%%%%%%%%%%%%%%%%%%%%%%%%%%%%%%%%%%%%%%%%%%%%%%%%%%%%%%%%%%%%%%%%%%%%%%%%%%%%%%%%%%%%
%\subsubsection{Grid Sensitivity Analysis: Comparison of Flow Behavior between L2-L4}
%The fluid behavior of the FSI simulations from the three grids are compared to ensure they properly capture the appropriate fluid dynamic behavior for initial transients and fully-started conditions. The FSI simulations is run for a time of $0.4$ $s$ for $1001$ timesteps, which are then averaged for obtaining a time-averaged set of flow parameters for each individual grid and compared in 
Figure \ref{fig:gridflows} shows the time-averaged density contours overlaid by time-averaged streamlines for cases L2, L3 and L4. % Accuracy of the simulations is determined through the flow features described in the schematic, figure \ref{fig:bojan_schematic} and also captured by the experiment for fully-started or steady conditions, figure \ref{fig:fullfluidexpt}b.
All three grids are able to capture the expansion fan, reattachment shock, shear layer reattachment, and recovering boundary layer. %, as confirmed by the velocity streamlines overlaid on the density field in figure \ref{fig:gridflows}. 
The recirculation region is also captured by the three grids; however, the vortex structure attached to the outside lower edge of the thin-flexible plate and just below the shear layer captured using L2 and L3 are very similar, while that from L4 differs from the other two. %better refined by the $L2$ grid which shows 2 vorticular structures at that location. This is because the smaller gridsize of $L2$ is able to better capture the smaller flow scales as compared to other - $L3$ and $L4$ grids. $L3$ and $L4$ capture a single vortex at the same location, which is significantly larger in comparison to the $L2$ results. These results show that $L2$ grid is able to resolve the smaller flow scales better. This prompts us to prefer the $L2$ grid for the accurate capture of spatiotemporal scales of flow for FSI simulations.

\begin{figure*}[htbp!]
    \centering
    \begin{subfigure}[h]{0.48\textwidth}
        \centering
        \includegraphics[width=\textwidth]{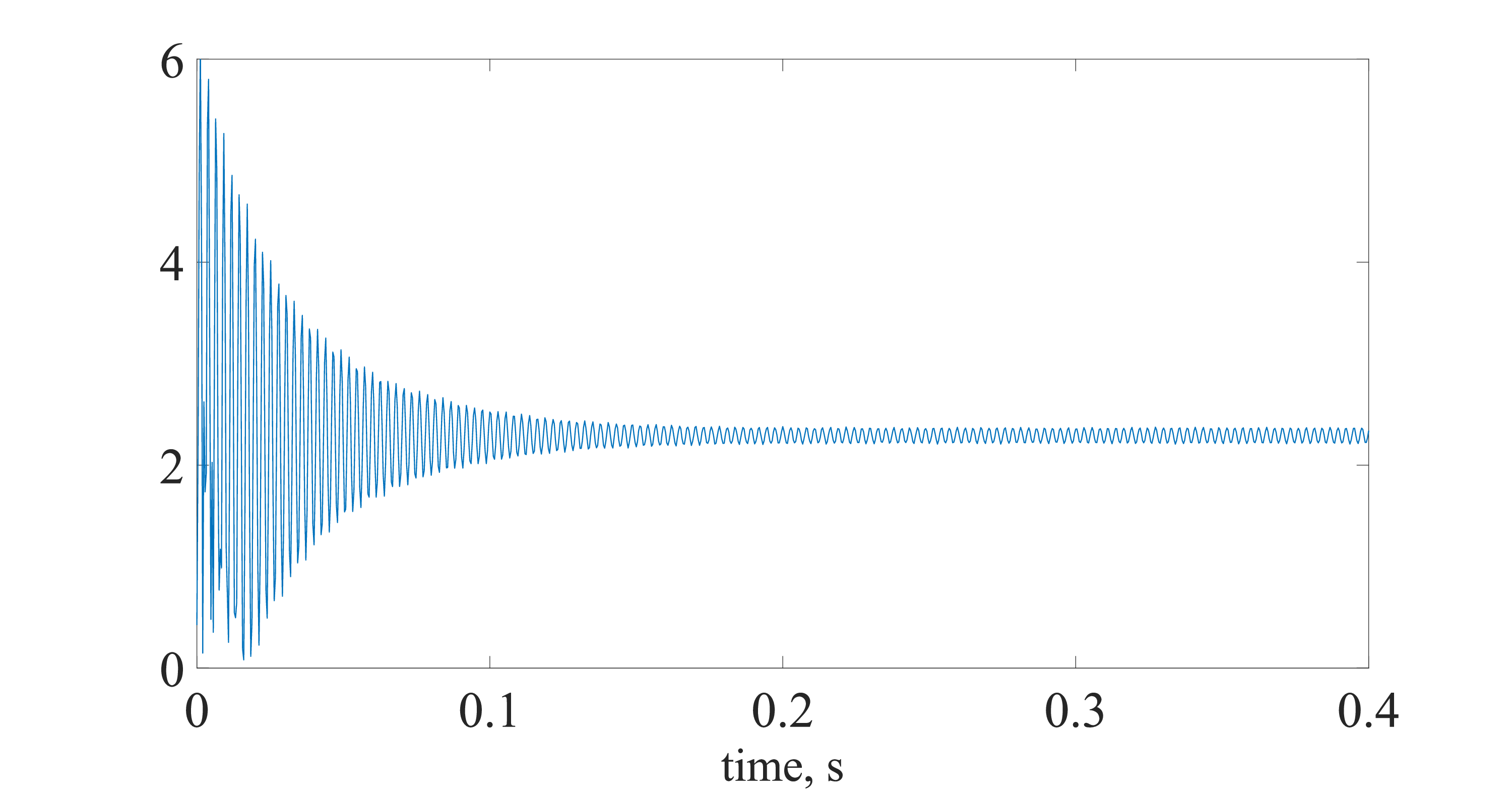}
        \caption{displacement-time plot for $L2$ grid}
        \label{fig:steadyl2}
    \end{subfigure}
    \hspace{2mm}
    \begin{subfigure}[h]{0.48\textwidth}
        \includegraphics[width=\textwidth]{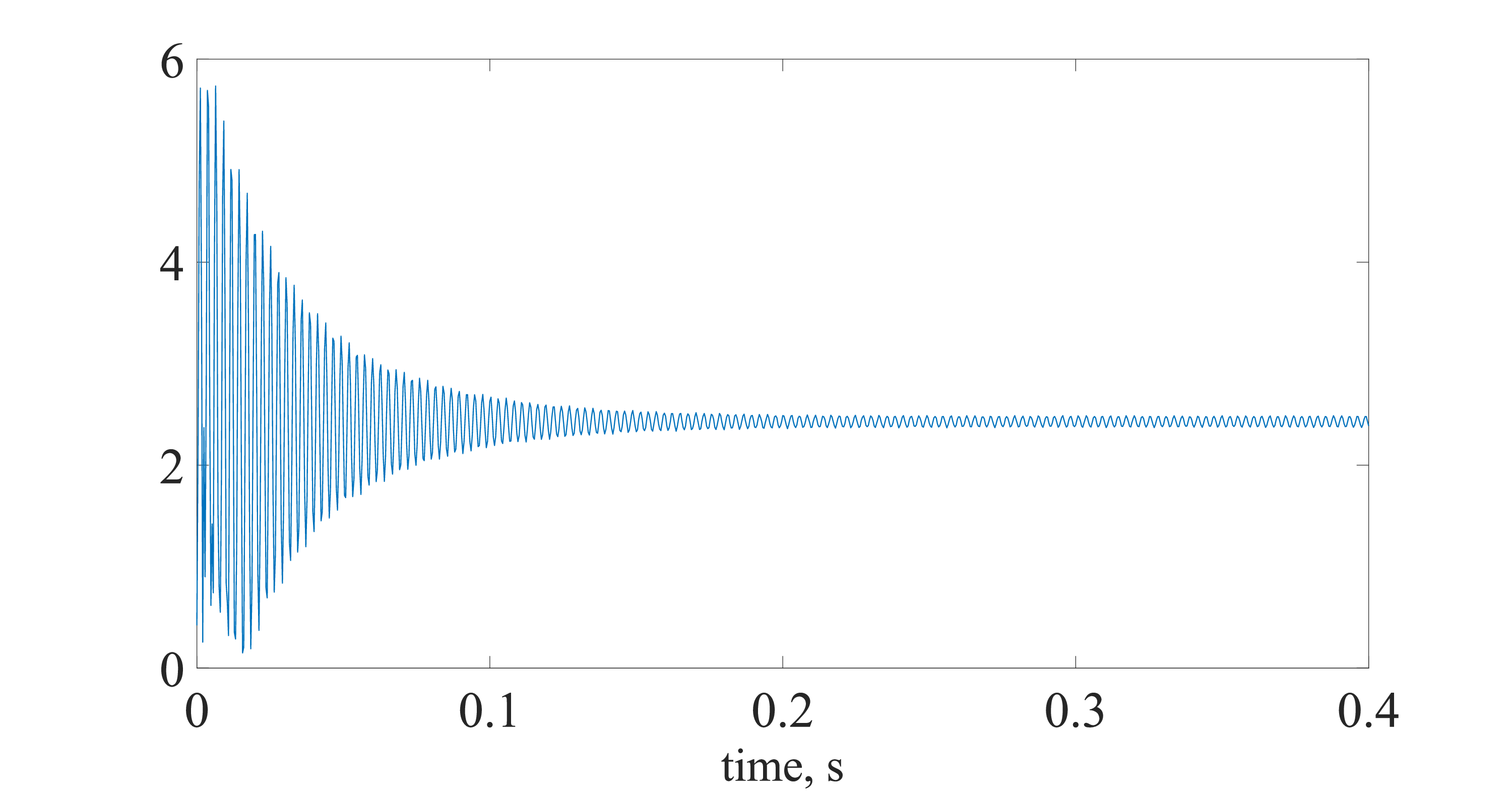}
        \caption{displacement-time plot for $L3$ grid}
    \end{subfigure}
    \hspace{2mm}
    \begin{subfigure}[h]{0.48\textwidth}
        \centering
        \includegraphics[width=\textwidth]{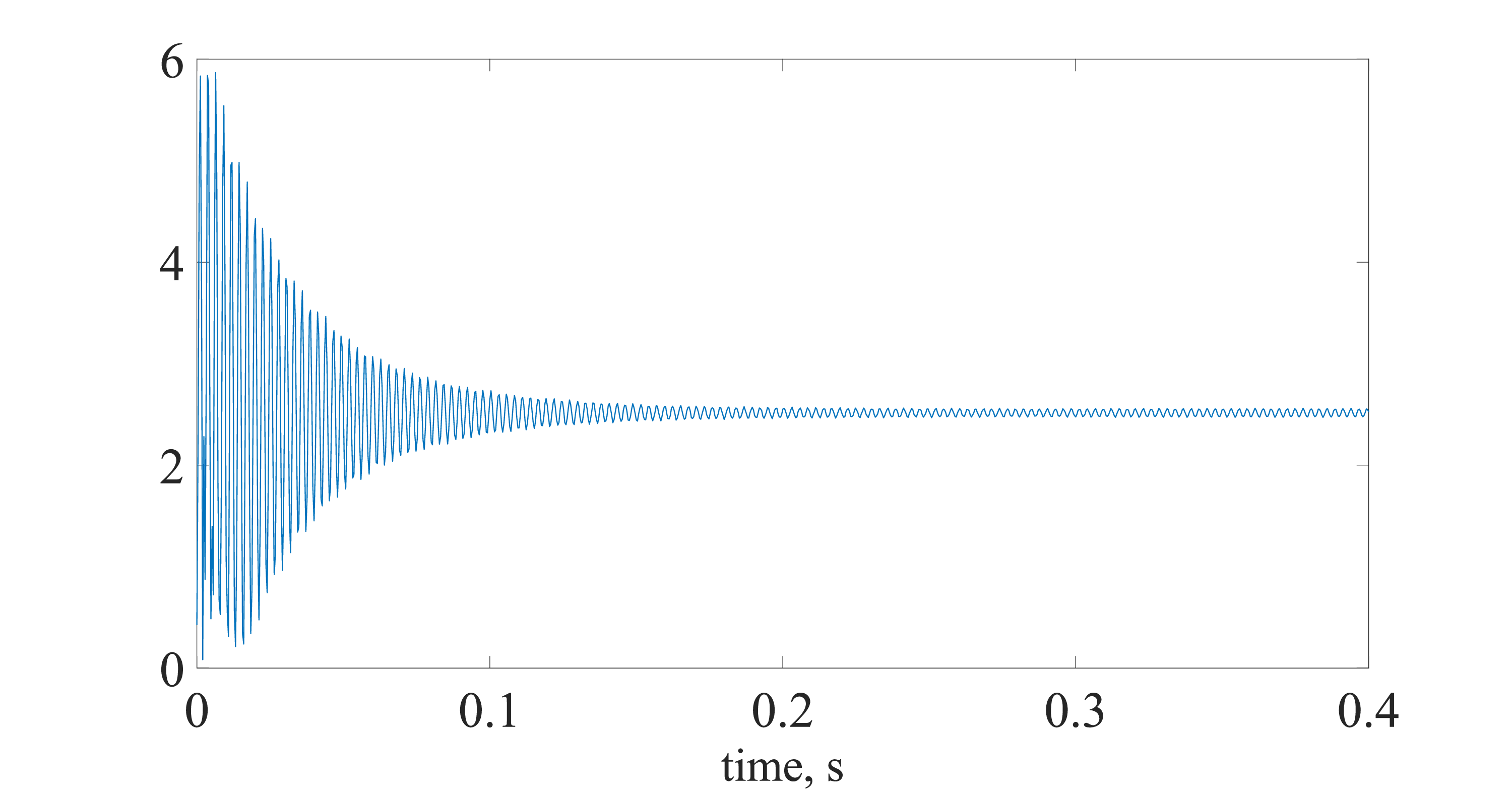}
        \caption{displacement-time plot for $L4$ grid}
    \end{subfigure}
    \hspace{2mm}
    \begin{subfigure}[h]{0.48\textwidth}
        \centering
        \includegraphics[width=\textwidth]{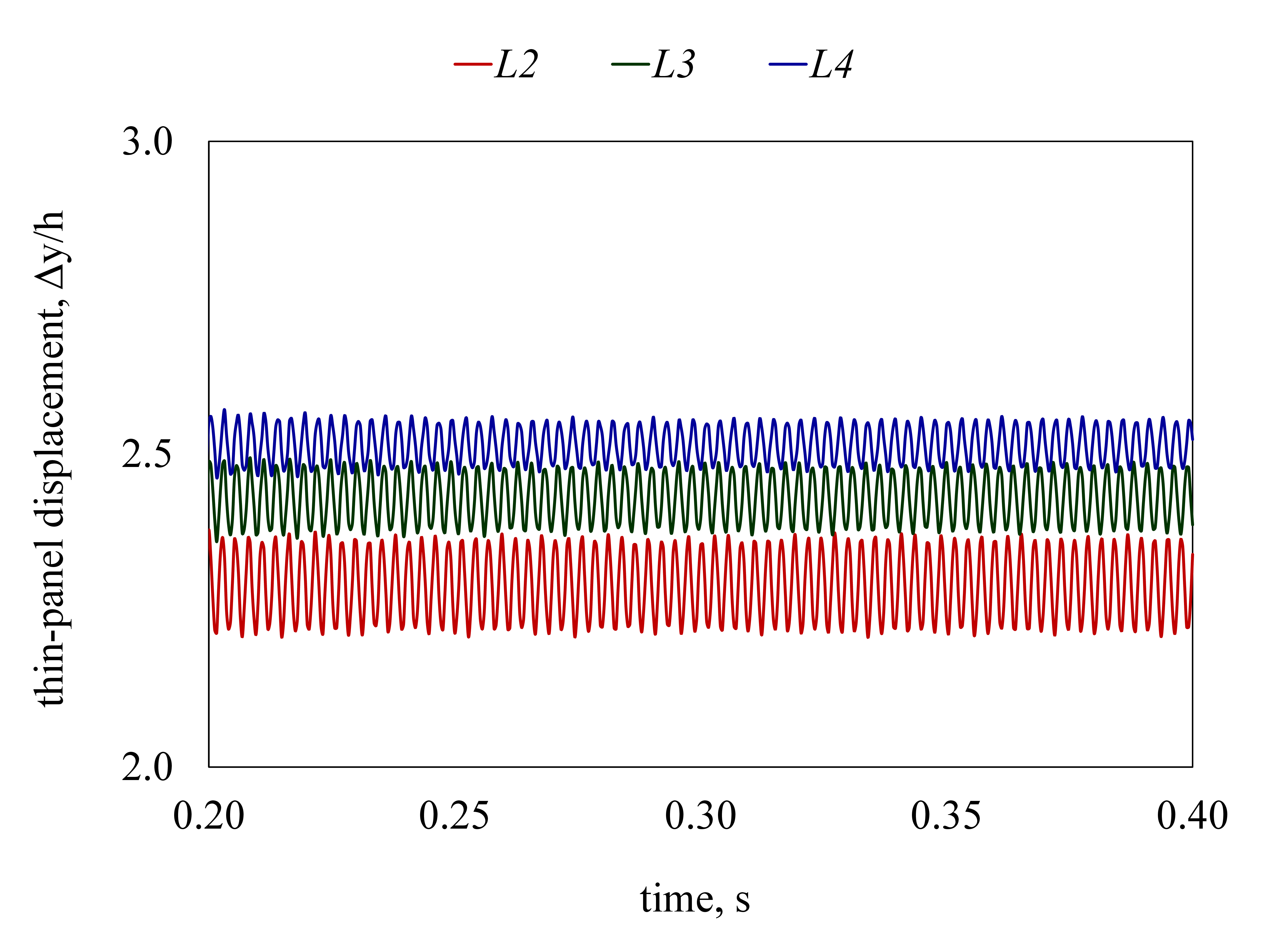}
        \caption{comparison of displacements from the three grids}
    \end{subfigure}
    \caption{Displacement-time plots and their comparison after initial transients are stabilized for the three grids, $L2$, $L3$ and $L4$.}
    \label{fig:griddisps}
\end{figure*}

%%%%%%%%%%%%%%%%%%%%%%%%%%%%%%%%%%%%%%%%%%%%%%%%%%%%%%%%%%%%%%%%%%%%%%%%%%%%%%%%%%%%%%%%%%%%%%%%%%%
%\subsubsection{Comparison of Structural Behavior}
%The preference for $L2$ grid established in the previous paragraph for capturing flow behaviors is further reinforced by the oscillatory behavior of the panel shown in 
Figures \ref{fig:griddisps}a, b, c show the displacement of the tip of the panel with time. All three grids capture the initial transients in the first $0.2$ $s$, which are then stabilized and sustained in amplitude from $0.2$ to $0.4$ $s$. The stabilized oscillations from the three grids beyond $0.2$ $s$ are shown in figure \ref{fig:griddisps}d. Note that the displacement of the tip is normalized by the height of the panel. Although there are clear differences in the amplitude when different grids are used, the frequency of oscillation calculated by taking the FFT is 367 Hz. So, while the detailed behaviors are dictated by the grid size, the characteristics do not change for grids L2-L4. Therefore, given that the flow features under the thin panel are going to be important as it oscillates, even though L3 is presumably sufficient to resolve them, since we intend to investigate flows up to Mach 3 (i.e., higher Reynolds number), we will use L2 for all cases described in the rest of this paper. 

%\noindent The amplitude reported by $L2$ is larger then the amplitude reported by $L3$, which in turn is again larger than the amplitude reported by $L4$. This is because a smaller gridsize is able to transfer the fluid forces to the solid and capture displacements of the solid more accurately. The accurate capture of solid behavior by the $L2$ grid is further supported by comparison with experimental results for transient conditions, figure \ref{fig:transsolidcomp}, and fully started conditions, figure \ref{fig:fullsolidcomp}. FFT was conducted on the thin-plate displacement-time signals from the three grids and they all reported the same peak frequency at $367$ $Hz$. The ability of the $L2$ grid to accurately transfer fluid forces and capture the solid deformations and oscillations better leads us to prefer the $L2$ grid for the FSI simulations of this study.

%%%%%%%%%%%%%%%%%%%%%%%%%%%%%%%%%%%%%%%%%%%%%%%%%%%%%%%%%%%%%%%%%%%%%%%%%%%%%%%%%%%%%%%%%%%%%%%%%%%
\subsection{Comparison between Computations and Measurements}%: Initial Transient Conditions}
\begin{figure*}[htbp!]
    \centering
    \begin{subfigure}[h]{0.64\textwidth}
        \centering
        \includegraphics[width=\textwidth]{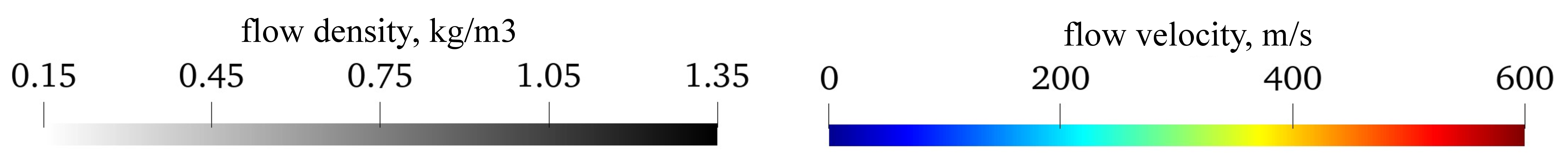}
    \end{subfigure}
    \hspace{2mm}
    \begin{subfigure}[h]{0.48\textwidth}
        \centering
        \includegraphics[width=\textwidth]{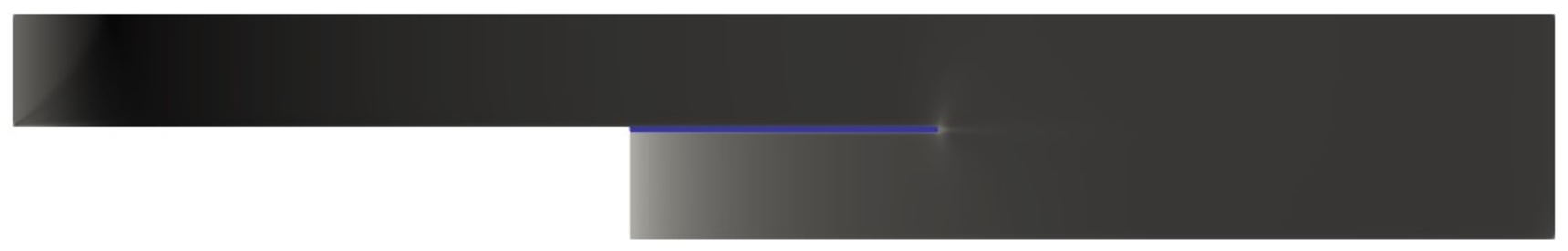}
        \caption{t = $0.0$ ms}
    \end{subfigure}
    \hspace{2mm}
    \begin{subfigure}[h]{0.48\textwidth}
        \includegraphics[width=\textwidth]{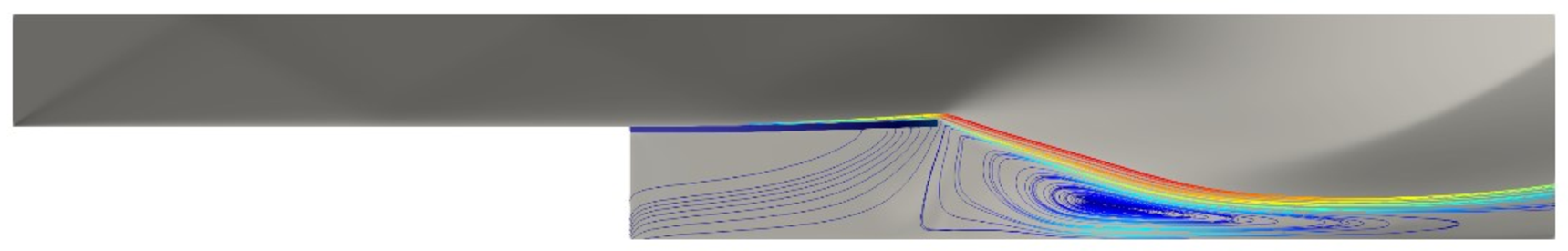}
        \caption{t = $2.0$ ms}
    \end{subfigure}
    \hspace{2mm}
    \begin{subfigure}[h]{0.48\textwidth}
        \centering
        \includegraphics[width=\textwidth]{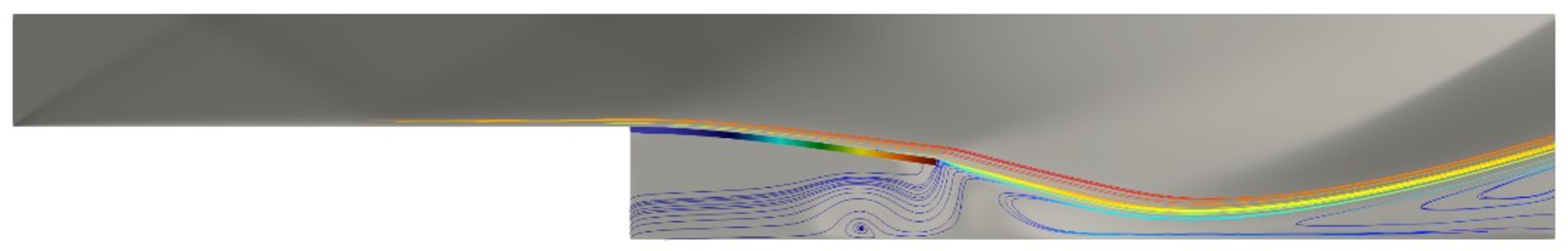}
        \caption{t = $4.0$ ms}
    \end{subfigure}
    \hspace{2mm}
    \begin{subfigure}[h]{0.48\textwidth}
        \centering
        \includegraphics[width=\textwidth]{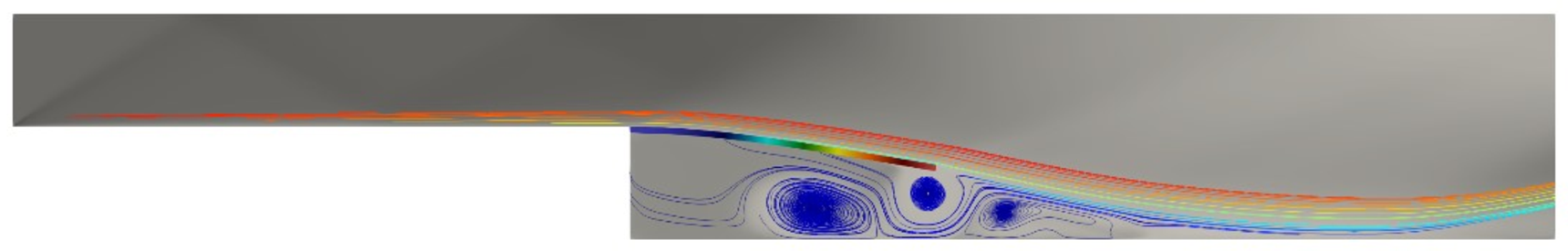}
        \caption{t = $6.0$ ms}
    \end{subfigure}
        \hspace{2mm}
    \begin{subfigure}[h]{0.48\textwidth}
        \centering
        \includegraphics[width=\textwidth]{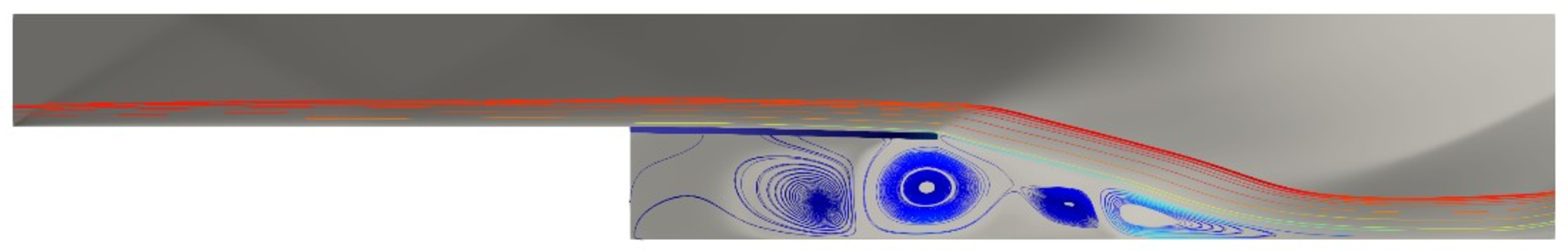}
        \caption{t = $8.0$ ms}
    \end{subfigure}
    \hspace{2mm}
    \begin{subfigure}[h]{0.48\textwidth}
        \centering
        \includegraphics[width=\textwidth]{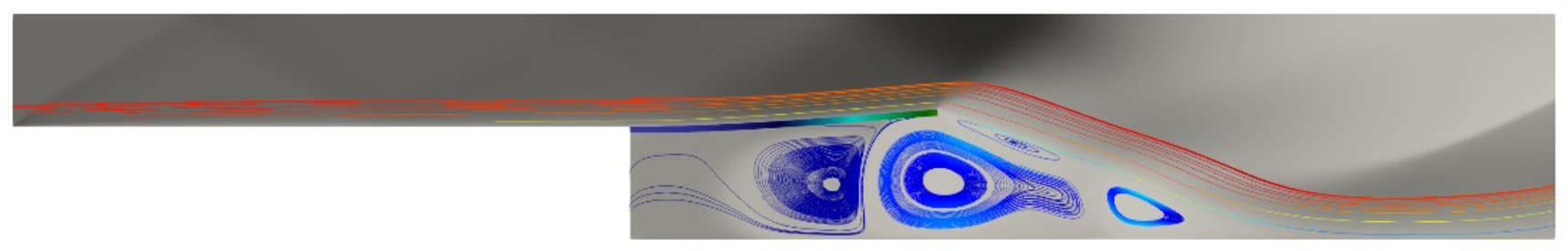}
        \caption{t = $10.0$ ms}
    \end{subfigure}
    \hspace{2mm}
    \begin{subfigure}[h]{0.32\textwidth}
        \centering
        \includegraphics[width=\textwidth]{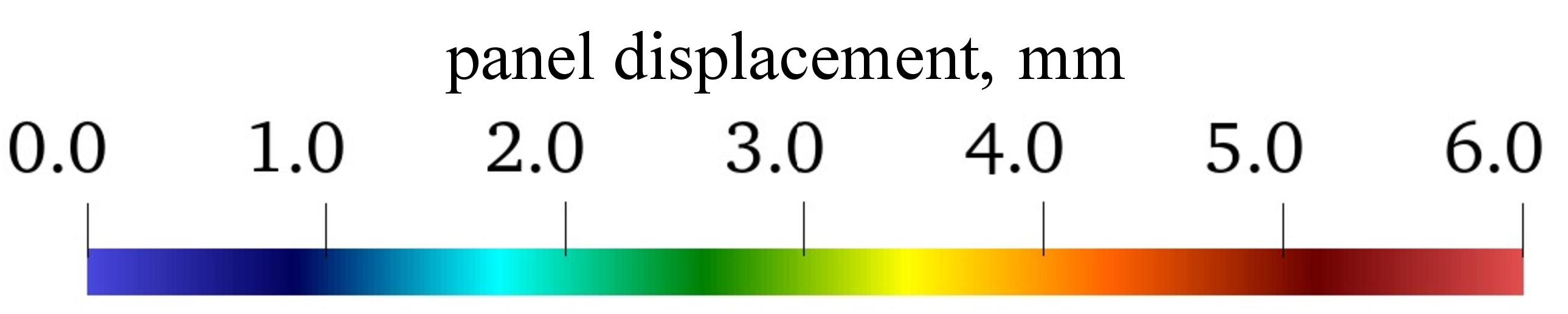}
    \end{subfigure}
    \caption{Temporal evolution of Ma $2.11$ flow, showing velocity streamlines on density contours and compliant panel displacement for initial transient conditions.}
    \label{fig:Mach2EvolveTrans}
\end{figure*}

Figure \ref{fig:Mach2EvolveTrans} shows the instantaneous flowfield and panel deformation at $0.0$, $2.0$, $4.0$, $6.0$, $8.0$, and $10$ $ms$ corresponding to the configuration described in the previous section. A shear layer is formed at the tip of the panel, which extends to the lower wall. As the panel oscillates, shown in the figures \ref{fig:Mach2EvolveTrans} a-f, the shear layer oscillates with it. Other flow features, such as the expansion fan when the panel moves downwards and the reattachment shock owing to the interactions between supersonic flow and the shear layer, are also observed. Tip vortices are formed due to the motion of the panel; the shear layer prevents them from convecting downstream. While all the vortices interact with the shear layer, the one that is furthest away from the wall stretches due to this interaction, eventually dissipating at the bottom wall. All these flow features were also observed in the experiment. % The shear layer, separating the high-speed flow from the stationary air along the bottom section descends to the bottom wall by $2.0$ $ms$, figure \ref{fig:Mach2EvolveTrans}b, and traps a pocket of air behind it and under the thin panel by $6.0$ $ms$, figure \ref{fig:Mach2EvolveTrans}d. Vortex shedding underneath the shear layer is not longer possible after $6.0$ $ms$, as the shear layer attaches to the bottom section wall leading to boundary-layer recovery. This leads to recirculation that increases the number of trapped vortices. For subsequent times up to $10$ $ms$, figure \ref{fig:Mach2EvolveTrans}d-f, these vortices remain trapped without significant viscous dissipation which follows as the simulation is allowed to run for longer times as will be discussed later in this section.

\begin{figure}[htbp!]
    \centering
        \includegraphics[width=0.48\textwidth]{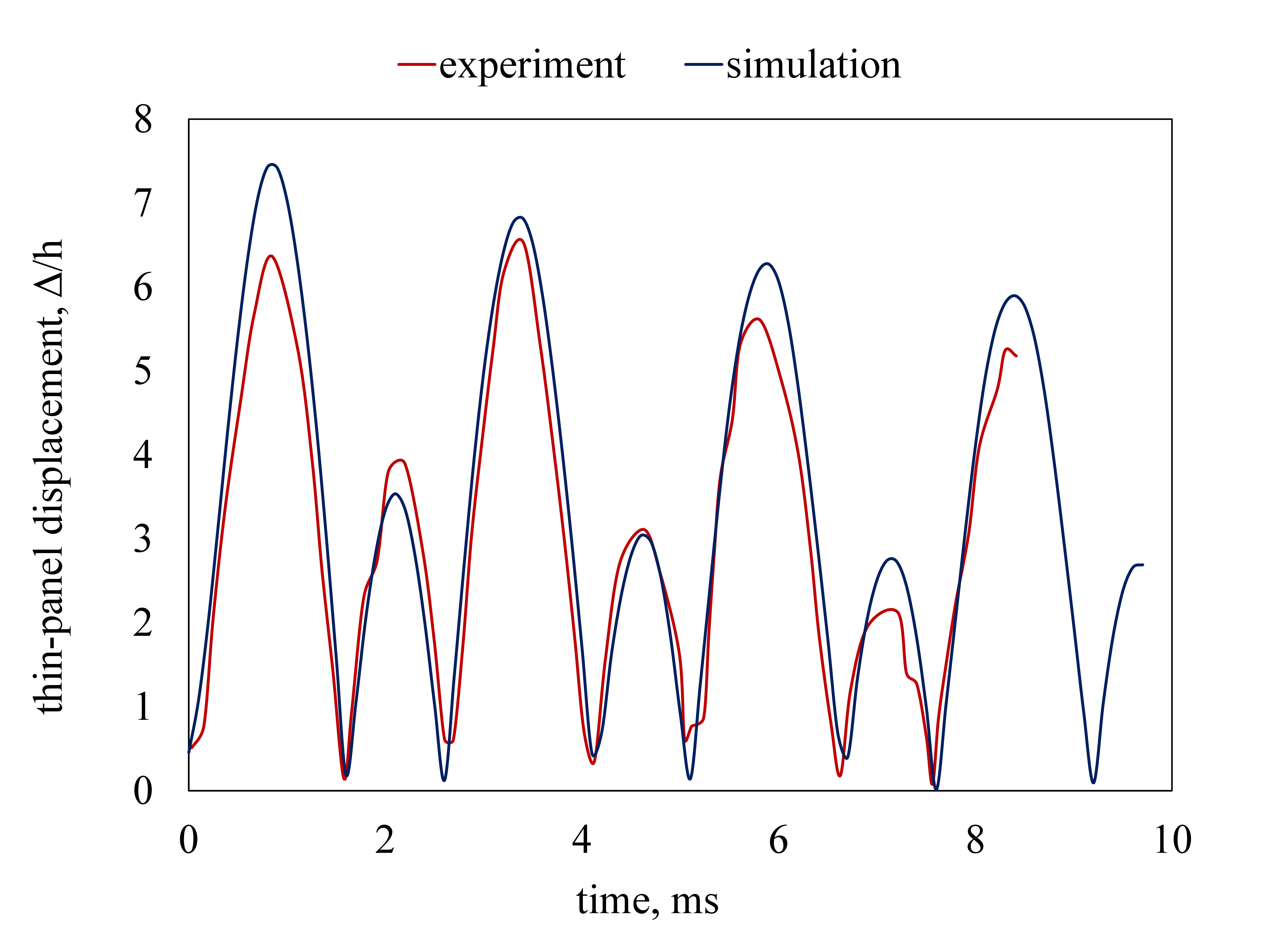}
    \caption{Comparison of panel tip displacement between simulations and experiments \citep{RN19} for the first 10 ms.}
    \label{fig:transsolidcomp}
\end{figure}

The qualitative time-averaged flow features observed in the experiments of \citet{RN19} were successfully captured by our calculations using the L2 grid -- see figure \ref{fig:gridflows}. Quantitatively, we compare the panel tip displacement and the corresponding oscillation frequency measured in the experiment for the first 10 ms and the fully started conditions (described as the time after 10 ms up to 0.4 s). %To capture the transient flow features, the FSI simulation was allowed to run for $10$ $ms$ for $101$ timesteps. The inlet flow was allowed to expand from the total pressure to the freestream pressure to achieve a Mach number of $2.11$ before entering the fluid domain. The flow is supersonic, and the shock-wave boundary-layer interactions (SWBLI) result in the thin-flexible aluminum plate oscillations from pressure loading inside the boundary layer. The flowfield from the experiment, figure \ref{fig:transfluidcomp}, shows the expansion fan, shear layer and the recirculation regions identified in figure \ref{fig:bojan_schematic}. The simulation captures shock reattachment and the recovering boundary layer in addition to the flow features observed in the experiment. Velocity streamlines are overlaid on the density contour for  visualizing of the shear layers, recirculation regions and boundary layer recovery. The results from the flow solver of the FSI simulation captures the flow behavior reported in the experiment for initial transients, thus validating the fluid flow simulation with the experiment. 
%\begin{figure*}[htbp!]
 %   \centering
  %      \includegraphics[width=\textwidth]{figures/05-ExptValGridSense/transsimulation.eps}
  %  \caption{Comparison of flow features captured by the simulation for transient flow features.}
   % \label{fig:transfluidcomp}
%\end{figure*}
This comparison is shown in figure \ref{fig:transsolidcomp} for the first 10 ms. The displacement, $\Delta$, is normalized by the thin-plate thickness, $h$. 

\begin{figure}[htbp!]
    \centering
    \begin{subfigure}[t]{0.5\textwidth}
        \centering
        \includegraphics[width=\textwidth]{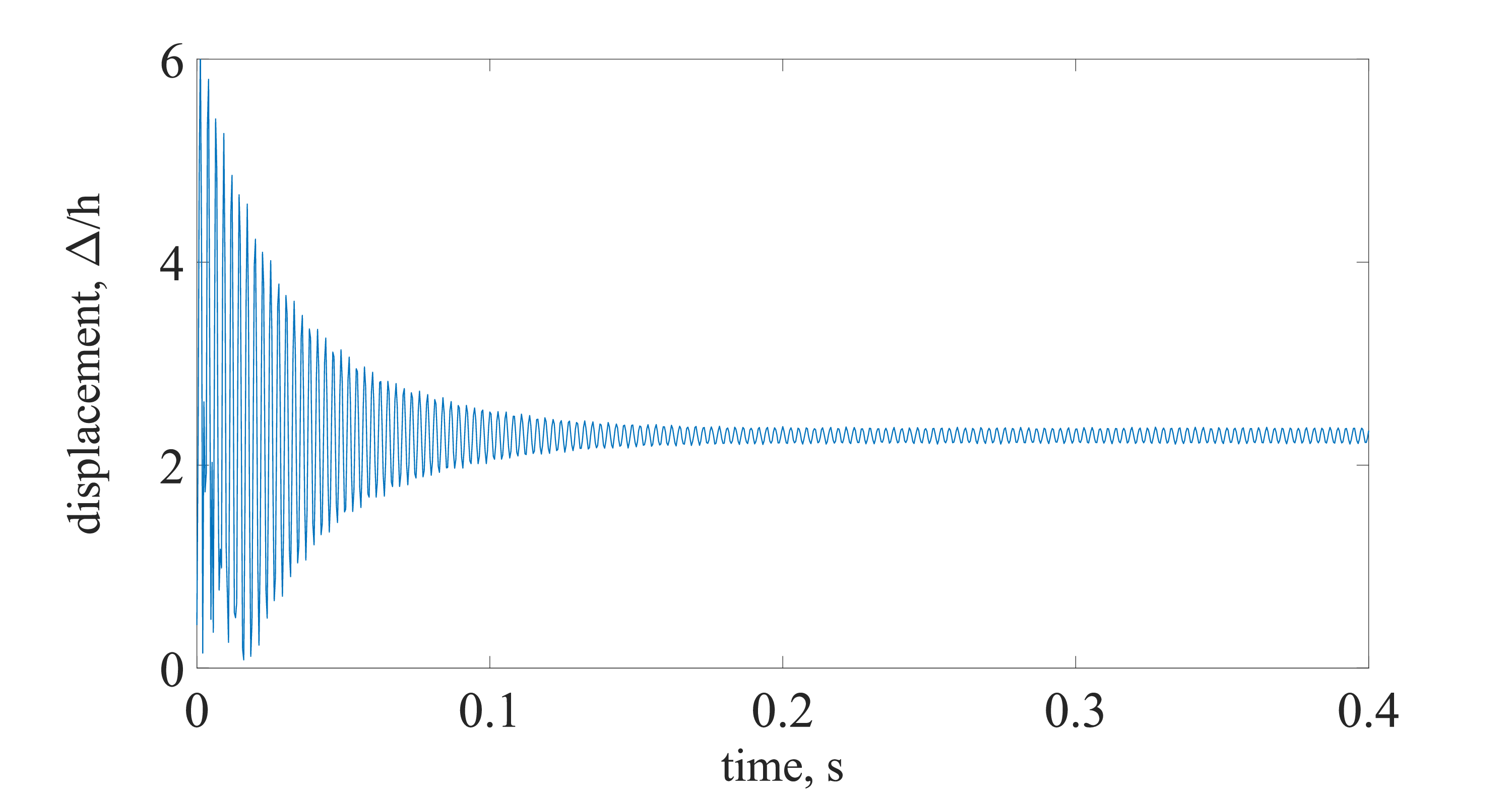}
        \caption{time evolution of tip displacement}
        \label{fig:tipdispfull}
    \end{subfigure}
    \begin{subfigure}[t]{0.5\textwidth}
        \centering
        \includegraphics[width=\textwidth]{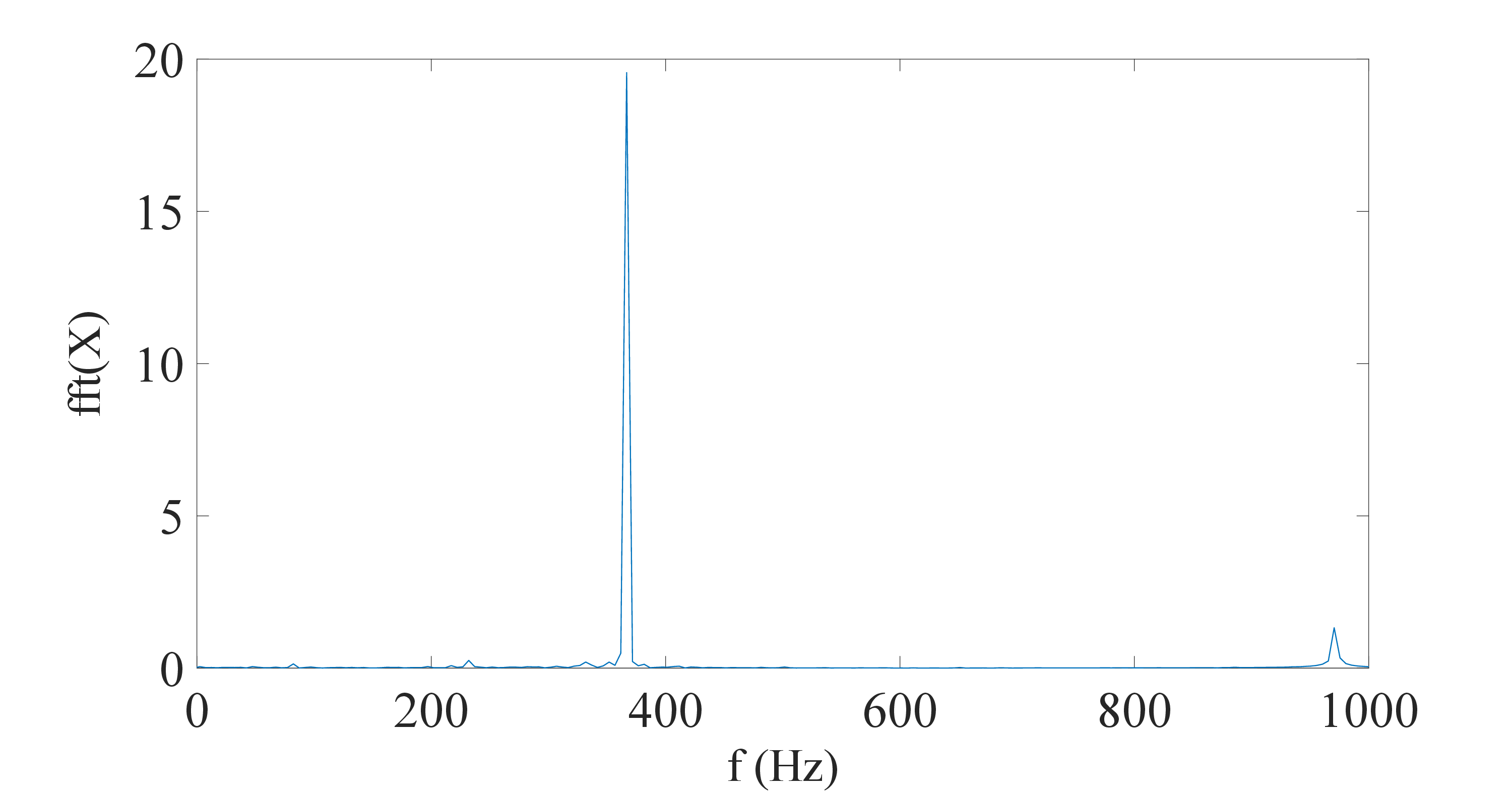}
        \caption{FFT of tip displacement over time shown in figure \ref{fig:tipdispfull} for $0.2 \le t \le 0.4$.}
    \end{subfigure}
    \caption{Numerical simulation results for steady or fully started conditions.}
    \label{fig:fullsolidcomp}
\end{figure}

The displacement amplitudes and phases compare well. The oscillation frequency calculated from this chart is 395 Hz, which is very close to the experimentally measured frequency of 397 Hz. After the initial transients, once the tip displacements reach a stationary state between 0.2 to 0.4 s, the frequency of oscillation from our computations is 367 Hz, in good agreement with the measured frequency of 366 Hz. The details are shown in figure \ref{fig:fullsolidcomp}.% These results also indicate that, at low supersonic Mach numbers, the thermal loading may not have a significant effect on fluid structure interactions. %, which gives us confidence that the structural solver is capturing the solid behavior properly. The peak displacement amplitudes between the experiment and the simulation line up, figure \ref{fig:transsolidcomp}, reaching as high as $8$ times the thin-plate thickness. The plate oscillation frequency was estimated from averaging the time-periods between corresponding larger peaks at $395$ $Hz$. The experiment captured the frequency for the thin-plate oscillations by measuring the oscillations in the cavity pressure in the recirculation region and noting the observation that the thin-plate position tracks the cavity pressure oscillations at $397$ $Hz$. The oscillation frequency from the experiment is $397$ $Hz$, and from the simulation is $395$ $Hz$, which are almost identical. The results from the structural solver of the FSI simulation captures the solid behavior reported in the experiment very well for the initial transient conditions, which validates the structural solver. 

% The oscillation frequencies from the experiment and simulation are almost identical with a difference of only $1$ $Hz$, which validates the structural solver for fully-started conditions. 

% Results from the simulation for fluid and solid behavior compare very well with the experiment of Bojan \cite{RN19} for transient and fully-started conditions, validating the fluid and solid solvers reliably for FSI simulations in supersonic flow regimes. 

%%%%%%%%%%%%%%%%%%%%%%%%%%%%%%%%%%%%%%%%%%%%%%%%%%%%%%%%%%%%%%%%%%%%%%%%%%%%%%%%%%%%%%%%%%%%%%%%%%%

%% file: 06results.tex
%%%%%%%%%%%%%%%%%%%%%%%%%%%%%%%%%%%%%%%%%%%%%%%%%%%%%%%%%%%%%%%%%%%%%%%%%%%%%%%%%%%%%%%%%%%%%%%%%%%
% Results & Discussion:
%%%%%%%%%%%%%%%%%%%%%%%%%%%%%%%%%%%%%%%%%%%%%%%%%%%%%%%%%%%%%%%%%%%%%%%%%%%%%%%%%%%%%%%%%%%%%%%%%%%
To identify similarities and differences in the FSI behaviors in sub-, trans- and supersonic flows, in addition to the validation case at Mach number $2.11$, three additional simulations were conducted at Mach numbers of $0.5$, $0.95$ and $3$. The cases are set up such that the static temperature and pressures are $146.53$ K and $29.89$ kPa, respectively, at the entrance of the conduit. All cases use the L2 grid. In the next two subsections, we discuss the effect of Mach number on the flow and structural dynamics for the initial transients (time up to 10 ms) and the fully started conditions (times up to 0.4 s). These discussions will be based on time-averaged variables. %Two sets of parametric simulations were conducted to investigate the effects of Mach number on the FSI configuration described in figure \ref{fig:configschematic} - one for transient and another for fully-started or steady conditions. Four different Mach numbers were selected to cover subsonic, transonic and supersonic flow regimes, table \ref{tab:paramprops}. Air is allowed to expand fully to the flow Mach number and the expanded flow conditions are imposed as inlet conditions for all four cases. The total pressure, total temperature and the Reynolds number of the fully expanded flow are listed in table \ref{tab:paramprops} after the Mach number of the case.
%The static pressure and static temperature for all four cases are kept constant at $29.886$ $kPa$ and $146.53$ $K$.

% The parametric sets of runs were conducted with the unstructured $L2$ grid, details of which are given in table \ref{tab:allgrids}. The two different parametric sets consists of simulations for the four Mach numbers to capture - 
% \begin{itemize}
%   \item[a.] the initial transient conditions for $10$ $ms$ with $101$ timesteps or snapshots;
%   \item[b.] the fully-started conditions for $0.4$ $s$ with $1001$ timesteps or snapshots;
% \end{itemize}
% \noindent For both sets of simulations, transient and steady as detailed above, the FSI behavior with changing Mach number is investigated and reported.

%%%%%%%%%%%%%%%%%%%%%%%%%%%%%%%%%%%%%%%%%%%%%%%%%%%%%%%%%%%%%%%%%%%%%%%%%%%%%%%%%%%%%%%%%%%%%%%%%%%
\subsection{Initial Transient Conditions (up to 10 ms)}
\input{06atrans10ms}
\subsection{Fully Started Conditions (up to 400 ms)}
\input{06bfull400ms}

%% file: 06atrans10ms.tex
%%%%%%%%%%%%%%%%%%%%%%%%%%%%%%%%%%%%%%%%%%%%%%%%%%%%%%%%%%%%%%%%%%%%%%%%%%%%%%%%%%%%%%%%%%%%%%%%%%%
% Initial Transient Conditions - 10 ms:
%%%%%%%%%%%%%%%%%%%%%%%%%%%%%%%%%%%%%%%%%%%%%%%%%%%%%%%%%%%%%%%%%%%%%%%%%%%%%%%%%%%%%%%%%%%%%%%%%%%
%It was identified during experimental validation and grid sensitivity studies in the previous sections that transients for the FSI configurations occur for the first $0.2$ $s$ for a Mach $2.11$. Hence, to capture the initial transients, the simulation is run for a physical time of $10$ $ms$ or $101$ snapshots, with a timestep of $1$ $\times$ $10^{-4}$ $s$.

%%%%%%%%%%%%%%%%%%%%%%%%%%%%%%%%%%%%%%%%%%%%%%%%%%%%%%%%%%%%%%%%%%%%%%%%%%%%%%%%%%%%%%%%%%%%%%%%%%%
\subsubsection{Flow Behaviors}% vs. Mach Number}
\begin{figure*}[htbp!]
    \centering
    \begin{subfigure}[h]{0.96\textwidth}
        \centering
        \includegraphics[width=\textwidth]{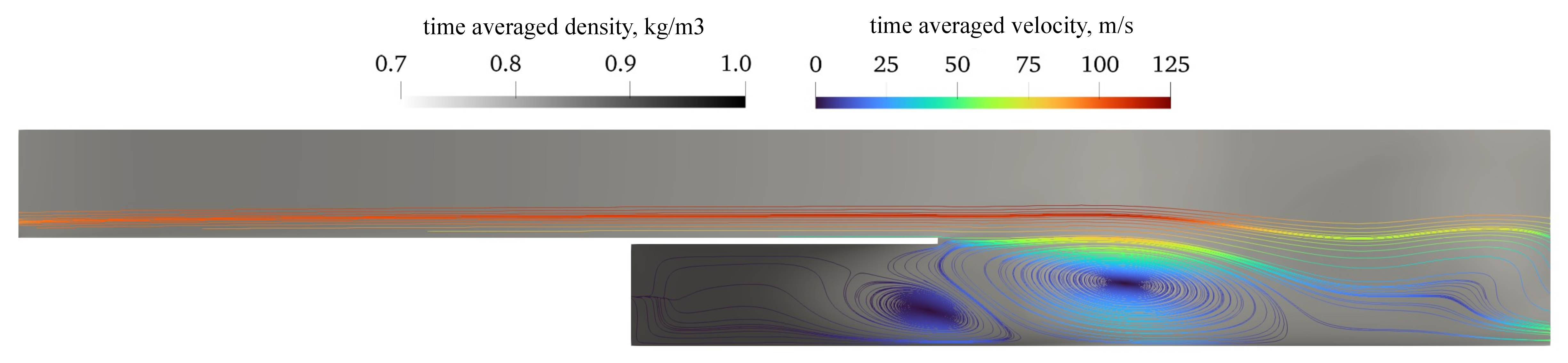}
        \caption{M = $0.50$}
    \end{subfigure}
    \hspace{2mm}
    \begin{subfigure}[h]{0.96\textwidth}
        \includegraphics[width=\textwidth]{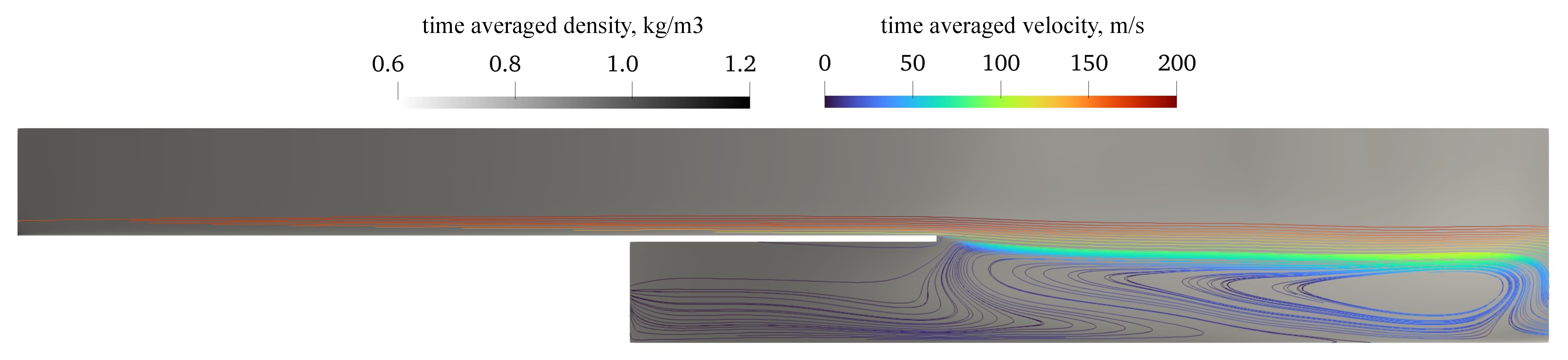}
        \caption{M = $0.95$}
    \end{subfigure}
    \hspace{2mm}
    \begin{subfigure}[h]{0.96\textwidth}
        \centering
        \includegraphics[width=\textwidth]{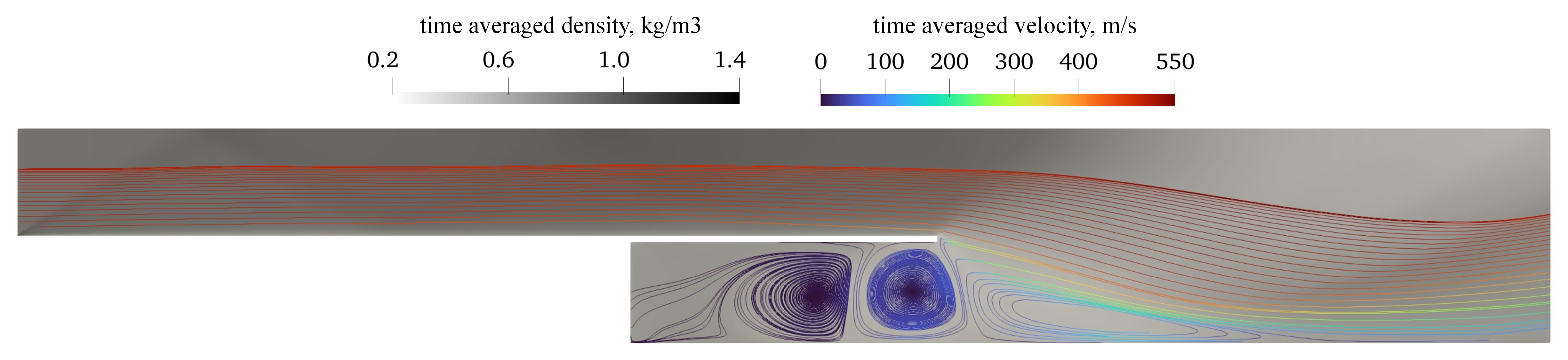}
        \caption{M = $2.11$}
    \end{subfigure}
    \hspace{2mm}
    \begin{subfigure}[h]{0.96\textwidth}
        \centering
        \includegraphics[width=\textwidth]{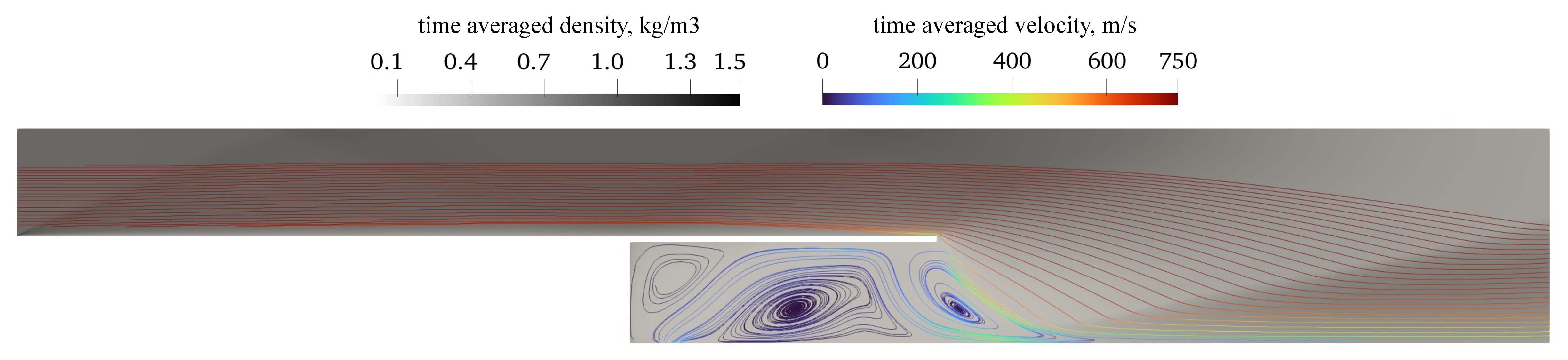}
        \caption{M = $3.00$}
    \end{subfigure}
    \caption{Time-averaged density fields, overlaid with velocity streamlines for initial transient conditions.}
    \label{fig:AllMachTransFlows}%
\end{figure*}

\begin{figure}[htbp!]
    \centering
    \begin{subfigure}[h]{0.5\textwidth}
        \centering
        \includegraphics[width=\textwidth]{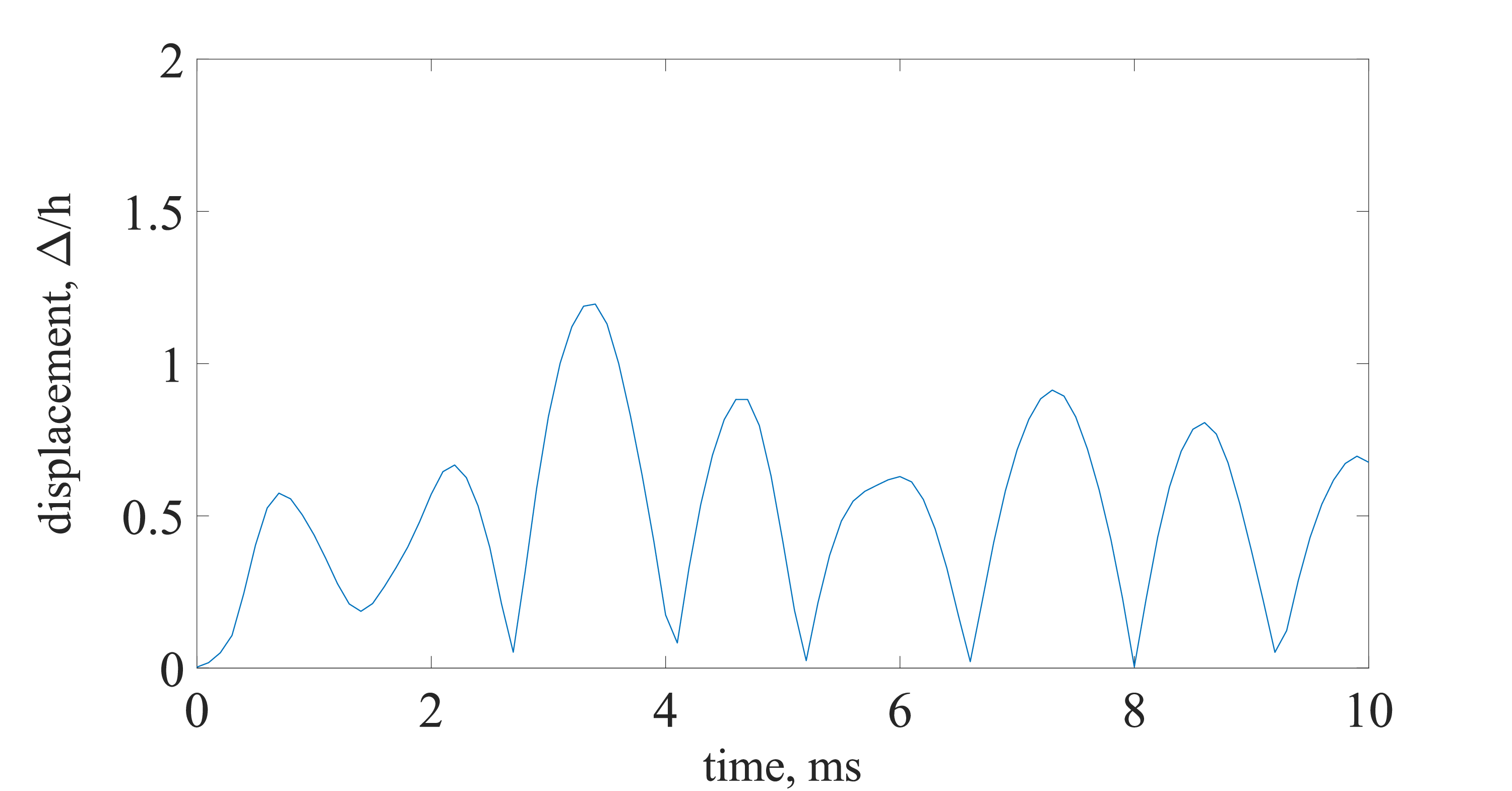}
        \caption{M = $0.50$}
    \end{subfigure}
    \hspace{2mm}
    \begin{subfigure}[h]{0.5\textwidth}
        \includegraphics[width=\textwidth]{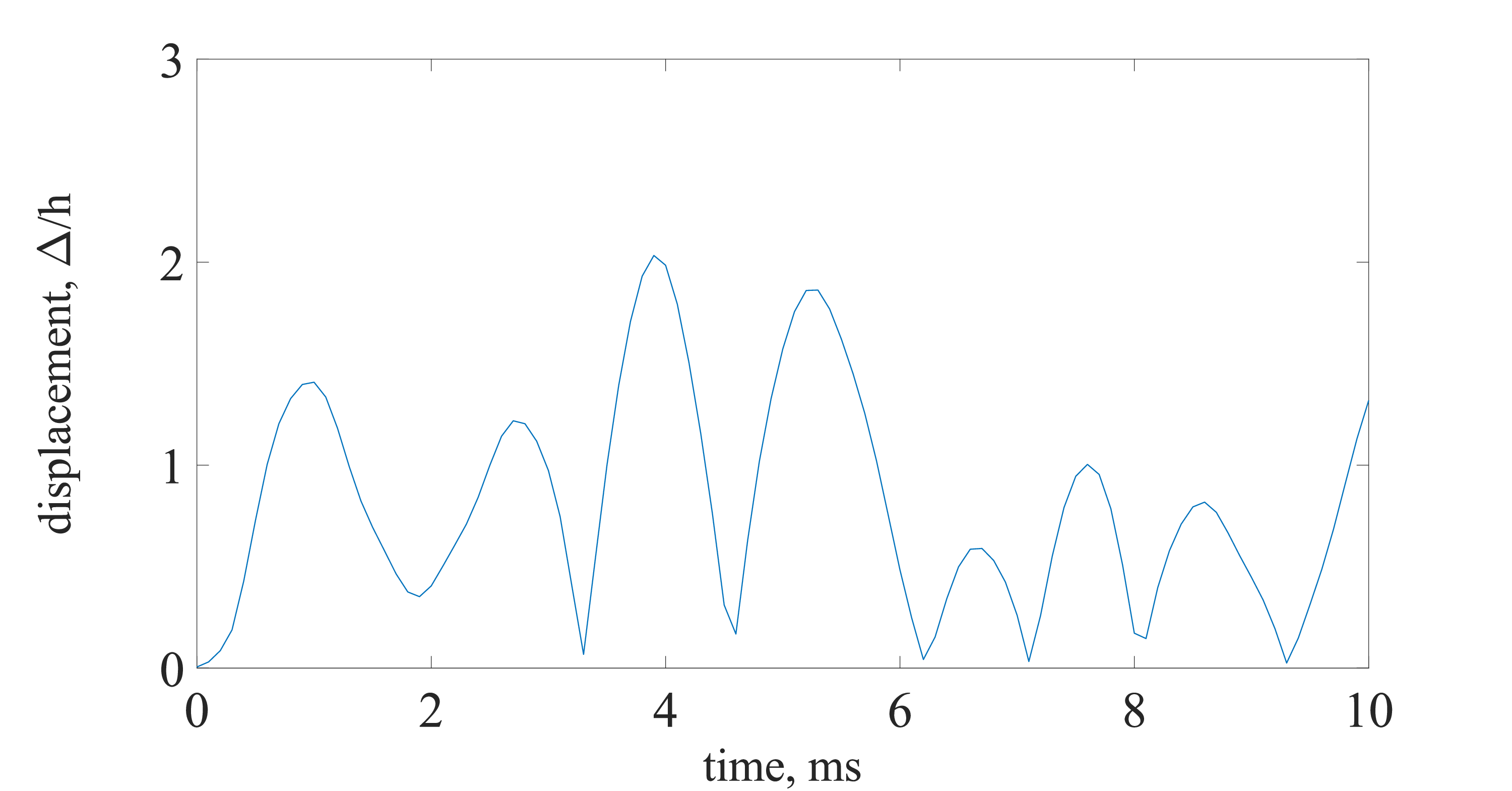}
        \caption{M = $0.95$}
    \end{subfigure}
    \hspace{2mm}
    \begin{subfigure}[h]{0.5\textwidth}
        \centering
        \includegraphics[width=\textwidth]{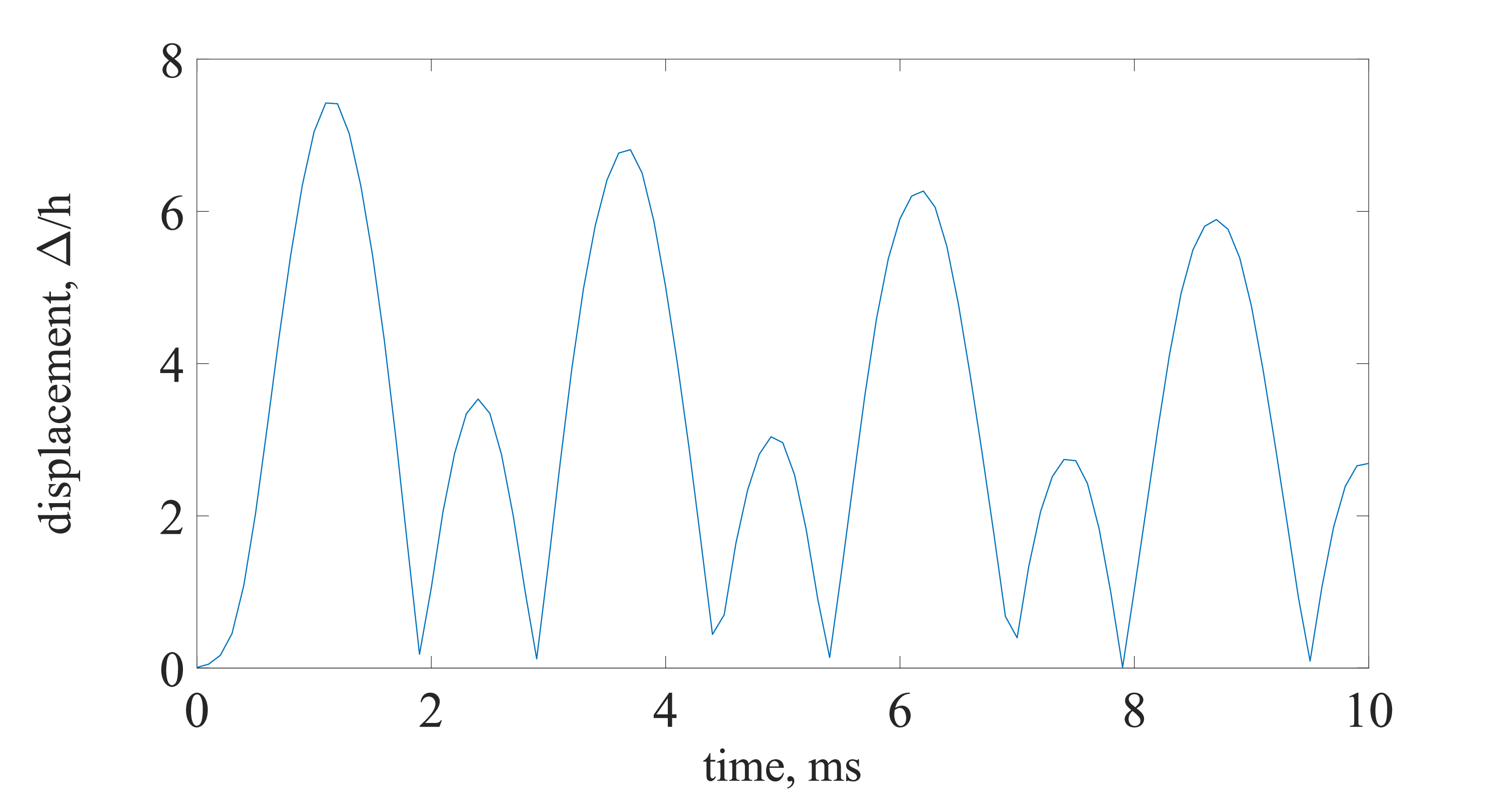}
        \caption{M = $2.11$}
    \end{subfigure}
    \hspace{2mm}
    \begin{subfigure}[h]{0.5\textwidth}
        \centering
        \includegraphics[width=\textwidth]{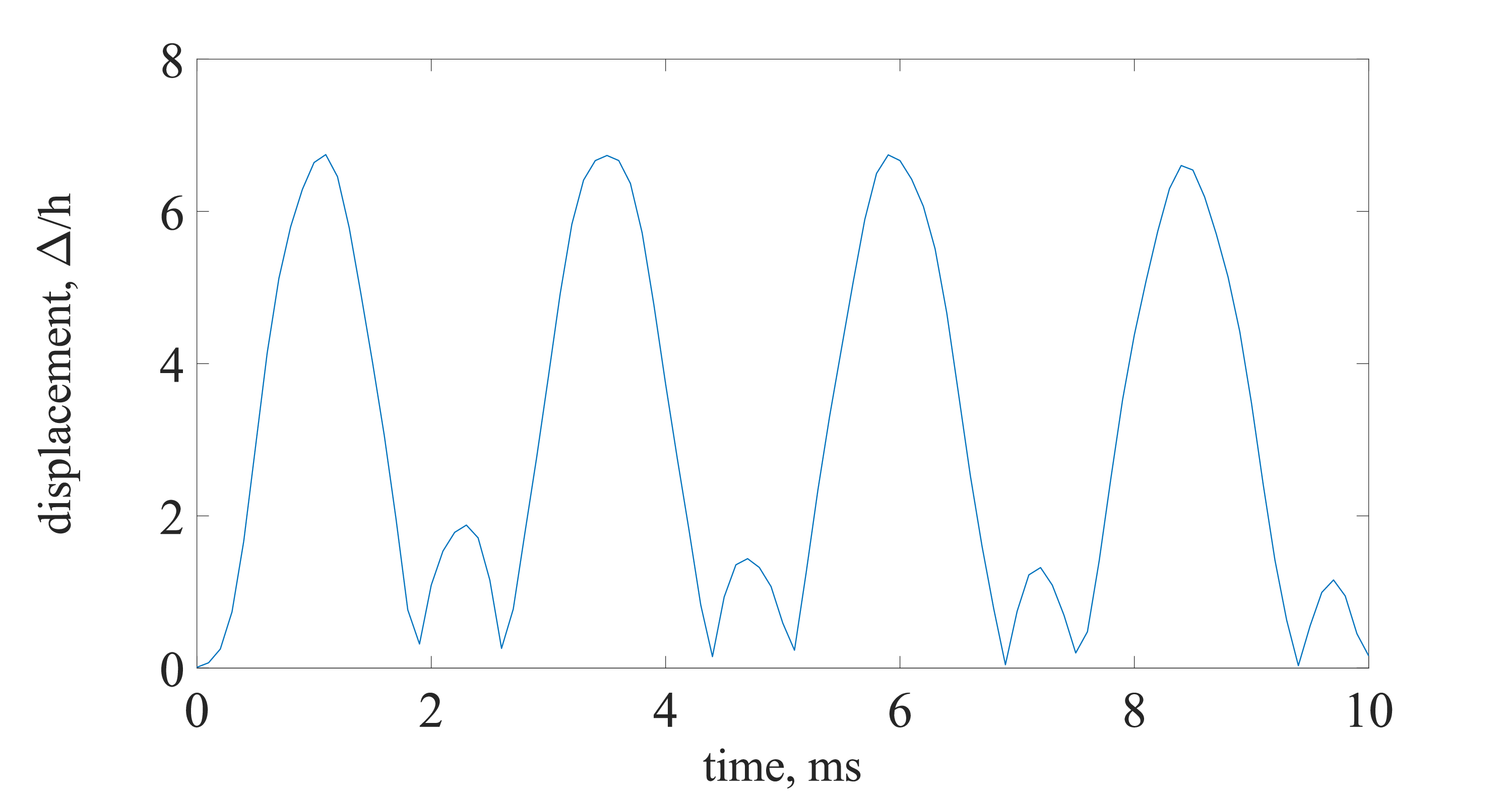}
        \caption{M = $3.00$}
    \end{subfigure}
    \caption{Tip displacement for the first 10 ms for Mach numbers of 0.5, 0.95, 2.11 and 3.}
    \label{fig:AllMachTransDisps}
\end{figure}

%Flow evolution and the corresponding thin panel deflections for Ma $2.11$ are shown in 
Figure \ref{fig:AllMachTransFlows} shows the time averaged (over the first 10 ms) density and streamlines for the four Mach numbers mentioned in the previous paragraph. As shown in figures \ref{fig:AllMachTransFlows}a and \ref{fig:AllMachTransFlows}b, at subsonic and transonic speeds, the shear layer does not adhere to the bottom wall downstream of the thin plate. This leads to stretching the tip vortex along the shear layer in both cases. Prominent vortices, trapped under the plate are observed for the two supersonic flow cases. The behaviors of the shear layer are dictated by the movement of the panel. To explain these behaviors, consider the tip displacement for the four cases shown in figure \ref{fig:AllMachTransDisps}. The displacements for Mach $0.50$ and $0.95$ show peaks at irregular intervals, whereas the displacements for Mach $2.11$ and $3.00$ look periodic. More importantly, the displacement amplitudes for the subsonic and transonic cases (up to 2 times the plate thickness) are at least 3-4 times smaller than those of the supersonic cases (up to 8 times the plate thickness). %are higher for the latter two cases reaching to almost $8$ times the plate thickness, $h$, as shown in figure \ref{fig:AllMachTransDisps}c, d, in comparison to the amplitudes of the other two cases at $2$ times the plate thickness. 
This is attributed to the higher fluid momentum for cases with higher Mach numbers leading to higher pressure loading on the plate. For Mach $2.11$ and Mach $3.00$, the displacement amplitude gradually decreases over time, indicating damping. These higher amplitudes of panel motion in case of higher Mach number flows lead to the motion of the shear layer, which when interacting with the boundary layer attaches to the bottom wall. This additionally enables the vortices formed at the plate's trailing edge to intensify and remain confined beneath the plate. These behaviors are absent for cases with Mach numbers of 0.5 and 0.95.

Another flow feature observed for the two supersonic cases is the weak shock train resulting from the interactions between the supersonic flow and the boundary layer; these are observed in the density contours in figures \ref{fig:AllMachTransFlows}c and d. 
%This damping behavior is more for Mach $2.11$, figure \ref{fig:AllMachTransDisps}c, when compared with Mach $3.00$, figure \ref{fig:AllMachTransDisps}d. Such behavior is not observed for Mach $0.50$ and $0.95$, figure \ref{fig:AllMachTransDisps}a, b, where the oscillations exhibit random behavior with irregular peaks.

%A large vortex is observed at the outer edge of the thin plate, which is not attached to the shear layer for Ma $0.50$, figure \ref{fig:AllMachTransFlows}a but gets attached for Ma $0.95$, figure \ref{fig:AllMachTransFlows}b. The shear layer completely restricts the high-speed flow to the upper region of the domain for both cases. The expansion fan and the reattachment shock are not observed and there is no recovery of the boundary layer due to the non-attachment of the shear layer to the bottom wall. The non-attachment of the shear-layer for Mach $0.50$ and $0.95$ further results in possible vortex shedding at low frequencies. 

%For Mach $2.11$ and $3.00$, figure \ref{fig:AllMachTransFlows}c, d, the shear layer gets attached to the bottom wall downstream of the thin-plate, leading to restriction of the translational or bulk movement of the vortices and formation of the recirculation regions underneath. The sudden expansion at higher Mach numbers results in a relatively smaller and more chaotic recirculation region for Ma $3.00$, figure \ref{fig:AllMachTransFlows}d, when compared to Ma $2.11$, figure \ref{fig:AllMachTransFlows}c, due to attachment of the shear layer to the bottom wall much closer to the thin-plate outer edge. The boundary layer is thicker and its recovery is clearly visible for Ma $2.11$, figure \ref{fig:AllMachTransFlows}c, in contrast to the Ma $3.00$, figure \ref{fig:AllMachTransFlows}d, which has a much thinner boundary layer, along the bottom wall. The effect of the fluid velocity on boundary layer is clearly visible for these two cases, with the higher flow Mach number leading to a thinner boundary layer along the bottom wall. These observations are supported by temperature contours, figure \ref{fig:AllMachTransTemps}. 
\begin{figure*}[htbp!]
    \centering
    \begin{subfigure}[h]{0.96\textwidth}
        \centering
        \includegraphics[width=\textwidth]{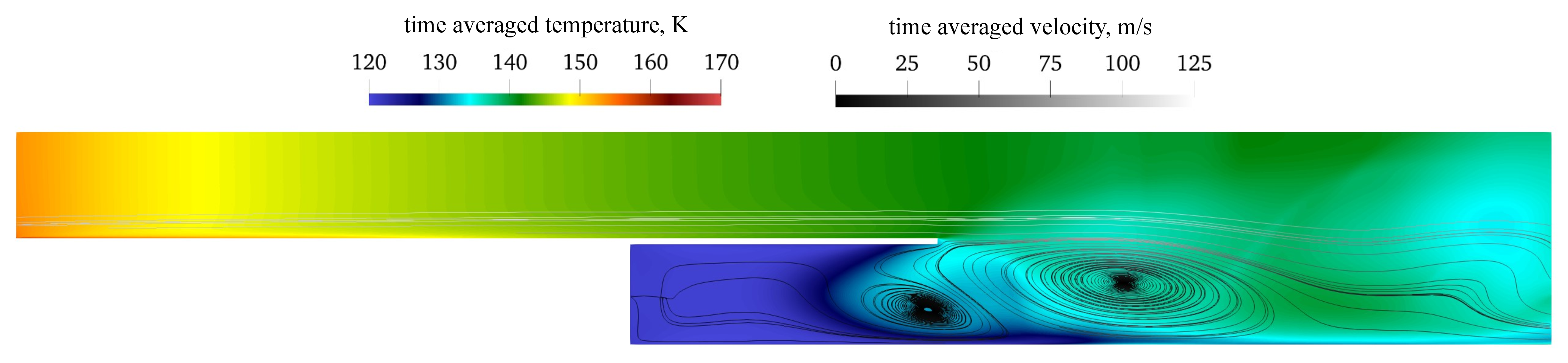}
        \caption{M = $0.50$}
    \end{subfigure}
    \hspace{2mm}
    \begin{subfigure}[h]{0.96\textwidth}
        \includegraphics[width=\textwidth]{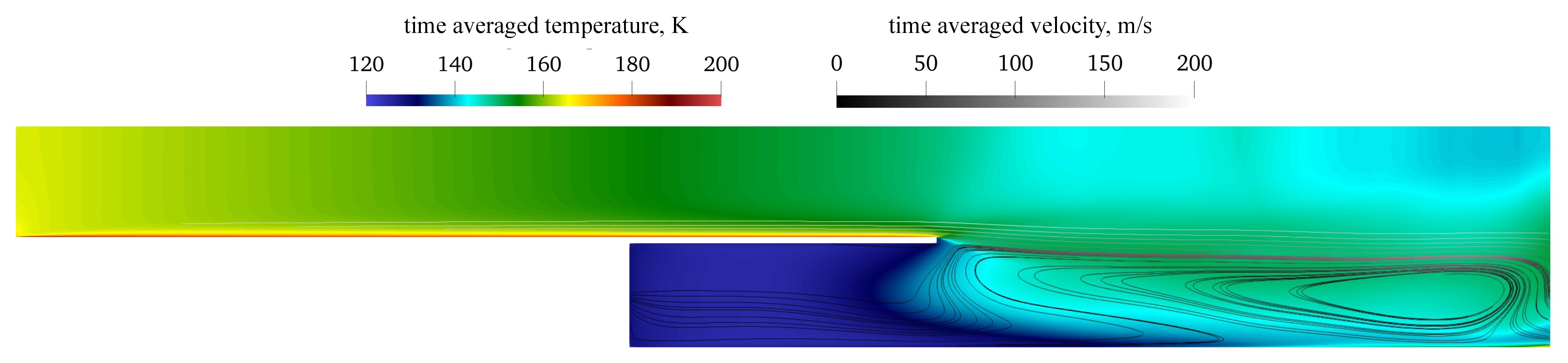}
        \caption{M = $0.95$}
    \end{subfigure}
    \hspace{2mm}
    \begin{subfigure}[h]{0.96\textwidth}
        \centering
        \includegraphics[width=\textwidth]{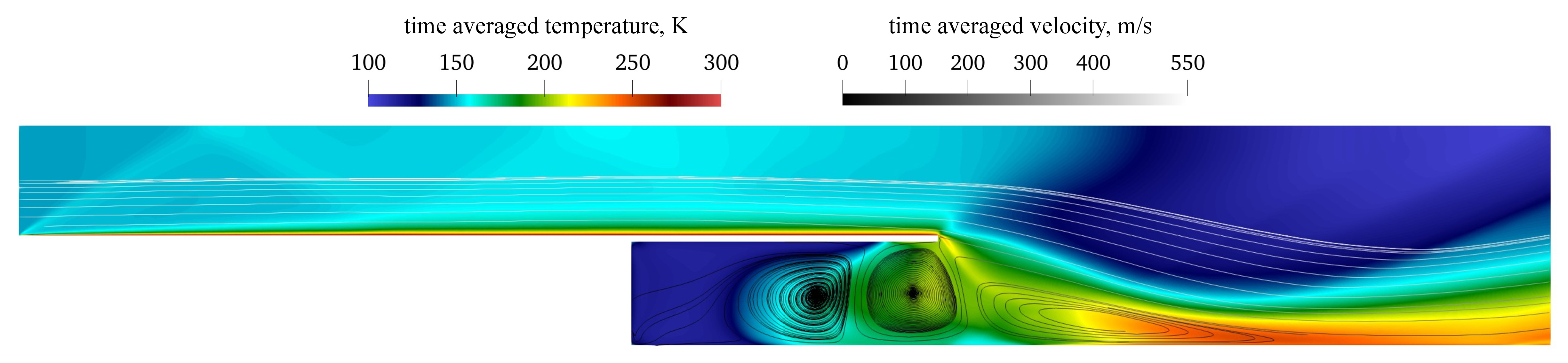}
        \caption{M = $2.11$}
    \end{subfigure}
    \hspace{2mm}
    \begin{subfigure}[h]{0.96\textwidth}
        \centering
        \includegraphics[width=\textwidth]{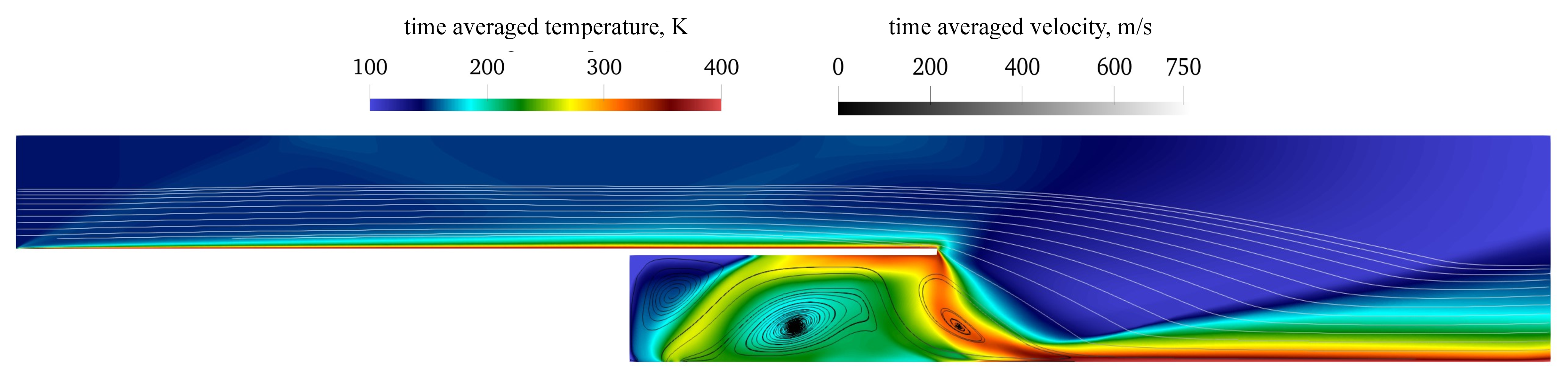}
        \caption{M = $3.00$}
    \end{subfigure}
    \caption{Time-averaged temperature fields, overlaid with velocity streamlines for initial transient conditions.}
    \label{fig:AllMachTransTemps}
\end{figure*}

Figures \ref{fig:AllMachTransTemps}a-d show the temperature contours overlaid with streamlines. Note that because of a wide variation in the range of temperature and velocities, to capture flow structures, the legends are different for each case. Similar to the velocity boundary layer near all the walls described in the previous paragraph, a thermal boundary layer is also observed with the highest temperatures near the wall, owing to dissipation, decreasing to the core flow value. The Mach 0.5 flow is weakly compressible and, consequently, there is no significant change in the core flow temperature except near the wall. For higher Mach number cases, other flow features, such as shock and expansion waves lead to significant variations in the core region's temperature field. For example, near the trailing edge of the panel, the temperature decreases across the expansion fan, and increases across the oblique shock waves, visible in figures \ref{fig:AllMachTransTemps}c and d. Similar to the boundary, there is significant dissipation in the shear layer, especially for the higher Mach number flows where velocity gradients are significant, resulting in higher temperature field in that region. %The highest temperatures for all cases are observed at the bottom wall of inlet section and the plate top surface, where a thin boundary layer appears with a large velocity gradient. Another region where relatively higher temperatures are observed is the region underneath the shear layer, which separates the high-speed flow from recirculating regions. The temperatures in this region are significantly lower for Mach $0.50$ and $0.95$, figure \ref{fig:AllMachTransTemps}a, b, where possible vortex shedding transports the heated gases out of this region as the shear layer does not get attached to the bottom boundary. When the flow speed is further increased to Mach $2.11$ and $3.00$, the shear layer gets reattached to the bottom wall leading to entrapment of the recirculation region vortices that eventually lead to viscous dissipation. Viscous dissipation elevates the temperatures within the shear layer to significantly higher values of $300$ $K$ for Mach $2.11$, figure \ref{fig:AllMachTransTemps}c, and $400$ $K$ for Mach $3.00$, figure \ref{fig:AllMachTransTemps}d. For both of these cases, the heated fluid transports the thermal energy outside the fluid domain along the recovering boundary layer at the bottom wall. The thermal boundary layer is observed to grow relatively thicker for Mach $2.11$ in comparison to Mach $3.00$ which has a larger velocity gradient.

%%%%%%%%%%%%%%%%%%%%%%%%%%%%%%%%%%%%%%%%%%%%%%%%%%%%%%%%%%%%%%%%%%%%%%%%%%%%%%%%%%%%%%%%%%%%%%%%%%%
\subsubsection{Panel Deformation and Stresses}
\begin{figure*}[htbp!]
    \centering
    \begin{subfigure}[h]{0.48\textwidth}
        \centering
        \includegraphics[width=\textwidth]{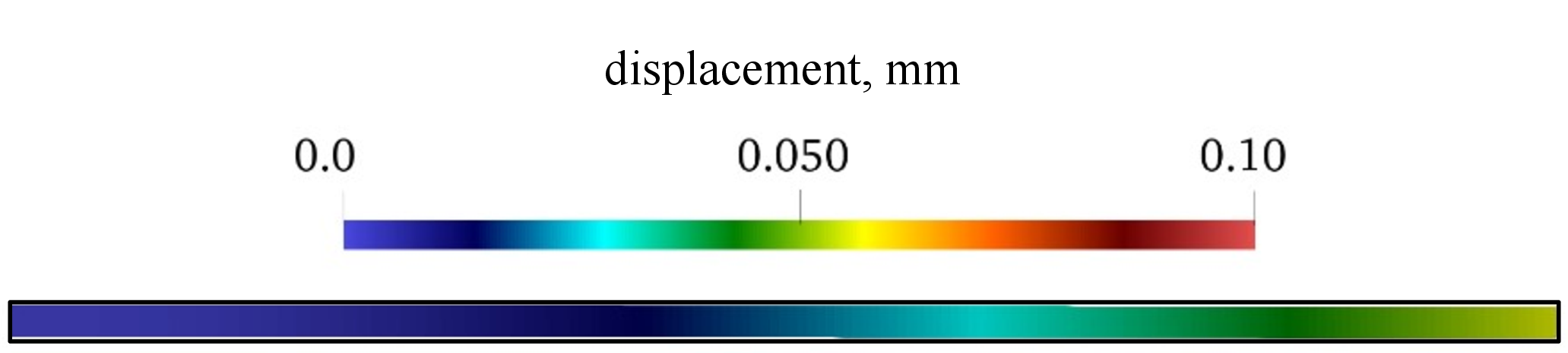}
        \caption{M $0.50$}
    \end{subfigure}
    \hspace{2mm}
    \begin{subfigure}[h]{0.48\textwidth}
        \includegraphics[width=\textwidth]{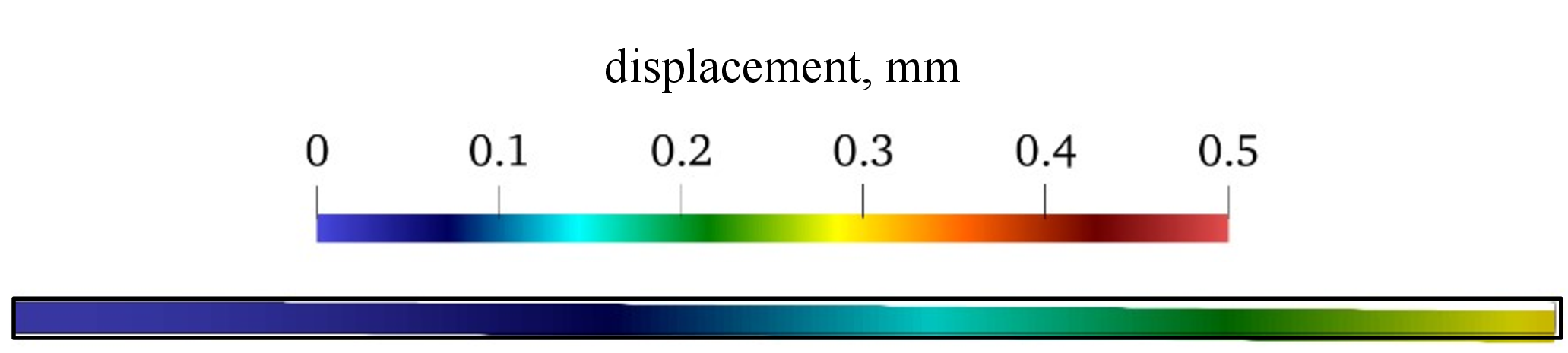}
        \caption{M $0.95$}
    \end{subfigure}
    \hspace{2mm}
    \begin{subfigure}[h]{0.48\textwidth}
        \centering
        \includegraphics[width=\textwidth]{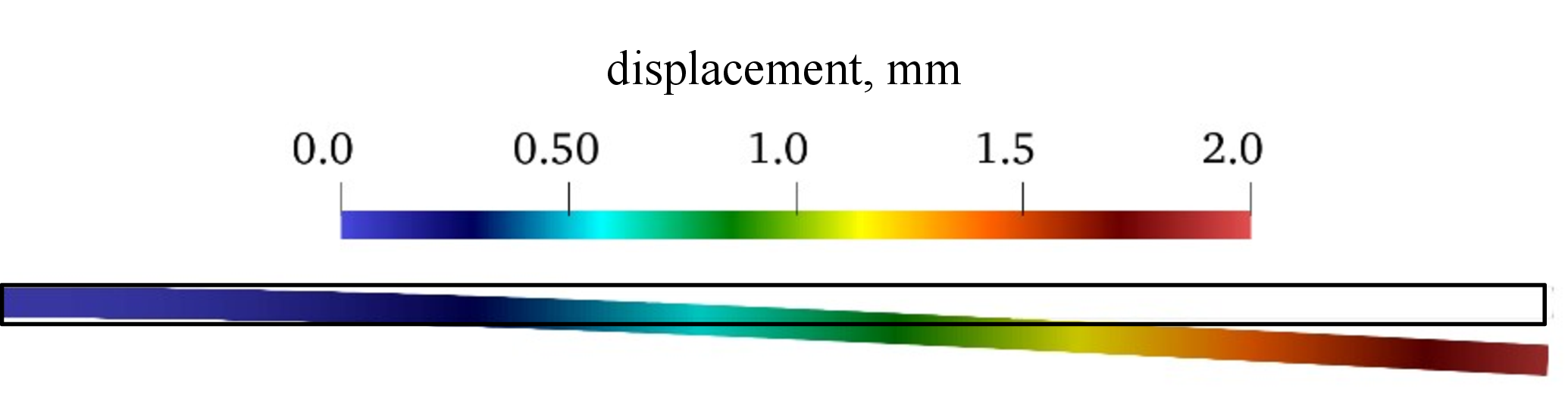}
        \caption{M $2.11$}
    \end{subfigure}
    \hspace{2mm}
    \begin{subfigure}[h]{0.48\textwidth}
        \centering
        \includegraphics[width=\textwidth]{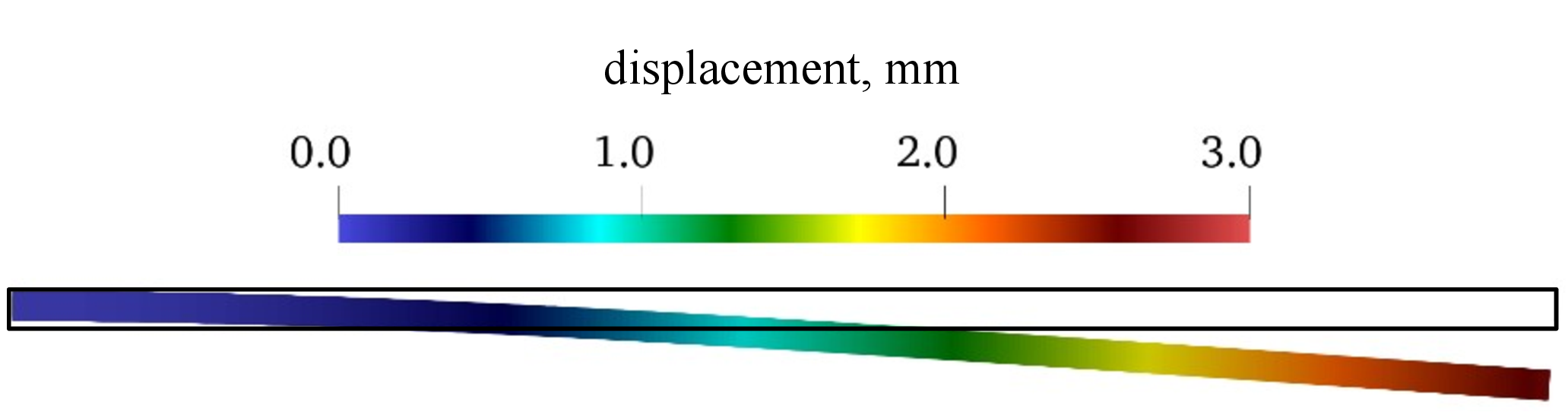}
        \caption{M $3.00$}
    \end{subfigure}
    \caption{Time averaged thin-panel displacements for all Mach numbers. Outline shows the initial undeformed panel.}
    \label{fig:AllMachTransTimeAvgDisps}
\end{figure*}

To understand the behavior of initial transient flow Mach number on the  panel behavior, time averaged panel deformations are compared for the four cases in figure \ref{fig:AllMachTransTimeAvgDisps}. A clear trend of increase in panel deformations with flow Mach number is observed. The deformation for the Mach numbers of $0.50$ and $0.95$ is significantly lower than the panel thickness of $1.02$ $mm$, indicating that the flow does not have enough momentum to significantly deform the panel. For Mach $2$ case, the deformation at the tip is two times the panel thickness which increases to three times when the flow Mach is increases to $3$. This is attributed to the higher momentum of the flow leading to higher pressure loading which leads to increased panel bending and deformation.

To characterize the motion of the panel, the dominant frequency of the oscillation was calculated by performing an FFT of the time evolution of the displacement of the panel tip, shown previously in figure \ref{fig:AllMachTransDisps}. The first peak in the frequency spectrum for each case is list in the table \ref{tab:transdispfreqs} below. %, The oscillation frequency was estimated by averaging the time between corresponding peaks in the displacement-time plot to obtain a time-averaged time-period for the thin-plate oscillations. The measured frequencies are shown below, in table \ref{tab:transdispfreqs}. 
\hl{A generic trend of increasing frequency with increasing Mach number is observed, with Mach $0.95$ reporting the same frequency as the Mach $2.11$.} Higher flow momentum at higher Mach numbers results in increased pressure fluctuations leading to higher frequency pressure loading of the panel.
\begin{table}[htbp!]
\caption{Thin-plate oscillation frequencies for initial transient conditions for the four Mach numbers under consideration.}
\begin{tabular}{ccc}
    \hline
    Mach & & $1^{st}$ Peak \\
    number & & Hz \\
    \hline \hline
    $0.50$ & & $380$ \\
    $0.95$ & & $395$ \\
    $2.11$ & & $395$ \\
    $3.00$ & & $411$ \\
    \hline
    \end{tabular}
    
    \label{tab:transdispfreqs}
\end{table}

\begin{figure*}[htbp!]
    \centering
    \begin{subfigure}[h]{0.48\textwidth}
        \centering
        \includegraphics[width=\textwidth]{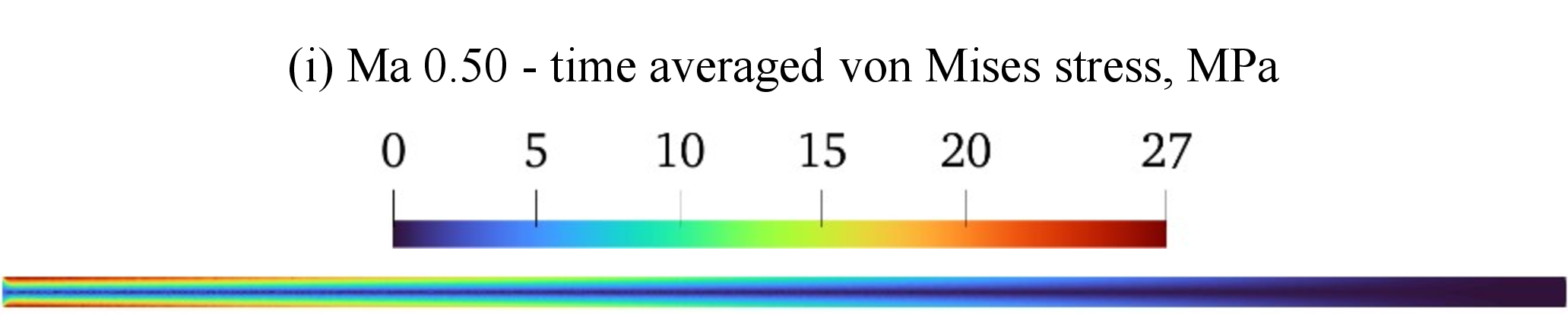}
        \includegraphics[width=\textwidth]{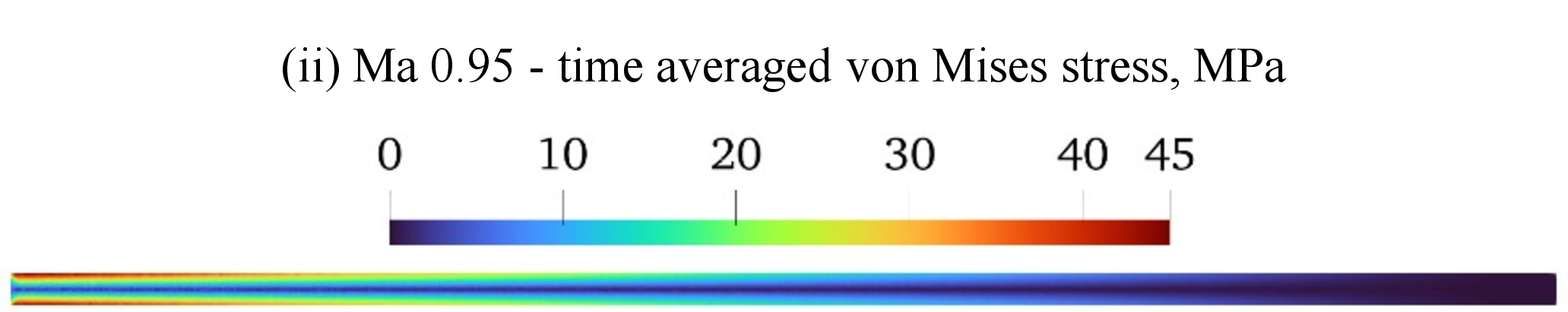}
        \includegraphics[width=\textwidth]{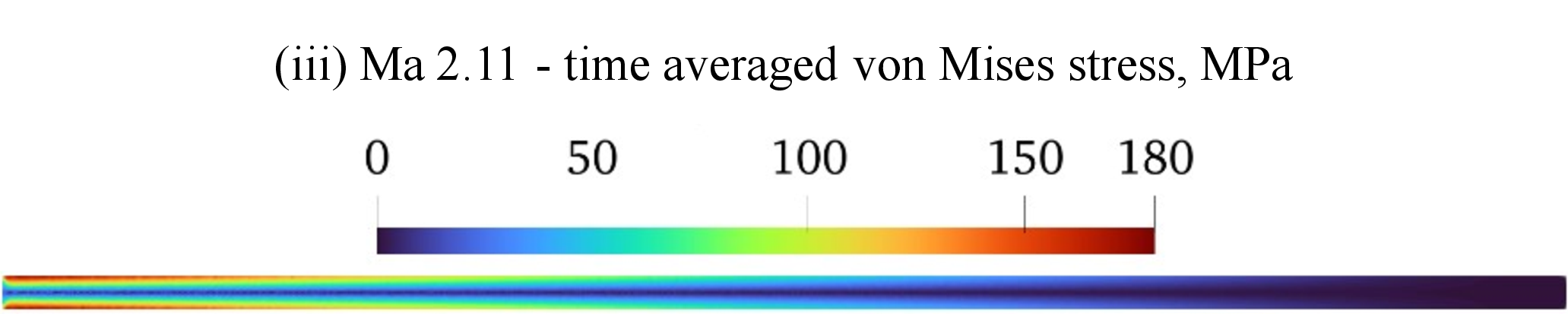}
        \includegraphics[width=\textwidth]{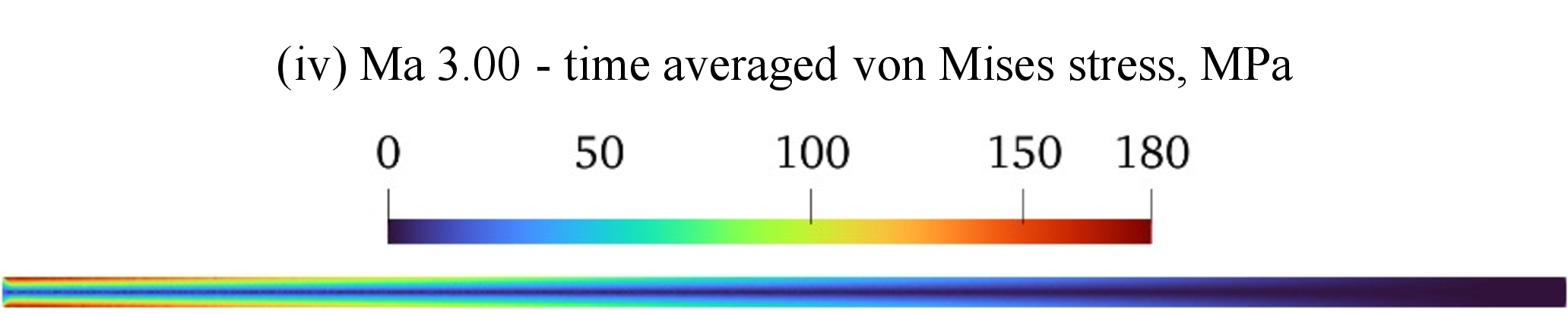}
        \caption{von Mises stress}
    \end{subfigure}
    \hspace{2mm}
    \begin{subfigure}[h]{0.48\textwidth}
        \includegraphics[width=\textwidth]{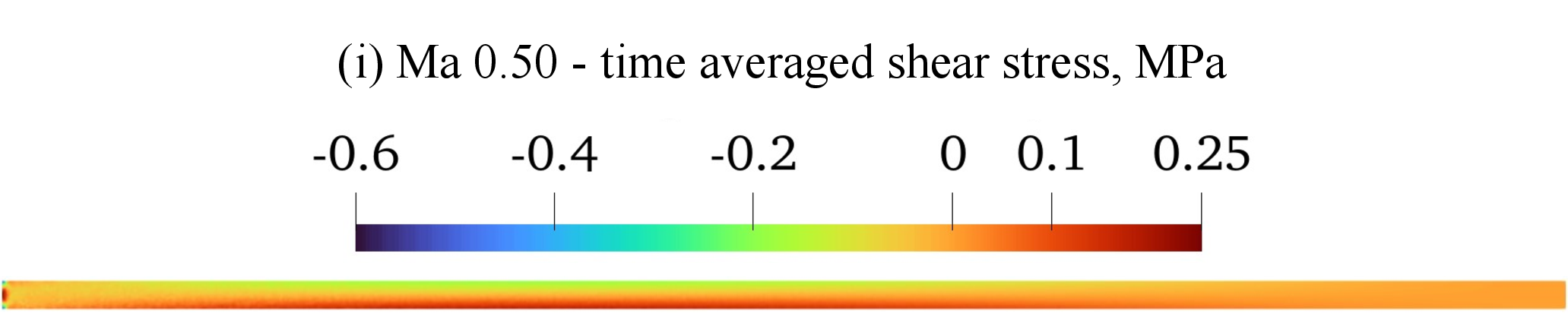}
        \includegraphics[width=\textwidth]{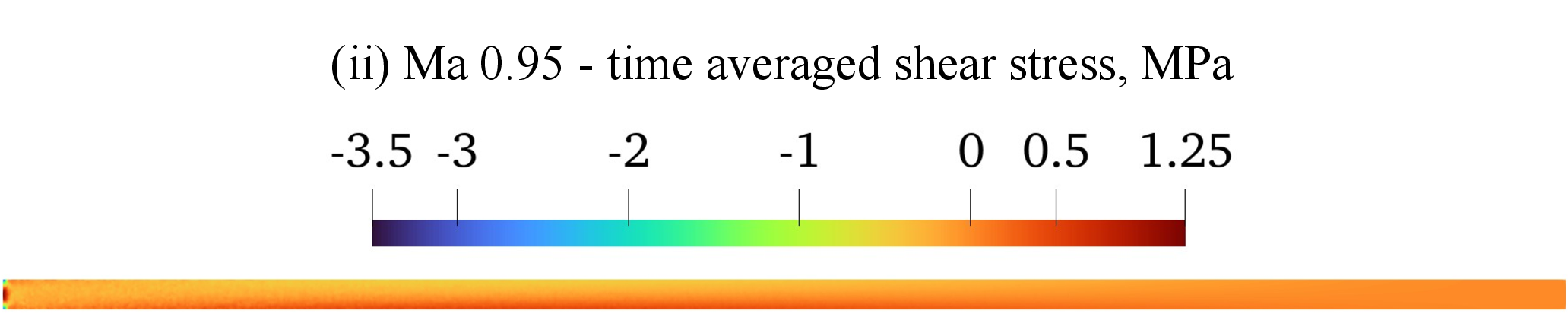}
        \includegraphics[width=\textwidth]{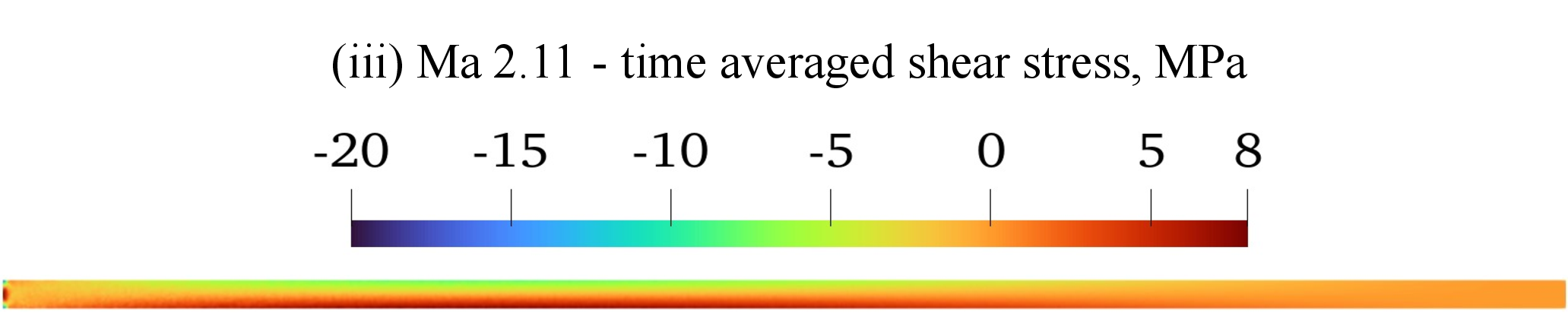}
        \includegraphics[width=\textwidth]{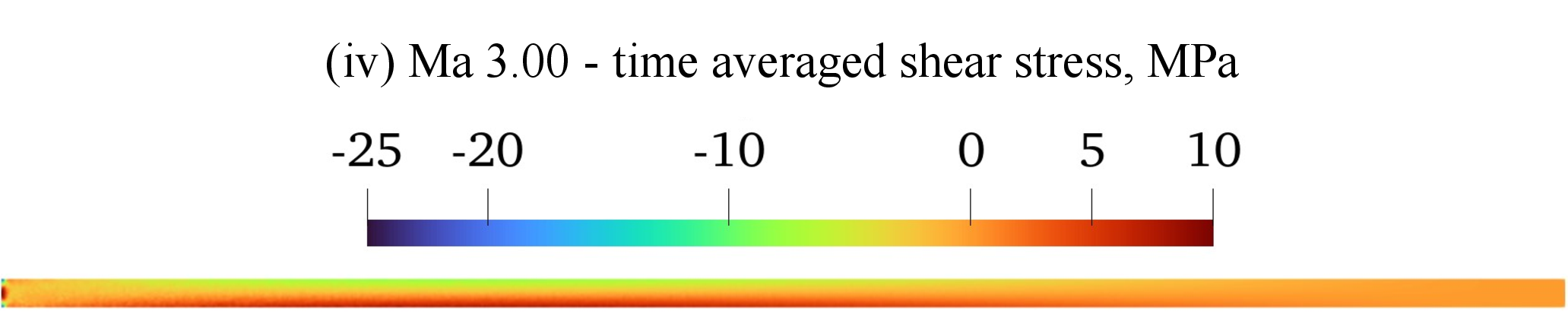}
        \caption{shear stress}
    \end{subfigure}
    \caption{Variation of von Mises and shear stress with Mach number for thin-aluminum plate for initial transient conditions.}
    \label{fig:AllMachTransStresses}
\end{figure*}

Figures \ref{fig:AllMachTransStresses}a and b show the von Mises and shear stresses averaged over time along the length of the panel, respectively, for the four Mach numbers investigated in the paper. , figure \ref{fig:AllMachTransStresses}a and the shear stress, figure \ref{fig:AllMachTransStresses}b, are compared for changes with Mach number. An increase is observed in the von Mises stress from Mach $0.50$ to $2.11$, figure \ref{fig:AllMachTransStresses}a: i, ii, iii, which then remains unchanged from $2.11$ to $3.00$, figure \ref{fig:AllMachTransStresses}a: iii, iv. As discussed previously, increase in the flow Mach number leads to increase in the pressure loading which causes the panel to bend more resulting in the increased von Mises stress at the clamped edge. This trend suggests that the plate is at its highest von Mises stress for Mach numbers $2.11$ and $3.00$ and it can lead to significant non-linear deformation with increased loading by increasing the flow Mach number. For all four cases, the von Mises stress has maximum values close to the face attached to the backward facing step in figure \ref{fig:bojan_schematic}. The maximum shear stress is observed at the base of the thin-plate, where it is attached to the backward facing step. The shear stress shows a consistent increase with increased loading, figure \ref{fig:AllMachTransStresses}b: i to iv, with the cantilevered plate experiencing higher shear stresses along the middle of the bottom edge. As discussed earlier, the increase in Mach number results in increased pressure loading of the plate which results in higher shear stresses.

%% file: 06bfull400ms.tex
%%%%%%%%%%%%%%%%%%%%%%%%%%%%%%%%%%%%%%%%%%%%%%%%%%%%%%%%%%%%%%%%%%%%%%%%%%%%%%%%%%%%%%%%%%%%%%%%%%%
% Fully Started Conditions - 400 ms:
%%%%%%%%%%%%%%%%%%%%%%%%%%%%%%%%%%%%%%%%%%%%%%%%%%%%%%%%%%%%%%%%%%%%%%%%%%%%%%%%%%%%%%%%%%%%%%%%%%%
To capture fully-started conditions, the simulation is run for a physical time of $400$ $ms$, or $1001$ snapshots, with a time step of $4$ $\times$ $10^{-4}$. Mach $0.50$, $2.11$ and $3.00$ run for the full simulation time while Mach $0.95$ case runs only for $0.1$ $s$ or $250$ snapshots.

%%%%%%%%%%%%%%%%%%%%%%%%%%%%%%%%%%%%%%%%%%%%%%%%%%%%%%%%%%%%%%%%%%%%%%%%%%%%%%%%%%%%%%%%%%%%%%%%%%%
\subsubsection{Flow Behaviors}
\begin{figure*}[htbp!]
    \centering
    \begin{subfigure}[h]{0.64\textwidth}
        \centering
        \includegraphics[width=\textwidth]{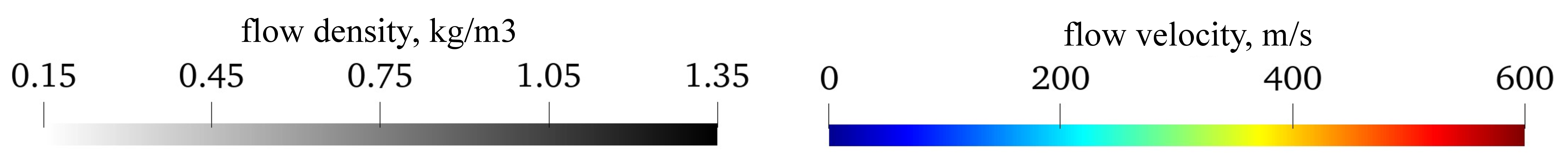}
    \end{subfigure}
    \hspace{2mm}
    \begin{subfigure}[h]{0.48\textwidth}
        \centering
        \includegraphics[width=\textwidth]{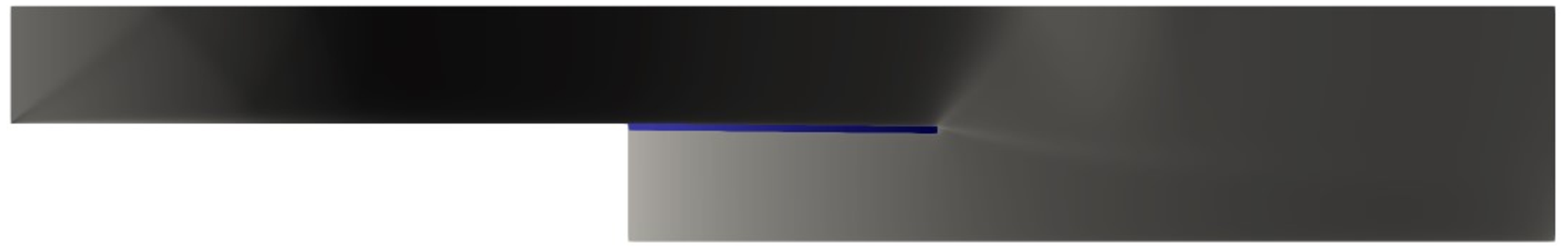}
        \caption{t = $0.0$ ms}
    \end{subfigure}
    \hspace{2mm}
    \begin{subfigure}[h]{0.48\textwidth}
        \includegraphics[width=\textwidth]{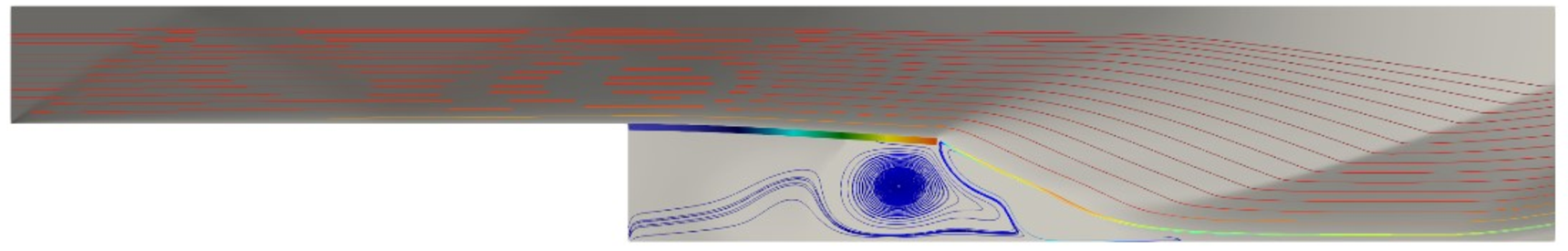}
        \caption{t = $50.0$ ms}
    \end{subfigure}
    \hspace{2mm}
    \begin{subfigure}[h]{0.48\textwidth}
        \centering
        \includegraphics[width=\textwidth]{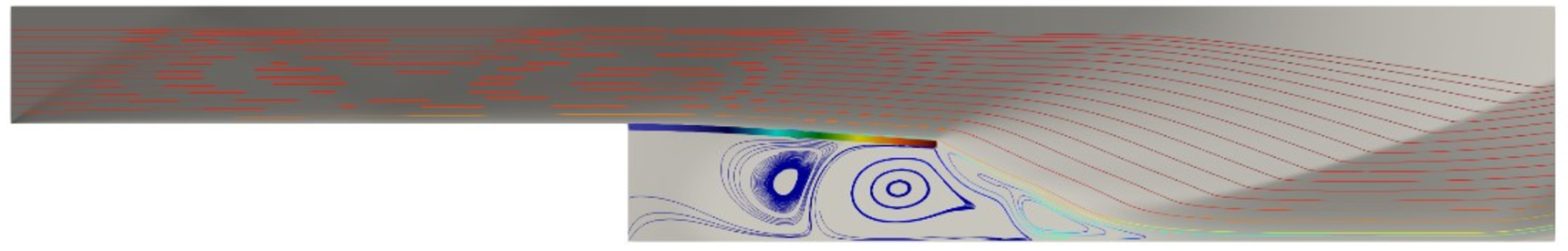}
        \caption{t = $100.0$ ms}
    \end{subfigure}
    \hspace{2mm}
    \begin{subfigure}[h]{0.48\textwidth}
        \centering
        \includegraphics[width=\textwidth]{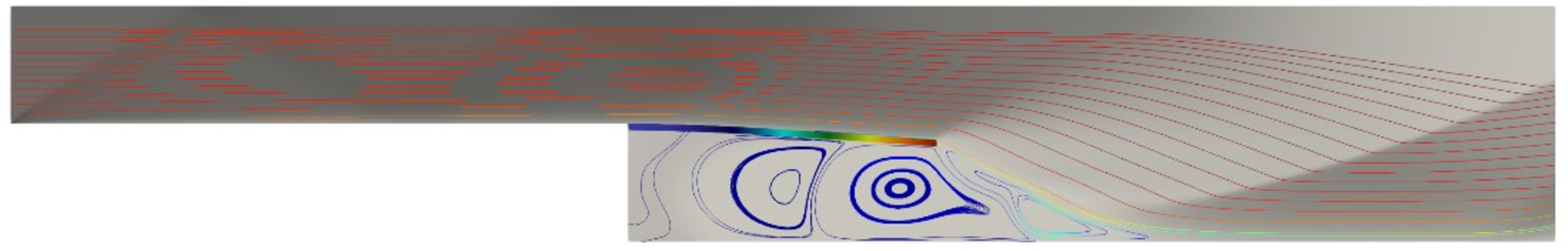}
        \caption{t = $200.0$ ms}
    \end{subfigure}
        \hspace{2mm}
    \begin{subfigure}[h]{0.48\textwidth}
        \centering
        \includegraphics[width=\textwidth]{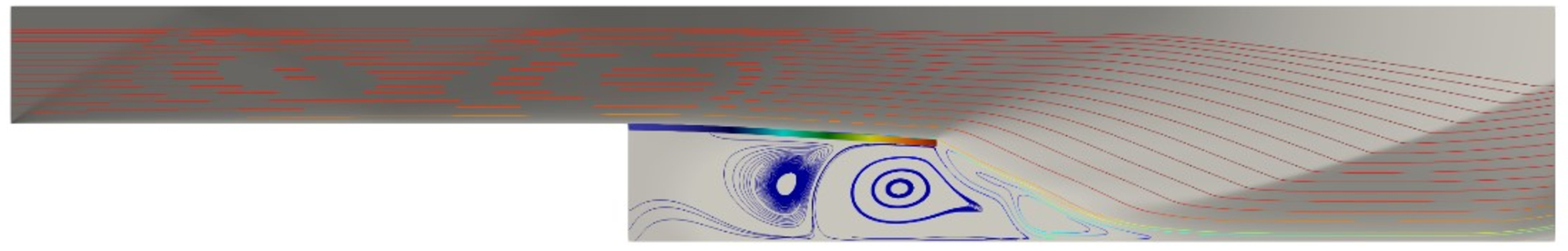}
        \caption{t = $300.0$ ms}
    \end{subfigure}
    \hspace{2mm}
    \begin{subfigure}[h]{0.48\textwidth}
        \centering
        \includegraphics[width=\textwidth]{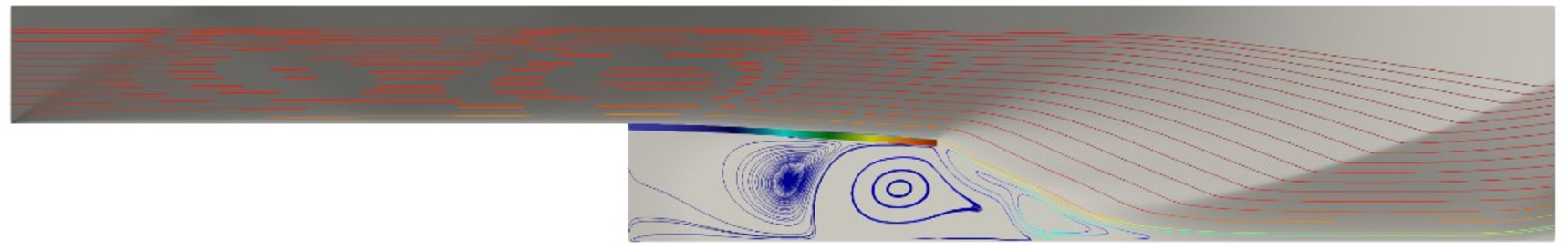}
        \caption{t = $400.0$ ms}
    \end{subfigure}
    \hspace{2mm}
    \begin{subfigure}[h]{0.32\textwidth}
        \centering
        \includegraphics[width=\textwidth]{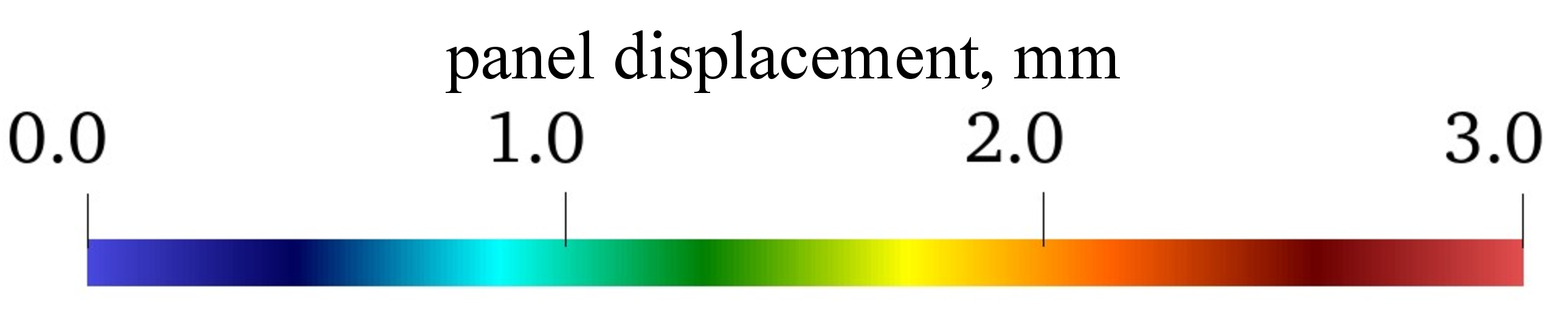}
    \end{subfigure}
    \caption{Temporal evolution of Ma $2.11$ flow, showing velocity streamlines on density contours and compliant panel displacement for fully started conditions.}
    \label{fig:Mach2EvolveFull}
\end{figure*}

Flow evolution and the corresponding thin panel deflections for Ma $2.11$ are shown in figure \ref{fig:Mach2EvolveFull}, for the times of $0$, $50$, $100$, $200$, $300$ and $400$ $ms$ after the flow is initialized. The shear layer, separating the high-speed flow from the stationary air along the bottom section descends to the bottom wall trapping a pocket of air behind it and under the thin panel by $50$ $ms$, figure \ref{fig:Mach2EvolveFull}b. This prevents vortex shedding underneath the shear layer and leads to recirculation that increases the number or vortices formed, figure \ref{fig:Mach2EvolveFull}c-f. For subsequent times up to $400$ $ms$, figure \ref{fig:Mach2EvolveFull}c-f, these vortices remain stationary and lead to viscous dissipation resulting in localized thermal hotspots that will be discussed later in this section. 

\begin{figure*}[htbp!]
    \centering
    \begin{subfigure}[h]{0.96\textwidth}
        \centering
        \includegraphics[width=\textwidth]{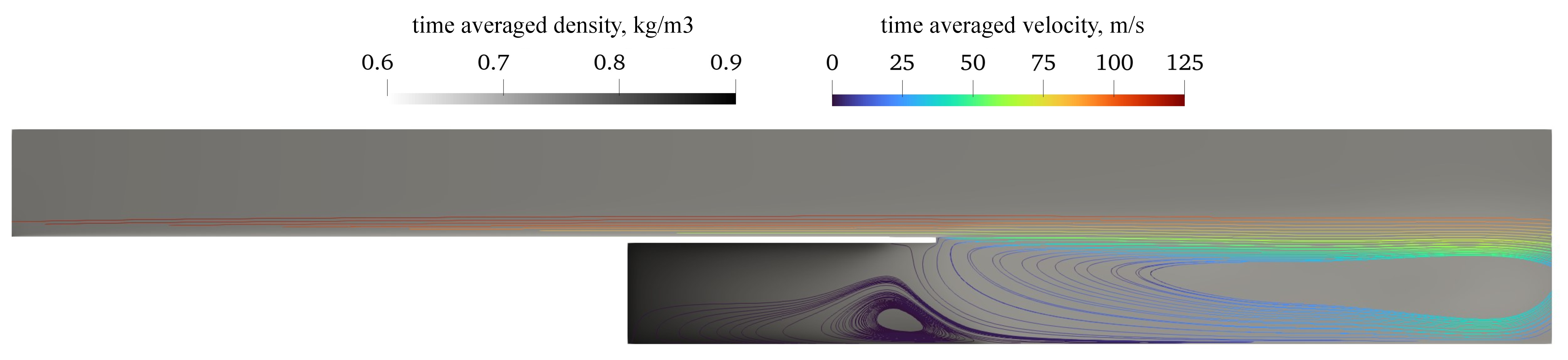}
        \caption{Ma $0.50$}
    \end{subfigure}
    \hspace{2mm}
    \begin{subfigure}[h]{0.96\textwidth}
        \includegraphics[width=\textwidth]{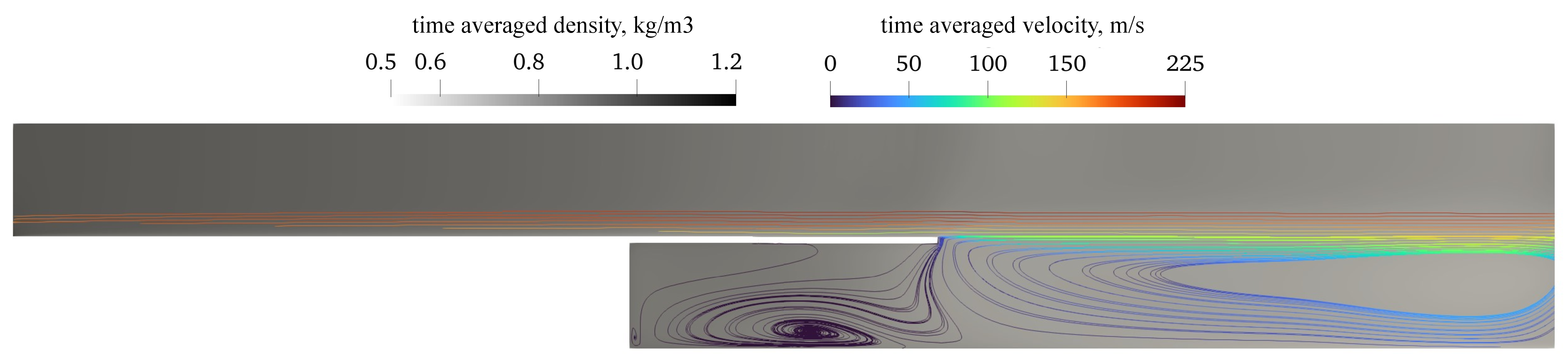}
        \caption{Ma $0.95$}
    \end{subfigure}
    \hspace{2mm}
    \begin{subfigure}[h]{0.96\textwidth}
        \centering
        \includegraphics[width=\textwidth]{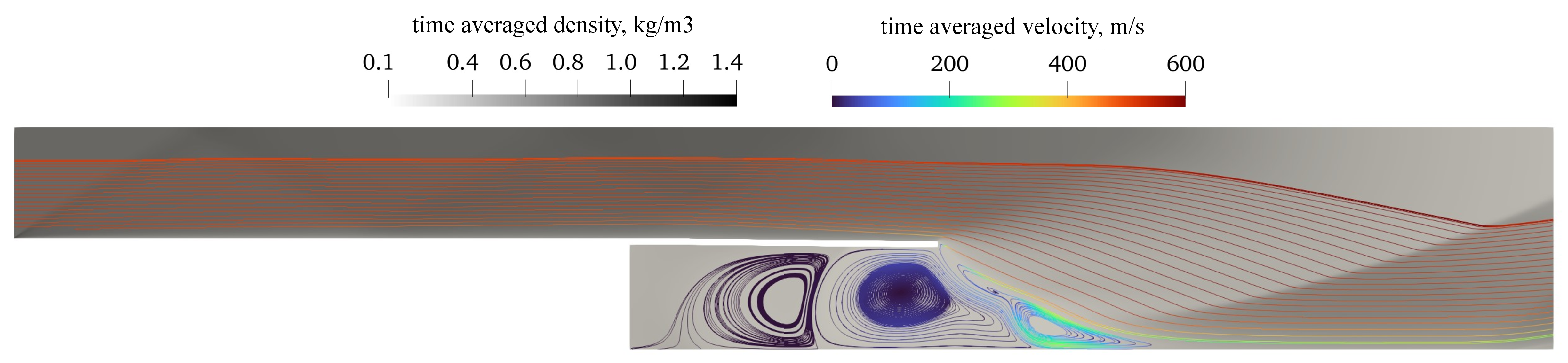}
        \caption{Ma $2.11$}
    \end{subfigure}
    \hspace{2mm}
    \begin{subfigure}[h]{0.96\textwidth}
        \centering
        \includegraphics[width=\textwidth]{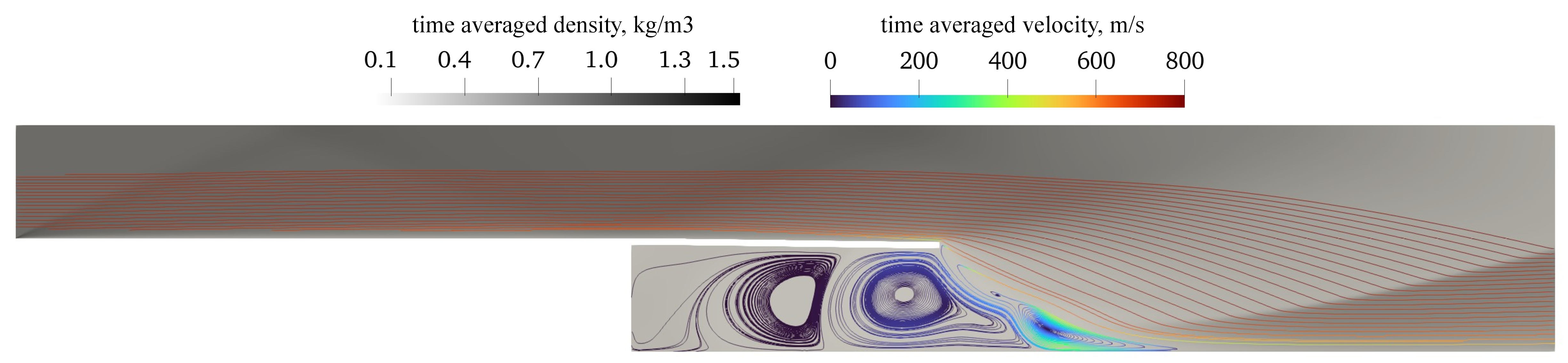}
        \caption{Ma $3.00$}
    \end{subfigure}
    \caption{Time-averaged density fields, overlaid with velocity streamlines for fully started conditions.}
    \label{fig:AllMachFullFlows}
\end{figure*}

The time averaged flow fields for fully started conditions are shown in figure \ref{fig:AllMachFullFlows}. The shear layer separates the high speed flow region at the top with near stagnant flow at the bottom for Mach $0.50$ and $0.95$ cases, figure \ref{fig:AllMachFullFlows}a, b. Large recirculating vortices are observed for these two cases, with possible low frequency vortex shedding due to no entrapment by the shear layer. The vortex that was attached to outer edge of the thin plate, but not to the shear layer for Mach $0.50$, figure \ref{fig:AllMachTransFlows}a, now gets attached to the shear layer, figure \ref{fig:AllMachFullFlows}a. The shear layer completely restricts the high-speed flow to the upper region of the domain for Mach $0.50$ and $0.95$, and features such as the expansion fan, reattachment shock and recovery of the boundary layer are not observed due to the non-attachment of the shear layer to the bottom wall, figure \ref{fig:bojan_schematic}, for either case.

\begin{figure*}[htbp!]
    \centering
    \begin{subfigure}[h]{0.96\textwidth}
        \centering
        \includegraphics[width=\textwidth]{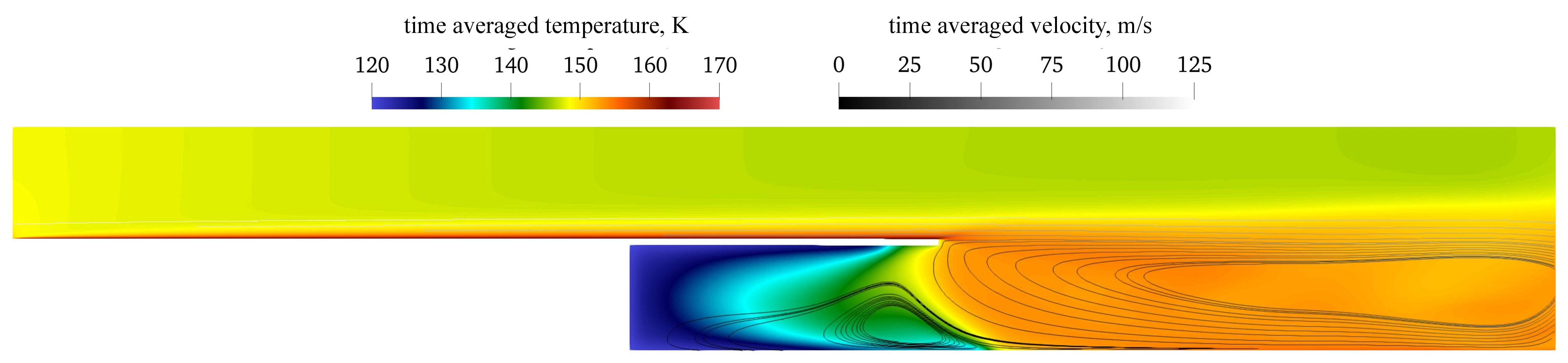}
        \caption{Ma $0.50$}
    \end{subfigure}
    \hspace{2mm}
    \begin{subfigure}[h]{0.96\textwidth}
        \includegraphics[width=\textwidth]{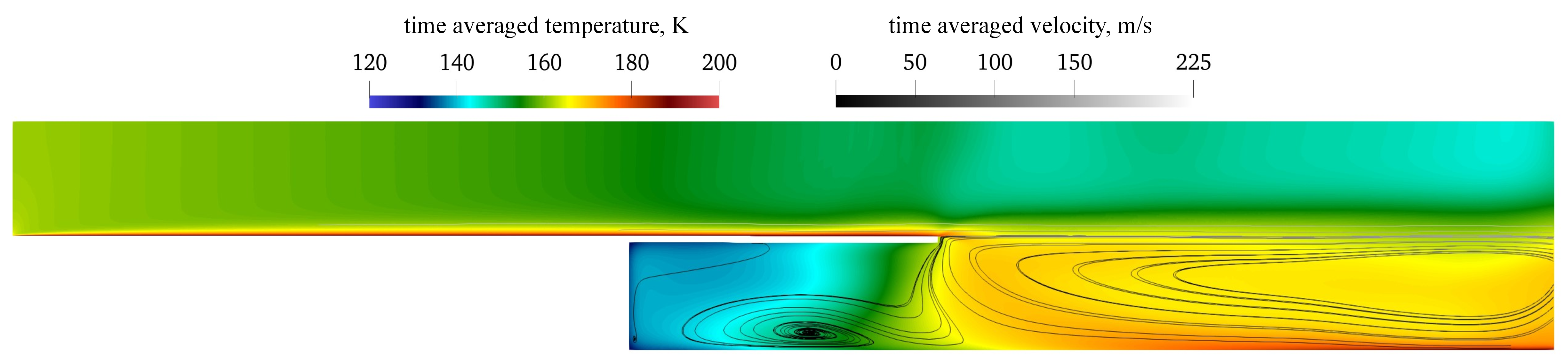}
        \caption{Ma $0.95$}
    \end{subfigure}
    \hspace{2mm}
    \begin{subfigure}[h]{0.96\textwidth}
        \centering
        \includegraphics[width=\textwidth]{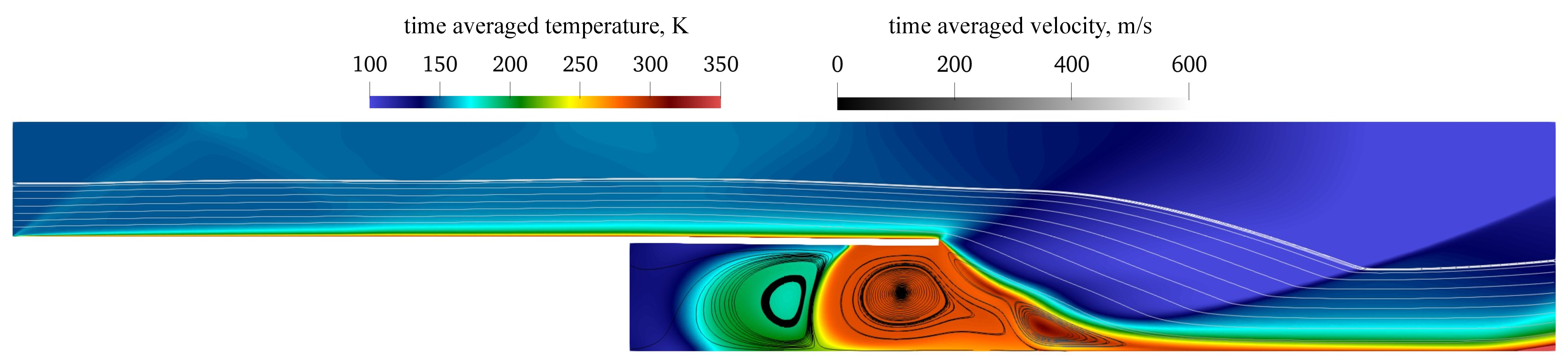}
        \caption{Ma $2.11$}
    \end{subfigure}
    \hspace{2mm}
    \begin{subfigure}[h]{0.96\textwidth}
        \centering
        \includegraphics[width=\textwidth]{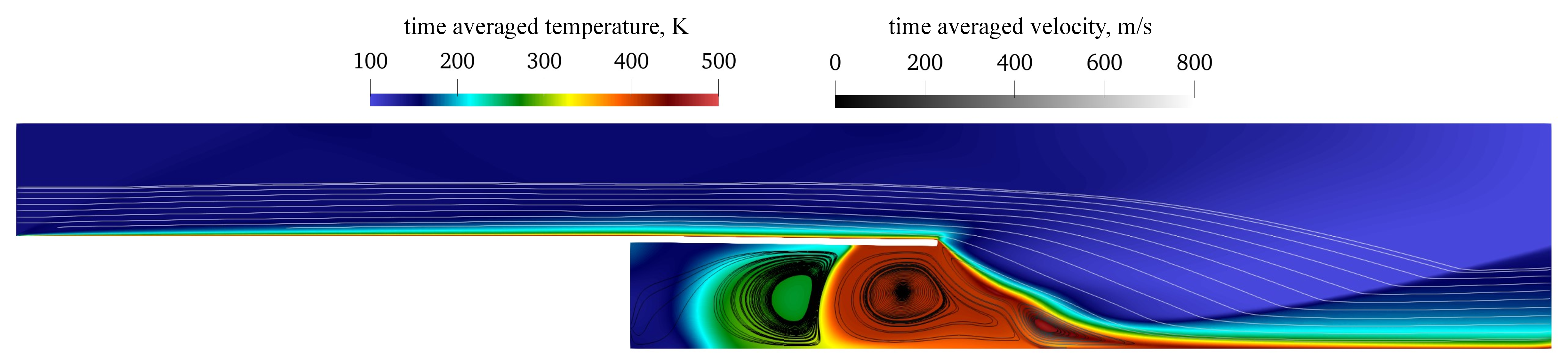}
        \caption{Ma $3.00$}
    \end{subfigure}
    \caption{Time-averaged temperature fields, overlaid with velocity streamlines for fully started conditions.}
    \label{fig:AllMachFullTemps}
\end{figure*}

The shear layer gets attached to the bottom wall for Mach $2.11$ and $3.00$ trapping the vortices, figure \ref{fig:AllMachFullFlows}c, d in an enclosed recirculation region as shown in figure \ref{fig:bojan_schematic}. The structure of the vortices remains relative unchanged between transient, figure \ref{fig:AllMachTransFlows}c, and fully-started, figure \ref{fig:AllMachFullFlows}c, conditions for Mach $2.00$. Additionally, two smaller vortices attached underneath the shear layer and before its reattachment to the bottom wall can be observed. For Mach $3.00$, vortex structures similar to the Mach $2.11$ case are observed, figure \ref{fig:AllMachFullFlows}d, which are localized to the recirculation region, with two smaller vortices attached underneath the shear layer prior to its reattachment. Flow features shown in the schematic, figure \ref{fig:bojan_schematic}, such as the expansion fan, reattachment shock, and recovering boundary layer are observed for Mach $2.11$ and $3.00$. The recovered boundary layer is thinner for the Mach $3.00$ when compared to Mach $2.11$ case, as the larger velocity magnitude after shear layer reattachment results in a larger velocity gradient for the Mach $3.00$ case. These observations are also supported by temperature contours, figure \ref{fig:AllMachFullTemps}. 

The subsonic Mach $0.50$, figure \ref{fig:AllMachFullTemps}a, and transonic Mach $0.95$, figure \ref{fig:AllMachFullTemps}b, cases see an increase in temperature of the flowfield, including the high speed and low speed regions separated by the shear layer as compared to the transient conditions, figure \ref{fig:AllMachTransTemps}a, b. The longer residence time for flow results in increased heating along the bottom wall of the inlet section and the thin plate top surface and a thicker thermal boundary layer. The sharp velocity and temperature gradients across the shear layer are present for fully-started conditions, with temperatures elevated due to viscous dissipation for longer times inside the recirculating vortices for both cases. While the temperature range remains unchanged for both Mach $0.50$ and Mach $0.95$ cases, the distribution is changed due to persistent viscous dissipation and thermal heating of the solid surfaces.

For the supersonic cases, Mach $2.11$ and $3.00$, the thermal boundary layer along the bottom wall of the inlet section experiences a much larger velocity and temperature gradient, resulting in increased temperatures, figure \ref{fig:AllMachFullTemps}a, b. The stratification in temperature contours confirm the presence of the expansion fan the thin plate outer tip, and the reattachment shock further downstream along the bottom wall. Significant temperature increase is observed inside the vortices attached to the underneath of the shear layer and the thin-plate and trapped inside the recirculation region for both Mach $2.11$ and $3.00$. This is due to sustained viscous dissipation and entrapment of the vortices, which lead to increased heating of the fluid inside the recirculation region. Sustained dissipation can lead to excessive heating of the trapped gases inside the recirculation region with very high temperature increase leading to possible thermal failure of the thin-plate. The Mach $3.00$ case has higher temperatures within the vortices inside the recirculation region, as compared to the Mach $2.11$ case. This can be attributed to the higher flow velocity of the Mach $3.00$ case, which has relatively higher energy content leading to higher heating and dissipation rates. The thermal boundary layer for the Mach $2.11$ case is recovers relatively earlier than the Mach $3.00$ case. This is also be attributed to the higher flow velocity for the Mach $3.00$ case which increases the length required for boundary layer recovery. Further increase in flow velocity can lead to extremely high temperatures, leading to ionization and chemical kinetic reactions that can lead to failure. These simulations are hence able to identify hot flow regions and zones for the flow configuration of figure \ref{fig:bojan_schematic}, which can lead to extremely high temperature rises leading to the possibility of solid structure failure. 

%%%%%%%%%%%%%%%%%%%%%%%%%%%%%%%%%%%%%%%%%%%%%%%%%%%%%%%%%%%%%%%%%%%%%%%%%%%%%%%%%%%%%%%%%%%%%%%%%%%
\subsubsection{Panel Deformation and Stresses}
\begin{figure*}[htbp!]
    \centering
    \begin{subfigure}[h]{0.48\textwidth}
        \centering
        \includegraphics[width=\textwidth]{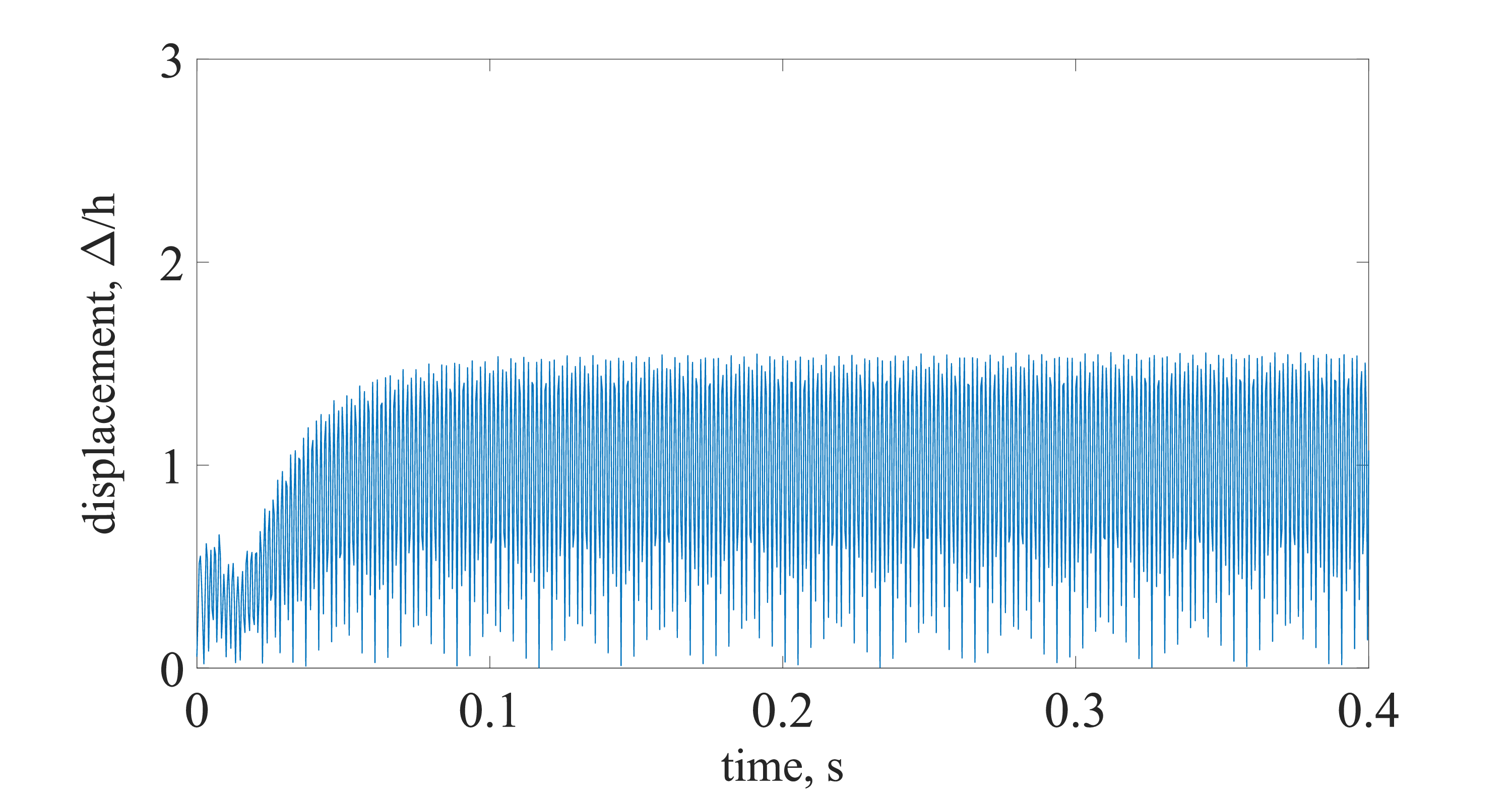}
        \caption{Ma $0.50$}
    \end{subfigure}
    \hspace{2mm}
    \begin{subfigure}[h]{0.48\textwidth}
        \includegraphics[width=\textwidth]{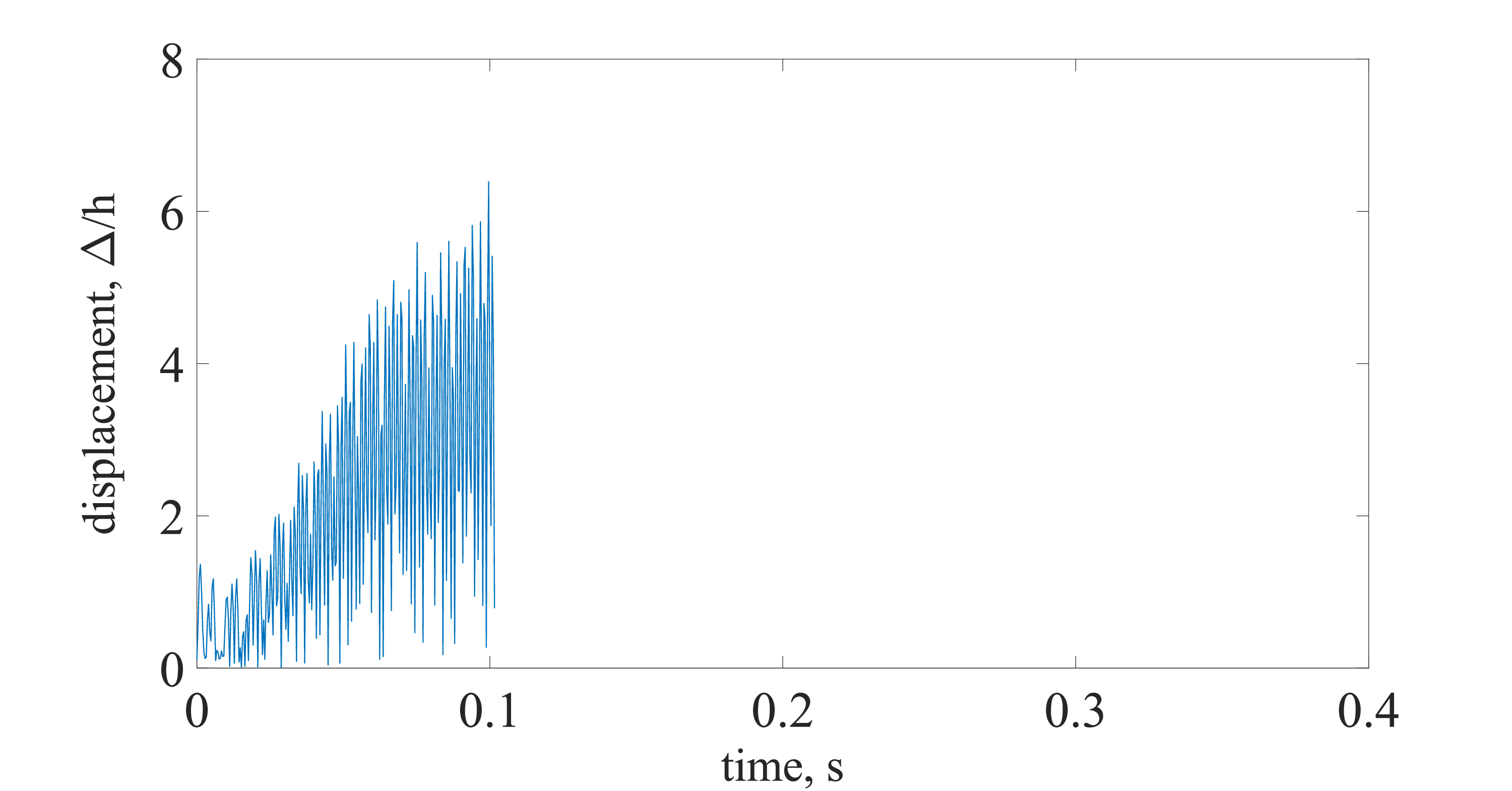}
        \caption{Ma $0.95$}
    \end{subfigure}
    \hspace{2mm}
    \begin{subfigure}[h]{0.48\textwidth}
        \centering
        \includegraphics[width=\textwidth]{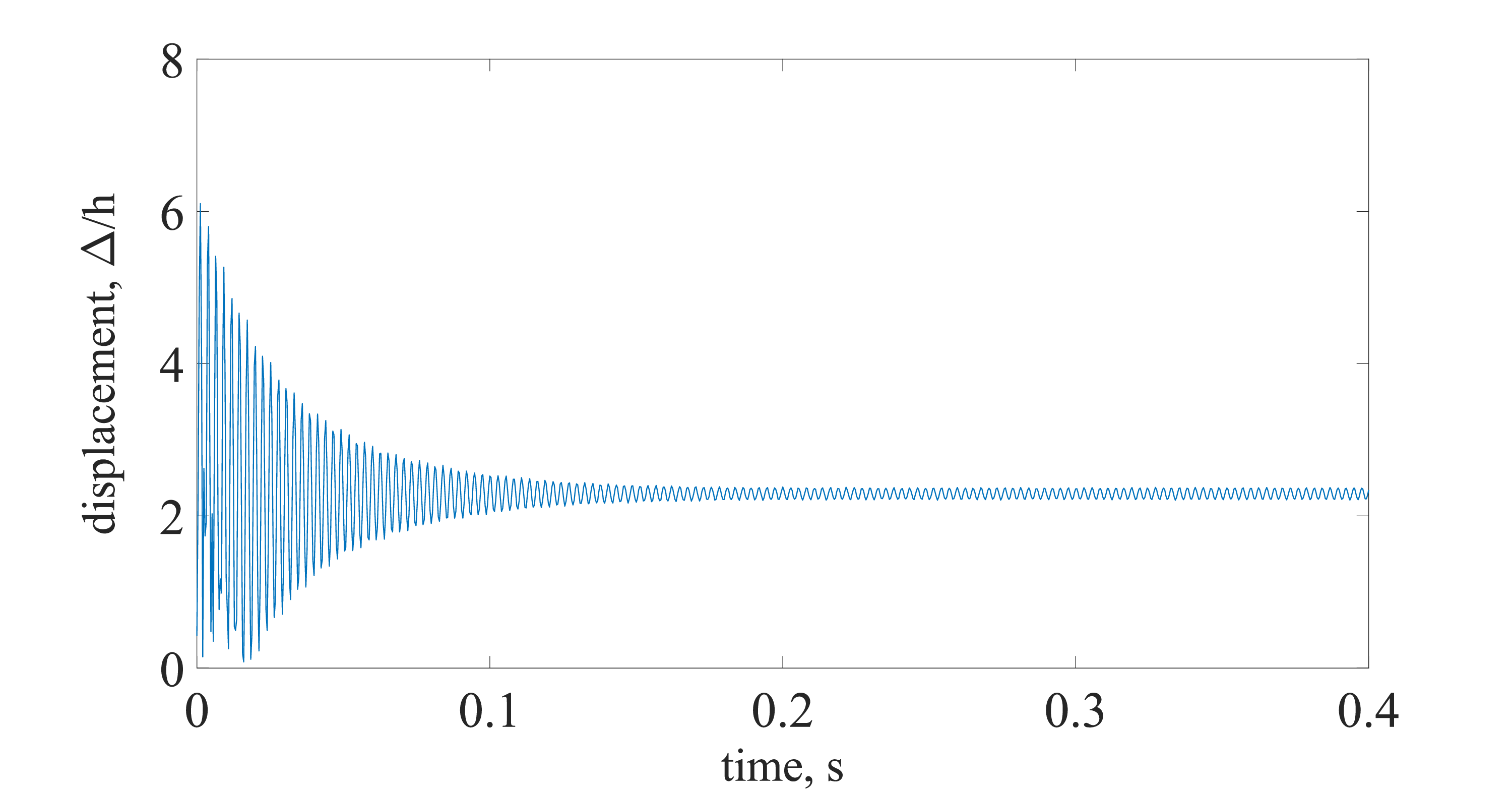}
        \caption{Ma $2.11$}
    \end{subfigure}
    \hspace{2mm}
    \begin{subfigure}[h]{0.48\textwidth}
        \centering
        \includegraphics[width=\textwidth]{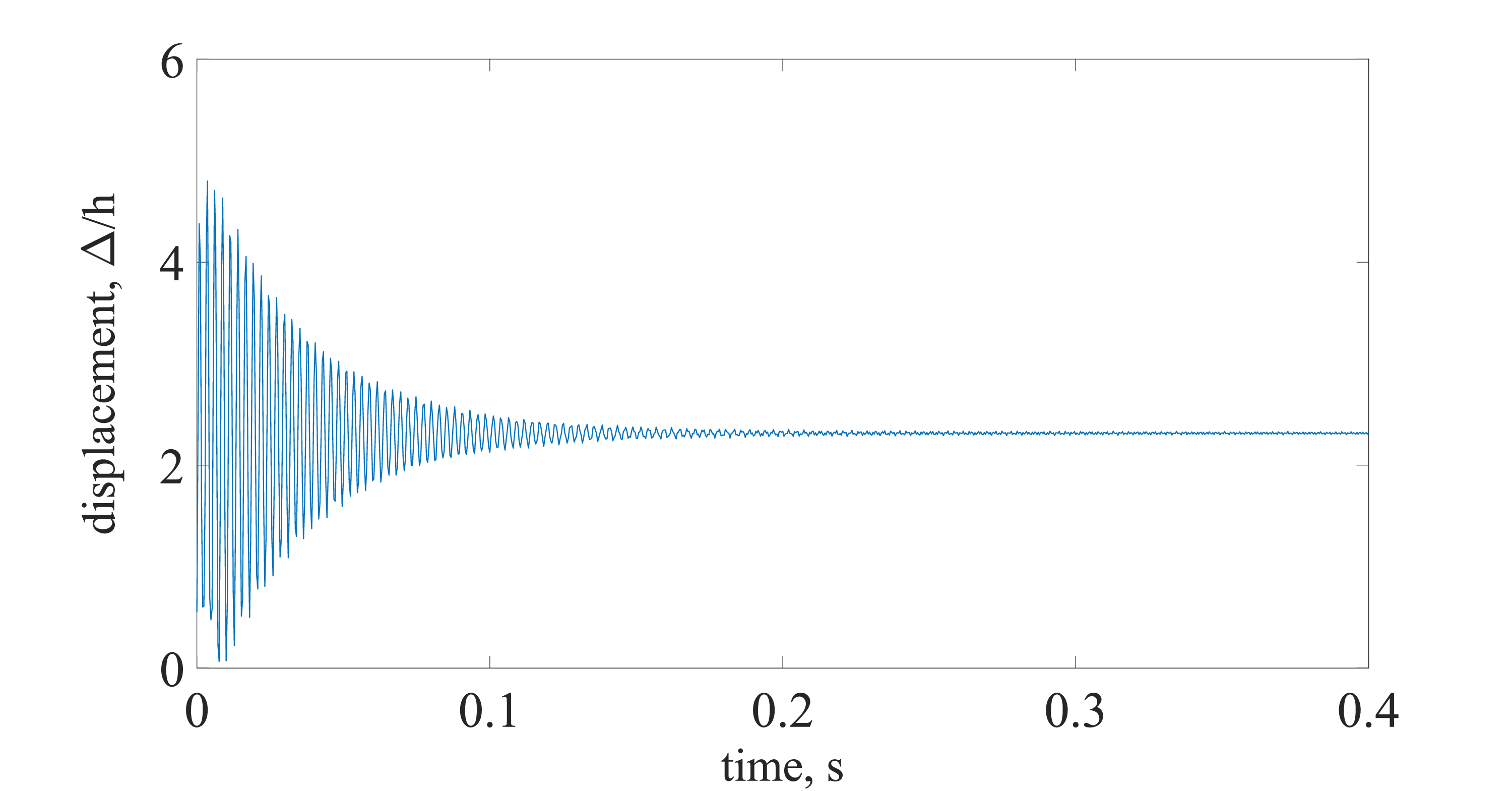}
        \caption{Ma $3.00$}
    \end{subfigure}
    \caption{Displacement-time plots for initial transient conditions for fully started conditions.}
    \label{fig:AllMachFullDisps}
\end{figure*}

The solid behavior is quantified with displacement-time plots to show the impact of  Mach number in figure \ref{fig:AllMachFullDisps}. The four cases exhibit different behavior for displacements with time and this is explored further with FFT analysis, stresses and modal decomposition to obtain dominant modes of vibration for analysis. The amplitude of displacement for the Mach $0.50$ case initially decreases, but around $0.02$ $s$ it starts to increase again to stabilize to a maximum amplitude close to $2$ times the plate thickness, figure \ref{fig:AllMachFullDisps}a, after a time of $0.1$ $s$. When the flow velocity is increased to Mach $0.95$, the amplitude of oscillation of the thin plate steadily increases to $8$ times the plate thickness, figure \ref{fig:AllMachFullDisps}b, after which the excessive deformation leads to divergence of the simulation. This is assumed as structural failure of the thin plate due to resonant behavior. For Mach $2.11$ case, the tip displacement amplitudes rise to around $6$ times the plate thickness, which are then stabilized around $0.2$ $s$, figure \ref{fig:AllMachFullDisps}c to $ \frac{1}{5} $ times the plate thickness. The Mach $3.00$ case shows a similar displacement-time plot, with the oscillations stabilized at smaller magnitudes than the Mach $2.11$ case at $0.2$ $s$, figure \ref{fig:AllMachFullDisps}d. FFT analysis was conducted on the displacement-time series for the stabilized oscillations between $0.2$ $s$ and $0.4$ $s$, for Mach $0.50$, $2.11$ and $3.00$ cases. 

\begin{figure*}[htbp!]
    \centering
    \begin{subfigure}[h]{0.48\textwidth}
        \centering
        \includegraphics[width=\textwidth]{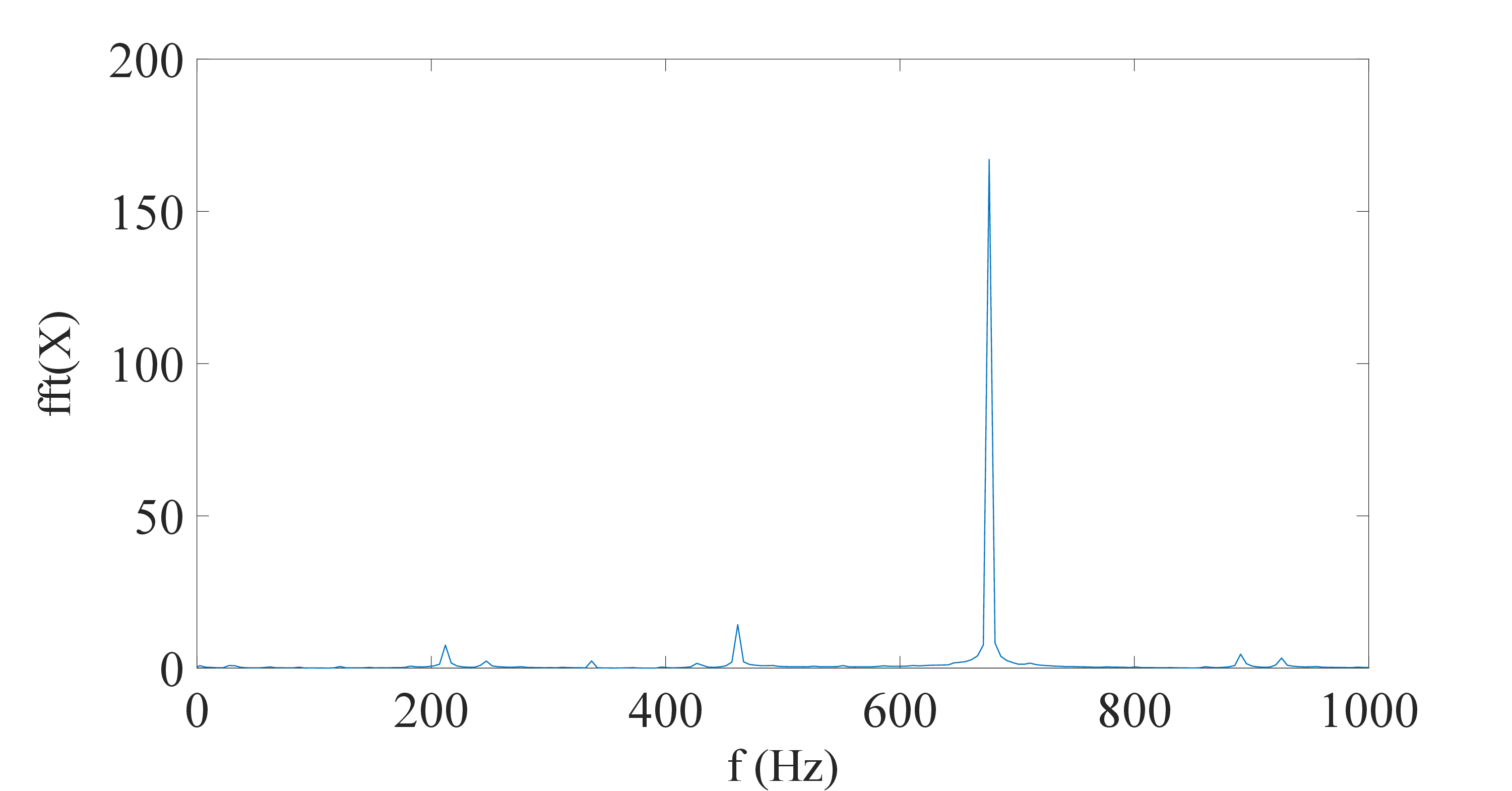}
        \caption{Ma $0.50$}
    \end{subfigure}
    \hspace{2mm}
    \begin{subfigure}[h]{0.48\textwidth}
        \includegraphics[width=\textwidth]{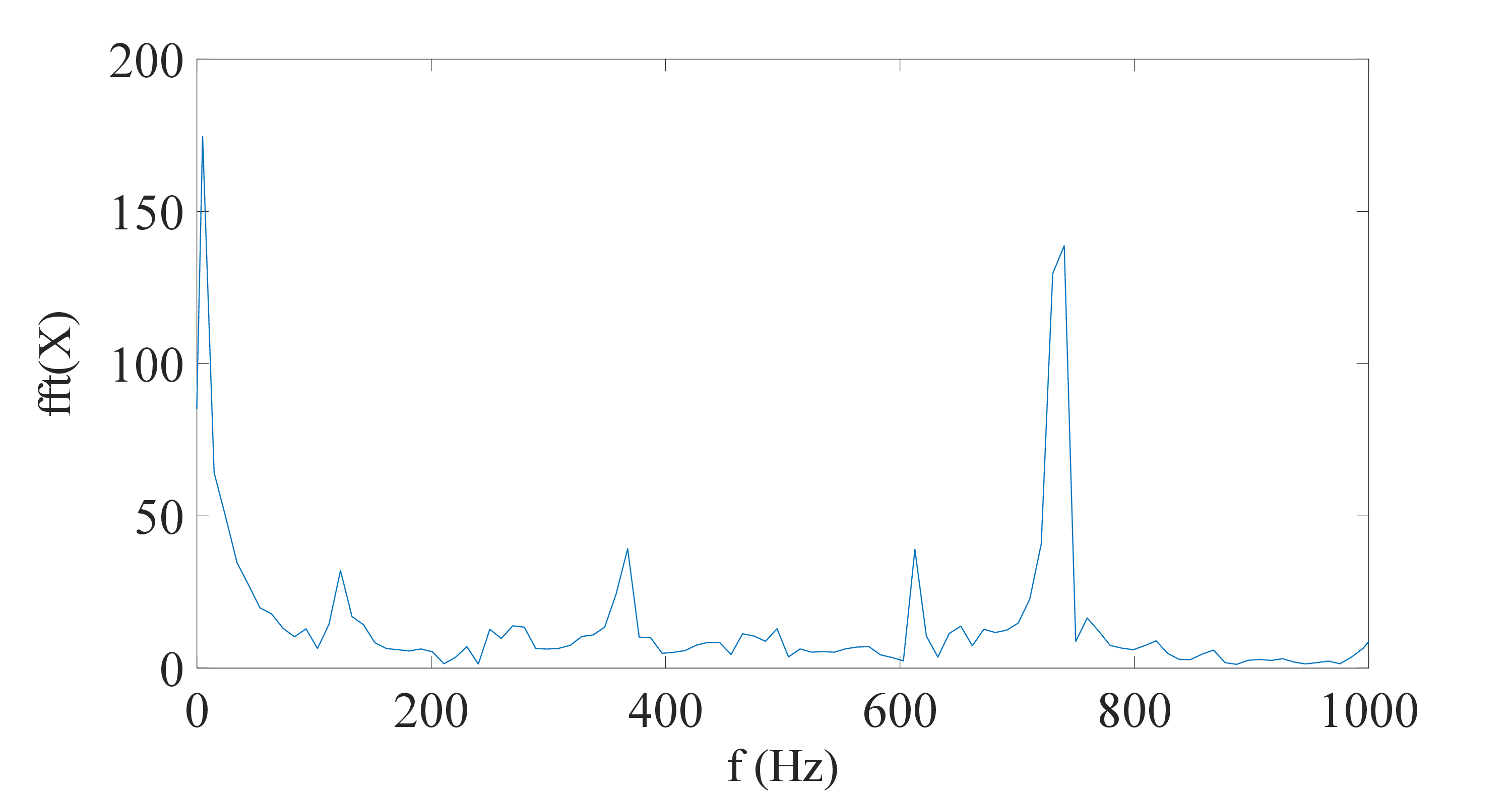}
        \caption{Ma $0.95$}
    \end{subfigure}
    \hspace{2mm}
    \begin{subfigure}[h]{0.48\textwidth}
        \centering
        \includegraphics[width=\textwidth]{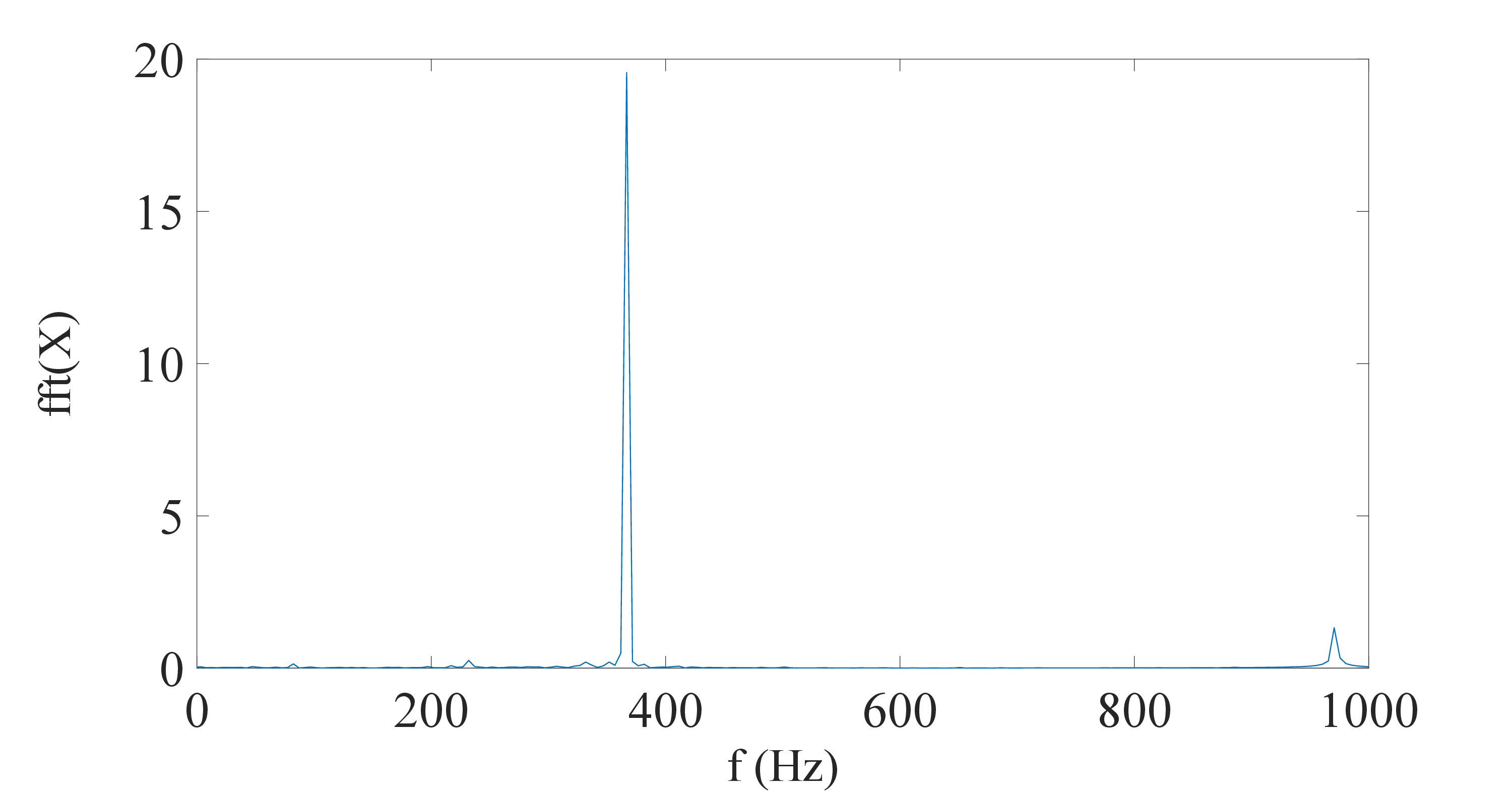}
        \caption{Ma $2.11$}
    \end{subfigure}
    \hspace{2mm}
    \begin{subfigure}[h]{0.48\textwidth}
        \centering
        \includegraphics[width=\textwidth]{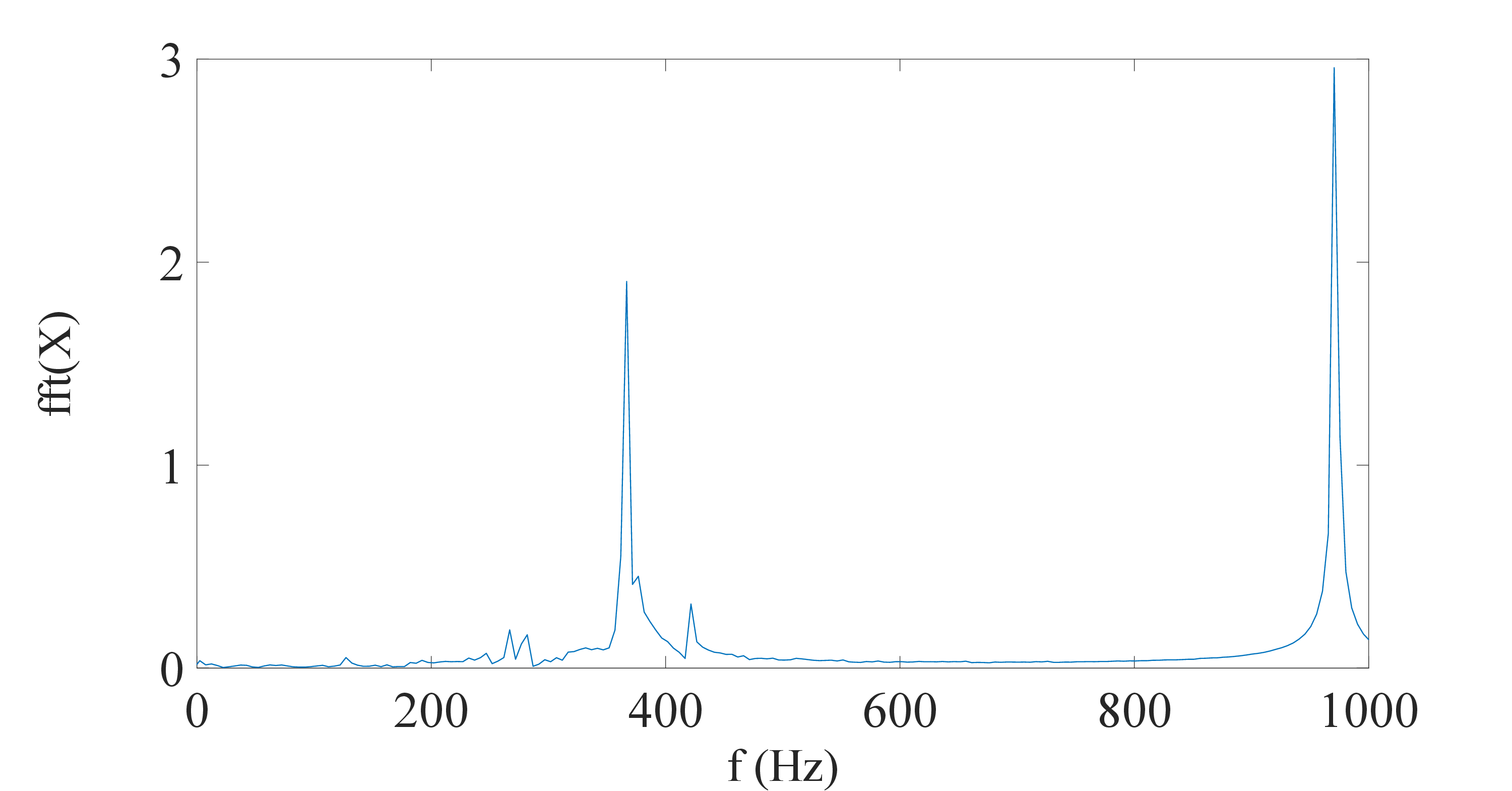}
        \caption{Ma $3.00$}
    \end{subfigure}
    \caption{FFT results for thin panel tip displacements for fully started conditions.}
    \label{fig:AllMachFullFFT}
\end{figure*}

FFT results are presented in figure \ref{fig:AllMachFullFFT}a, c, d for Mach $0.50$, $2.11$ and $3.00$ cases. For the Mach $0.95$ case, FFT was conducted on the displacement-time data without truncation of the first $0.2$ $s$, figure \ref{fig:AllMachFullFFT}b. The peak frequencies are listed in table \ref{tab:fulldispfreqs} and frequencies lower than $300$ $Hz$ are not included in the discussion as this is the lower than the first natural frequency of the thin-plate. Mach $0.50$ and $2.11$ show a single peak at $676$, figure \ref{fig:AllMachFullFFT}a and $367$ $Hz$, figure \ref{fig:AllMachFullFFT}c, respectively, while Mach $3.00$ shows two peaks with the first one at $971$ $Hz$ and a second peak at $367$ $Hz$, figure \ref{fig:AllMachFullFFT}d. This implies that increasing flow Mach number from $2.11$ to $3.00$ leads to a change in thin plate loading that results in it oscillating at a higher frequency. For Mach $0.95$ case, $3$ peaks are observed - at $740$, $613$ and $368$ $Hz$, figure \ref{fig:AllMachFullFFT}b. This suggests that a natural mode of vibration is activated for this case leading to resonance and is followed by failure of the thin plate through excessive deformation by oscillations as shown by the plate tip displacement-time plot, figure \ref{fig:AllMachFullDisps}d. This behavior is observed for transonic flows, where flutter phenomenon leading to resonance and failure has been frequently observed and reported in literature \cite{RN54, RN53, RN52, RN51}.
\begin{table}[htbp!]
\begin{tabular}{ccccc}
    \hline
    Mach & & $1^{st}$ Peak & & $2^{nd}$ Peak \\
     & & Hz & & Hz \\
    \hline \hline
    $0.50$ & & $676$ & & \\
    $0.95$ & & $740$ & & $613$, $368$ \\
    $2.11$ & & $367$ & &  \\
    $3.00$ & & $971$ & & $367$ \\
    \hline
    \end{tabular}
    \caption{Thin-plate oscillation frequencies for fully started conditions.}
    \label{tab:fulldispfreqs}
\end{table}

\begin{figure*}[htbp!]
    \centering
    \begin{subfigure}[h]{0.48\textwidth}
        \centering
        \includegraphics[width=\textwidth]{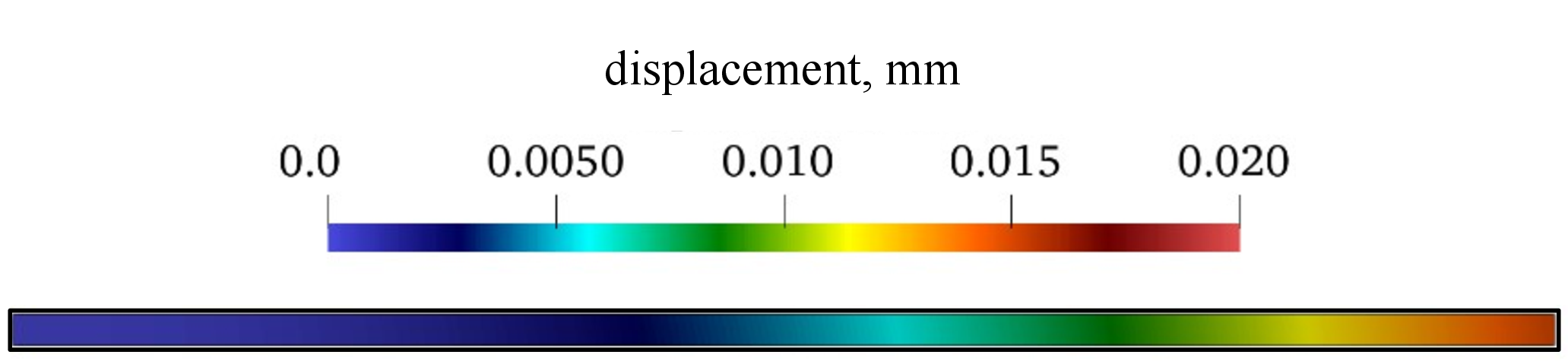}
        \caption{M $0.50$}
    \end{subfigure}
    \hspace{2mm}
    \begin{subfigure}[h]{0.48\textwidth}
        \includegraphics[width=\textwidth]{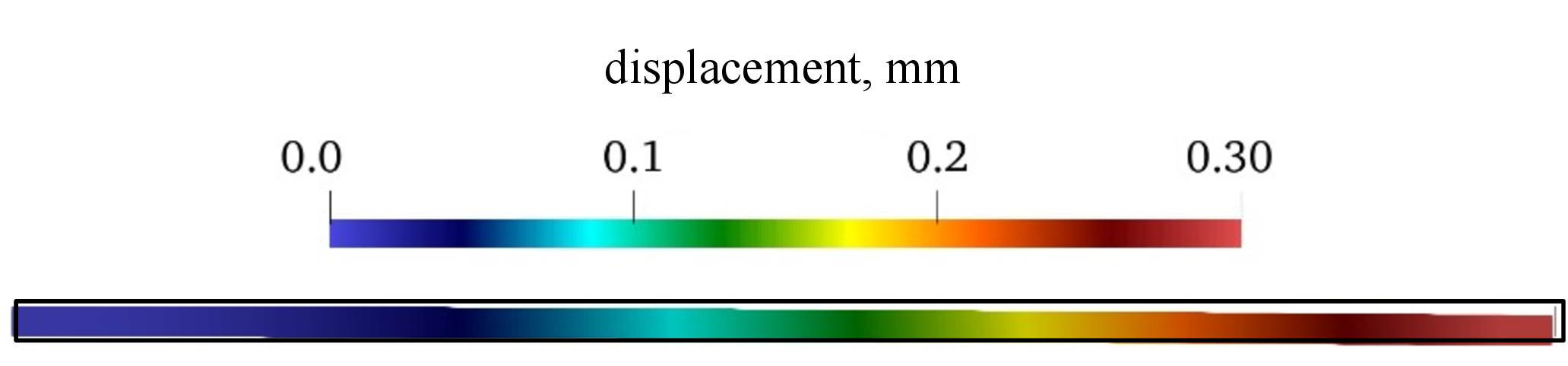}
        \caption{M $0.95$}
    \end{subfigure}
    \hspace{2mm}
    \begin{subfigure}[h]{0.48\textwidth}
        \centering
        \includegraphics[width=\textwidth]{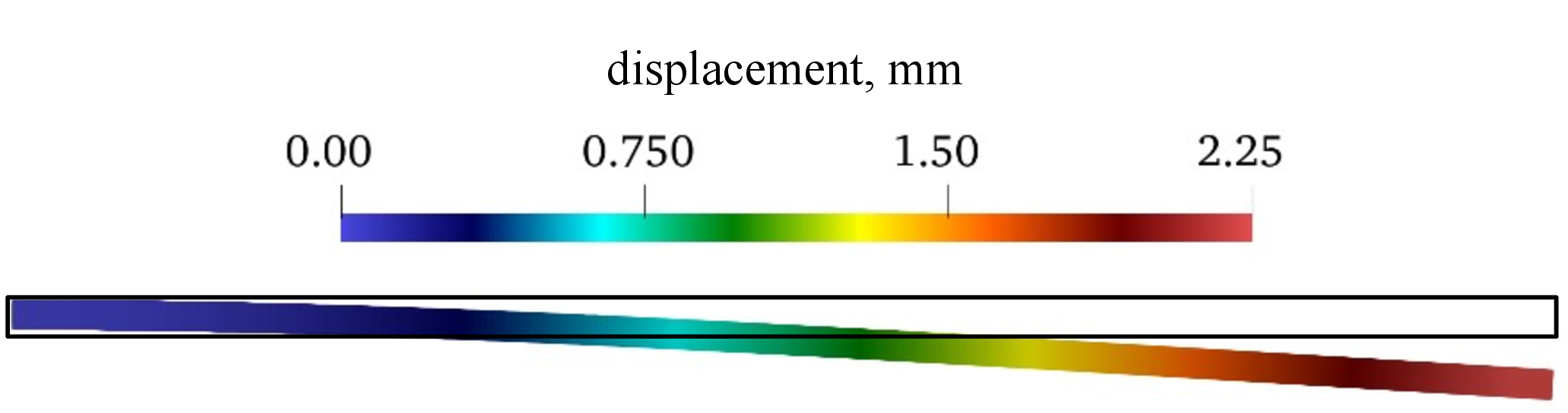}
        \caption{M $2.11$}
    \end{subfigure}
    \hspace{2mm}
    \begin{subfigure}[h]{0.48\textwidth}
        \centering
        \includegraphics[width=\textwidth]{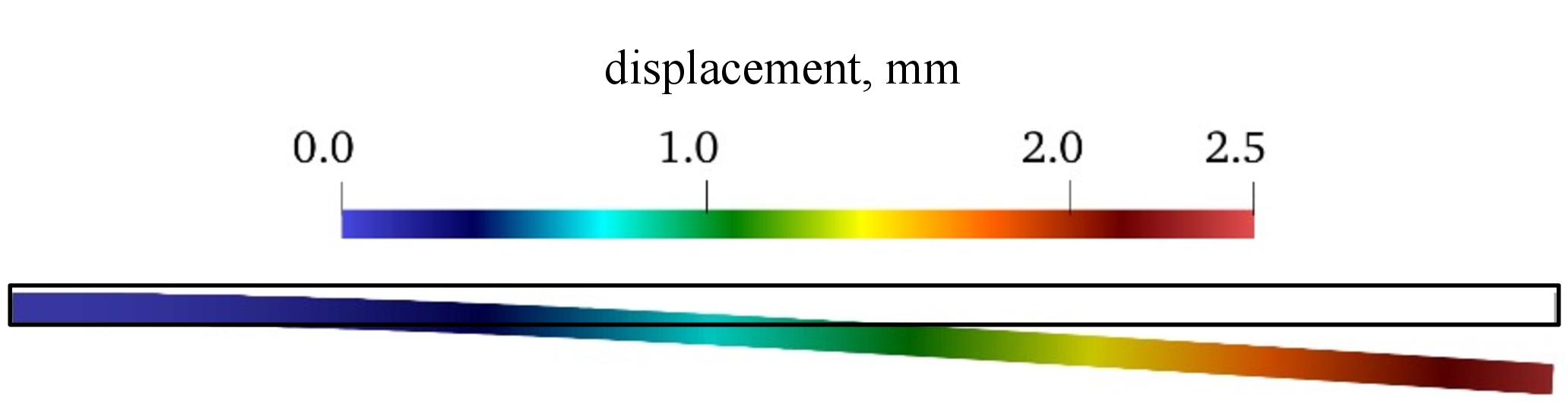}
        \caption{M $3.00$}
    \end{subfigure}
    \caption{Time averaged thin-panel displacements for all Mach numbers. Outline shows the initial undeformed panel.}
    \label{fig:fig:AllMachFullTimeAvgDisps}
\end{figure*}

\begin{figure*}[htbp!]
    \centering
    \begin{subfigure}[h]{0.48\textwidth}
        \centering
        \includegraphics[width=\textwidth]{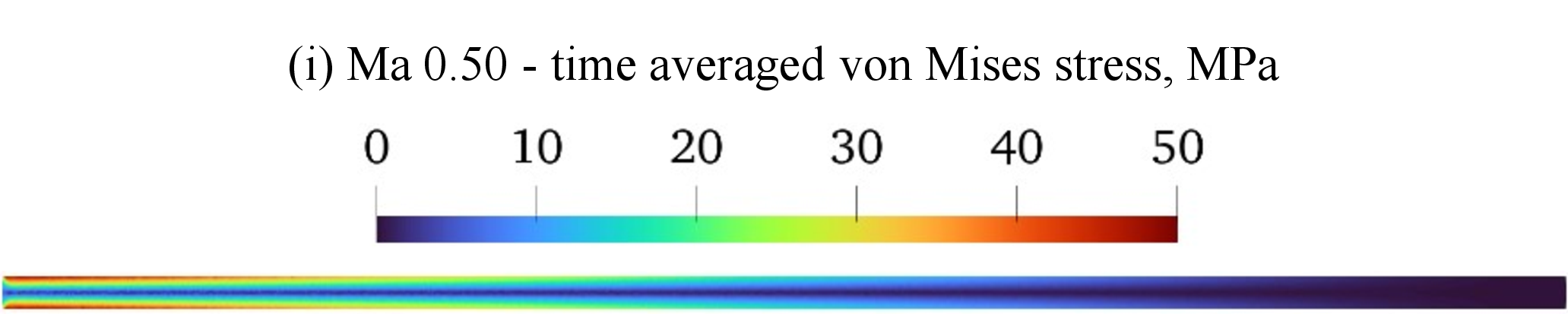}
        \includegraphics[width=\textwidth]{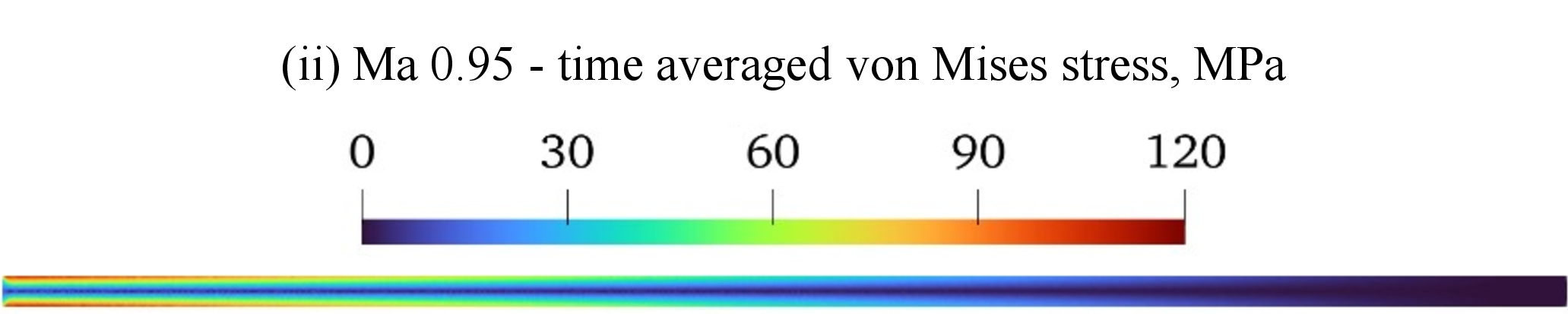}
        \includegraphics[width=\textwidth]{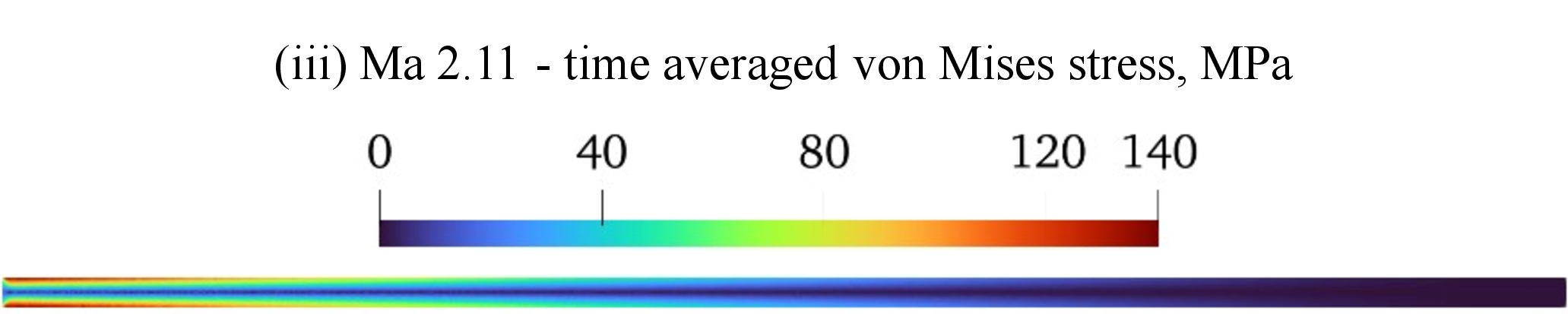}
        \includegraphics[width=\textwidth]{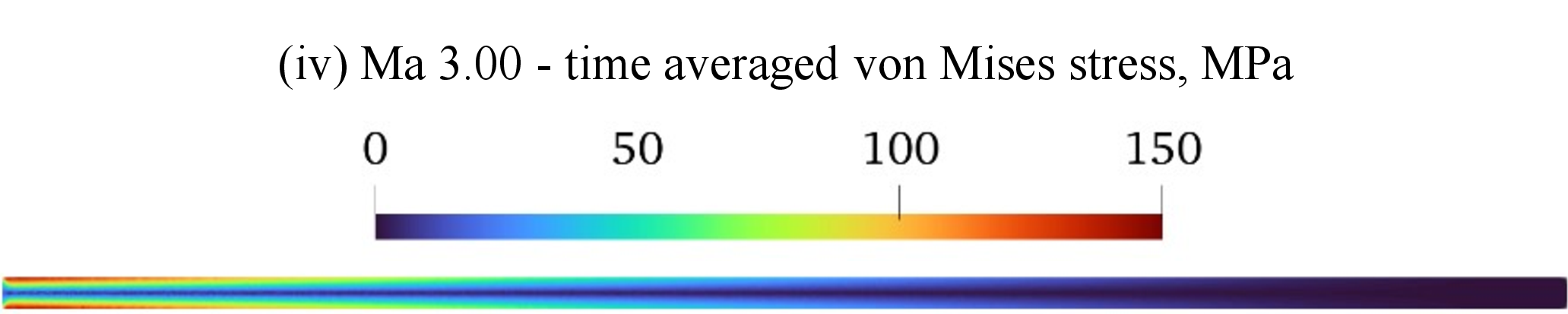}
        \caption{von Mises stress}
    \end{subfigure}
    \hspace{2mm}
    \begin{subfigure}[h]{0.48\textwidth}
        \includegraphics[width=\textwidth]{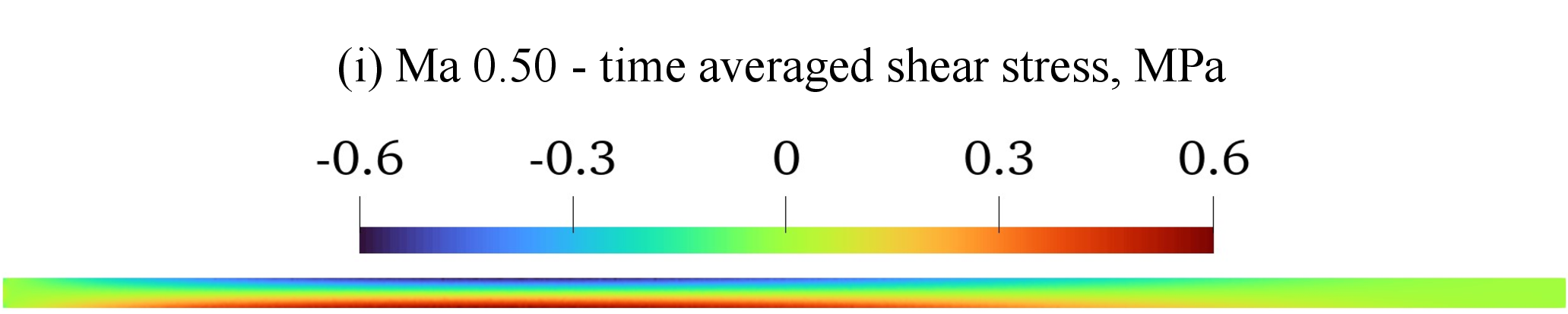}
        \includegraphics[width=\textwidth]{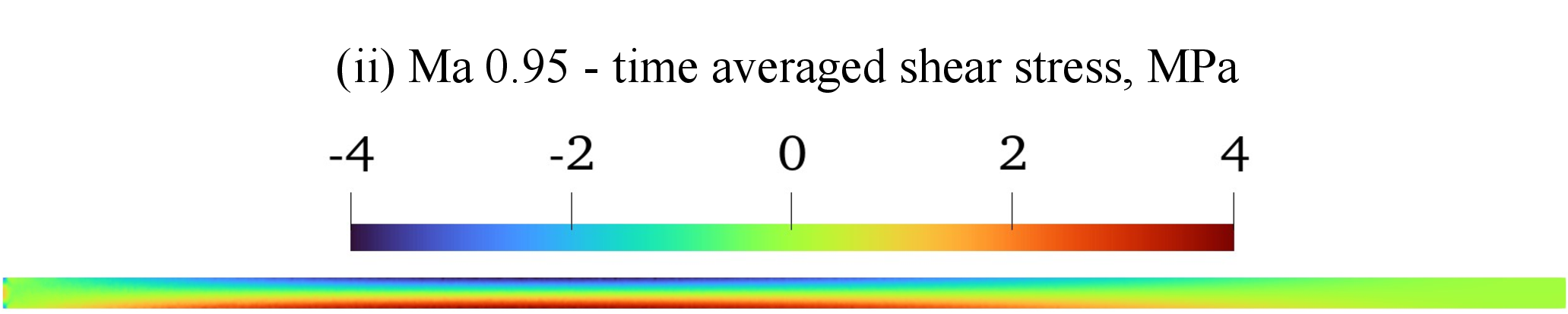}
        \includegraphics[width=\textwidth]{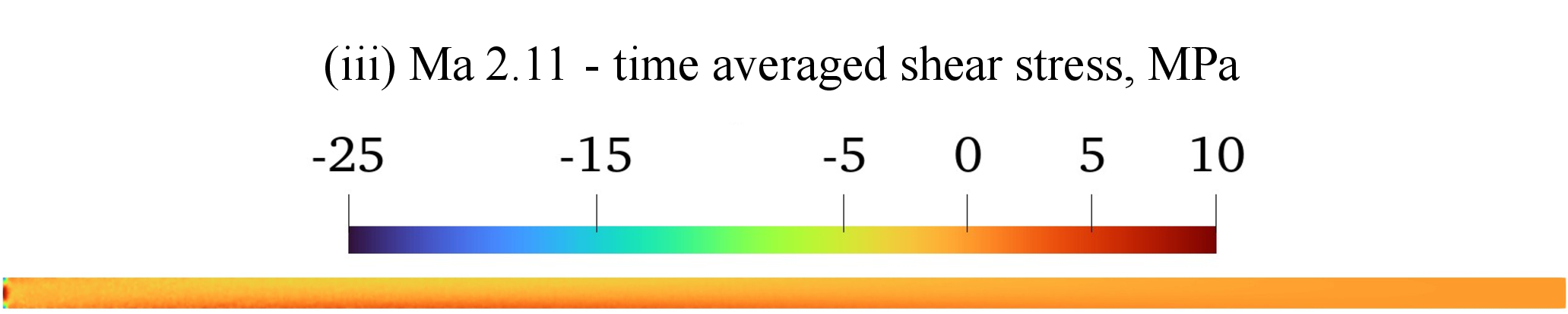}
        \includegraphics[width=\textwidth]{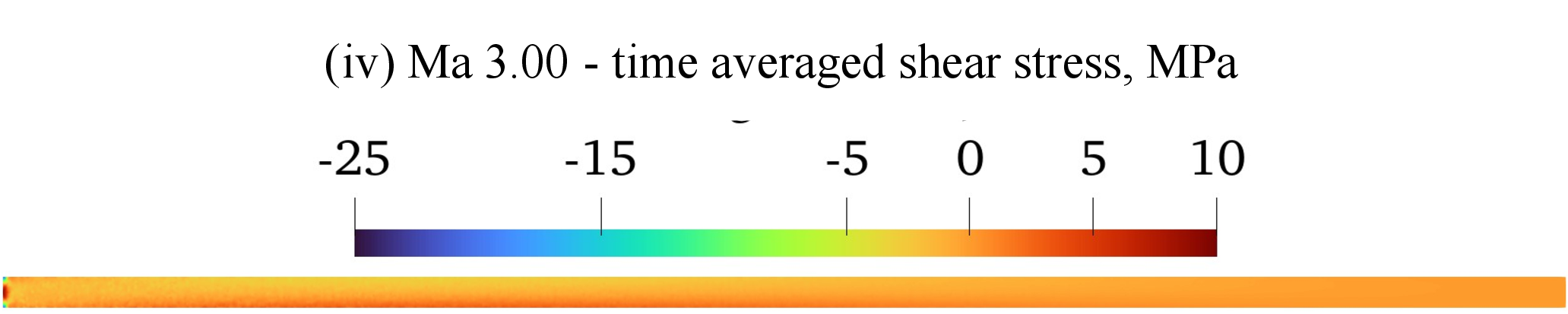}
        \caption{shear stress}
    \end{subfigure}
    \caption{Variation of von Mises and shear stress with Mach number for thin-flexible plate.}
    \label{fig:AllMachFullStresses}
\end{figure*}

Time averaged panel deformations are compared for all four Mach numbers in figure \ref{fig:fig:AllMachFullTimeAvgDisps}. An increase with increasing Mach number is observed. As discussed previously this is due to the higher pressure loading of the panel from the flow at higher Mach numbers. 

von Mises stress for Mach $0.50$, shows an increase in the peak value from $27$, figure \ref{fig:AllMachTransStresses}a: (i), to $50$ $MPa$, figure \ref{fig:AllMachFullStresses}a (i),. The maximum shear stress also increases from $0.25$ $MPa$, figure \ref{fig:AllMachTransStresses}b: (i), to $0.60$, figure \ref{fig:AllMachFullStresses}b (i). This indicates that the plate stresses increase when the simulation is allowed to run for a longer time for the initial-transients to stabilize resulting in the fully-started conditions for Mach $0.50$. However, for Mach $2.11$ and $3.00$, the peak value is reduced from $180$ $MPa$, figure \ref{fig:AllMachTransStresses}a: (iii), (iv), to $140$ $MPa$ for Mach $2.11$, figure \ref{fig:AllMachFullStresses}a: (iii), and to $150$ $MPa$, figure \ref{fig:AllMachFullStresses}a: (iv). This is due to the stabilization of the transient oscillations leading to relaxation in the plate.  The shear stresses show a different trend with the peak value remaining unchanged for Mach $3.00$ at $10$ $MPa$, figure \ref{fig:AllMachFullStresses}b: (iv), but lowering for Mach $2.11$ from $10$ $MPa$ for initial transient conditions, figure \ref{fig:AllMachTransStresses}b: (iii), to $8$ $MPa$ figure \ref{fig:AllMachFullStresses}b: (iii), for fully started conditions. This implies that the thin plate is undergoing relaxation when the initial transient oscillations are stabilized and fully started conditions take over. Mach $0.95$ is a unique case in this regard as the increase in von Mises stress from $45$ $MPa$ for the initial transient conditions, figure \ref{fig:AllMachTransStresses}a: (ii), to $120$ $MPa$ for fully-started conditions, figure \ref{fig:AllMachFullStresses}a: (ii), is comparatively much higher than the other three cases. As the thin-panel is oscillating at a natural frequency, the amplitude of oscillations keeps on consistently increasing as shown in figure \ref{fig:AllMachFullDisps}b and this increase in amplitude results in a significantly large rise in the von Mises stress which leads to large amplitude oscillations and catastrophic failure from excessive deformation for the thin plate.

%% file: 07conclusion.tex
%%%%%%%%%%%%%%%%%%%%%%%%%%%%%%%%%%%%%%%%%%%%%%%%%%%%%%%%%%%%%%%%%%%%%%%%%%%%%%%%%%%%%%%%%%%%%%%%%%%
% Conclusion:
%%%%%%%%%%%%%%%%%%%%%%%%%%%%%%%%%%%%%%%%%%%%%%%%%%%%%%%%%%%%%%%%%%%%%%%%%%%%%%%%%%%%%%%%%%%%%%%%%%%
FSI of a thin-flexible aluminum panel placed in high-speed air flow for initial-transient and fully-started conditions was investigated through high fidelity 2D numerical simulations. 

Grid resolution of the flow domain has a significant impact on the thin-panel displacement, due to the better resolution of the boundary layer with the wall-model. This leads to accurate loading of the thin-plate, with improvements to the resulting plate displacements. 

In this study, the standard logarithmic function was used for resolving the boundary layer flow, however there are other wall models that can significantly change the boundary layer behavior. A comparative study of the various wall models that resolve the boundary layer behavior effecting the thin-plate loading FSI, will be investigated in a future study. 

Initial transient behavior was consistent for flow features and solid behavior with experiments. Hotspots were identified inside the recirculation regions which show significant temperature rise for the supersonic cases with Mach $2.11$ and $3.00$, that could lead to even higher temperatures at those locations with continued high-speed flow.  

The large oscillations due to pressure loading from the high-speed flow, along with the increased thermal loading from friction heating and viscous dissipation can lead to excessive deformation of the thin-plate. These initial transient simulations can be used as indicators for thin-panel failure if the simulation is allowed to run for longer times. 

Running the simulations for times of $400$ $ms$ results in the stabilization of some of the observed transient behavior of the fluid and solid. The subsonic, Mach $0.50$ and transonic Mach $0.95$ cases see a significant rise in the thin-panel oscillation amplitude, with the oscillation amplitude getting stabilized and sustained for the subsonic case. However, for the transonic case, Mach $0.95$, the amplitude steadily increases with solver divergence after $0.1$ $s$ indicating thin-panel failure due to resonant behavior. For the supersonic cases of Mach $2.11$ and $3.00$, the transient oscillations are stabilized with sustained oscillatory behavior which decreases in amplitude from Mach $2.11$ to $3.00$. 

The supersonic Mach number cases see a very significant rise in temperatures in the fluid at the recirculation vortices, due to accumulation of thermal energy from viscous dissipation for a longer time. The high temperature hotspots in the fluid can result in significant thermal loading of the thin-panel which when sustained for longer times can lead to excessive panel deformation and ultimate failure when combined with pressure loading.

%%%%%%%%%%%%%%%%%%%%%%%%%%%%%%%%%%%%%%%%%%%%%%%%%%%%%%%%%%%%%%%%%%%%%%%%%%%%%%%%%%%%%%%%%%%%%%%%%%%